\documentclass[a4paper,11pt]{article}
\usepackage{jheppub} 
\usepackage[utf8]{inputenc}
\usepackage{physics}
\usepackage{caption}
\usepackage{xcolor}
\usepackage{comment}
\usepackage{multirow}
\usepackage{graphics}
\usepackage{float}
\usepackage{cancel}
\usepackage{soul}
\usepackage{cancel}
\usepackage{array}
\usepackage{slashed}
\title{\boldmath\boldmath A Spin on the Bulk Locality of Slightly Broken Higher Spin Theories}

\author{Sachin Jain and Dhruva K.S}

\affiliation{Indian Institute of Science Education and Research,\\ Dr Homi Bhabha Road, Pashan, Pune, India}

\emailAdd{sachin@iiserpune.ac.in}
\emailAdd{k.s.dhruva@students.iiserpune.ac.in}
\abstract{In this paper, we investigate if it is possible to express correlation functions in Large $N$ Chern-Simons (CS) matter theories/ Slightly Broken Higher Spin (SBHS) theories purely in terms of single trace twist conformal blocks (TCBs). For this, we first develop the machinery for spinning TCBs. We do this both by explicitly solving the spinning TCB eigenvalue equation taking into account consistency with the operator product expansion (OPE) and crossing symmetry, and also by employing weight shifting and spin raising operators and acting with them on scalar seeds. Using these results we show that spinning correlators in theories with exact higher spin symmetry can be entirely expressed in terms of single trace TCBs. However, when the higher spin symmetry is slightly broken at large-N, even though the scalar four-point function is given by single-trace TCBs, the spinning correlators in general, are not.
Our results suggest that it may be possible to identify a sub-sector of SBHS theory which has a local bulk dual. 
 }

\begin{document}
\maketitle
\flushbottom
\section{Introduction}
Three dimensional $SU(N)_k$ Chern-Simons (CS) theories coupled to matter in the fundamental representation form a very important class of quantum field theories \cite{Giombi:2011kc,Aharony:2011jz,Aharony:2012nh}. At large N, these theories are examples of slightly broken higher spin (SBHS) theories which are a class of conformal field theories (CFTs) \cite{Giombi:2011kc,Maldacena:2011jn,Maldacena:2012sf}. These theories are interesting from various perspectives such as exhibiting  non-supersymmetric field theory-field theory dualities \cite{Giombi:2011kc,Aharony:2011jz,Aharony:2012nh,Aharony:2012ns}, being holographically dual to a four dimensional parity violating higher spin Vasiliev theory in (A)dS \cite{Sezgin:2002rt,Klebanov:2002ja,Giombi:2009wh,Chang:2012kt} and even enabling several exact computations such as determining the partition function \cite{Giombi:2011kc,Aharony:2011jz,Jain:2012qi,Yokoyama:2012fa,Aharony:2012ns,Jain:2013py,Jain:2013gza,Yokoyama:2013pxa,Minwalla:2015sca,Geracie:2015drf,Wadia:2016zpd,Dey:2018ykx,Dey:2019ihe,Halder:2019foo, Minwalla:2023esg}, the S Matrix \cite{Jain:2014nza,Dandekar:2014era,Inbasekar:2015tsa,Yokoyama:2016sbx,Inbasekar:2020hla,Inbasekar:2017ieo,Inbasekar:2017sqp,Gabai:2022snc} and correlation functions at large $N$ \cite{Giombi:2011rz,Maldacena:2011jn,Maldacena:2012sf,Aharony:2012nh,Bedhotiya_2015,Chowdhury:2017vel,Turiaci:2018nua, Skvortsov:2018uru,Yacoby:2018yvy,Chowdhury:2018uyv,Li:2019twz,Kalloor_2020,Jain:2020rmw,Jain:2020puw,Jain:2021vrv,Jain:2021wyn,Jain2_2021,Jain4_2021,Silva_2021,Gandhi_2022,Gerasimenko:2021sxj,Jain:2021whr,Scalea:2023dpw}. Further, there has also been impressive progress involving super symmetric SBHS theories \cite{Gur-Ari:2015pca,Aharony:2019mbc,Inbasekar:2019wdw, Inbasekar:2019azv, Jain:2022izp}, the Hilbert space structure of the CS matter theories \cite{Minwalla:2022sef,Mehta:2022lgq}, their spectrum \cite{ GurAri:2012is,Takimi:2013zca,Giombi:2016zwa,Charan:2017jyc,Minwalla:2020ysu}, calculations of anomalous dimensions \cite{Gurucharan:2014cva,Giombi:2017rhm,Jain:2019fja}, see also  \cite{Bardeen:2014qua, Bardeen:2014paa,
  Frishman:2014cma, Moshe:2014bja,
 Gur-Ari:2016xff, Sezgin:2017jgm,Jensen:2017bjo,Chattopadhyay:2018wkp, Choudhury:2018iwf, Aitken:2018cvh, Aharony:2018pjn,
  Chattopadhyay:2019lpr,
  Jensen:2019mga, Ghosh:2019sqf, Mishra:2020wos,Gabai_2022,Gabai:2022mya,Skvortsov:2022wzo} for other interesting developments.  Various proposals away from large N have been put forth as well
\cite{Benini:2011mf,Aharony:2015pla,Radicevic:2015yla,Aharony:2015mjs,Radicevic:2016wqn,Karch:2016sxi,Hsin:2016blu,Murugan:2016zal,Seiberg:2016gmd,Giombi:2016ejx,Karch:2016aux,Aharony:2016jvv,Benini:2017dus,Gaiotto:2017tne,Jensen:2017xbs,Gomis:2017ixy,Cordova:2017vab,Benini:2017aed}. \\
  
One of the important quantities to compute in these theories are correlation functions of single trace operators. At the level of three-point functions, a lot of progress has been made. In \cite{Giombi:2011rz}, three-point functions in general three-dimensional CFTs involving higher spin currents contain three pieces, two of which are realized by the free bosonic and free fermionic theories and a third, which is parity violating and not realized by free fields. In \cite{Maldacena:2012sf} it was found that if one imposes in addition, the ward identities due to the slightly broken higher spin symmetry, the coefficients of these structures get determined in terms of only one parameter. This parameter was then related to the Chern-Simons matter theory coupling in \cite{Aharony:2012nh}. Subsequently, in \cite{Skvortsov_2019} it was shown that the parity-violating term in Chiral Higher Spin theory correlator appears from a certain EM duality. In \cite{Bedhotiya_2015,Turiaci:2018nua}, the scalar four point function was determined. Sub-leading corrections in the $\order{1/N}$ expansion to the scalar four-point function were investigated in \cite{Aharony:2018npf}. In \cite{Li:2019twz}, the correlator of one stress tensor with three scalars was computed by solving the slightly broken higher spin ward identities. In \cite{Jain:2020rmw,Jain:2020puw}, (slightly broken) higher spin ward identities in momentum space were considered which greatly simplified the analysis compared to position space. In \cite{Silva_2021}, correlators involving one arbitrary spinning operator and three scalars were also systematically computed both in position space and Mellin space. In \cite{Kalloor_2020}, using Schwinger Dyson equations, an all loop exact answer for some spinning four-point correlators was computed in the light-cone gauge in a very special kinematic regime.\\

An analysis of CFT correlators in momentum space also led to a variety of interesting insights, such as the parity odd structure getting determined in terms of the parity even ones \cite{Jain4_2021} via an \textit{epsilon} transformation. The three-point correlation functions were also considered in spinor helicity variables \cite{Jain:2021vrv,Jain:2021wyn,Jain2_2021} and it was discovered in \cite{Gandhi_2022} that in particular for CS matter theory, that the correlators take an interesting anyonic form which interpolates between the free fermionic and the critical bosonic theory which makes duality manifest. This relation between parity odd and parity even correlators eventually led to the understanding that the SBHS equations for CS matter theories can be mapped to the free fermionic and critical bosonic equations which helped to determine the four and higher point functions in \cite{Jain:2022ajd}.  The authors showed that a general spinning correlator in the SBHS theory can be determined in terms of the free fermionic and critical bosonic theory correlators. In principle, the results of \cite{Jain:2022ajd} render the SBHS theory solved, but it would nevertheless be nice to get explicit expressions for the correlation functions. One possible line of attack is to see if these correlators are given by single trace twist conformal blocks \cite{Alday_2017,Alday1_2017} since these theories do possess a large degeneracy in twist at leading order in $N$.\\

In \cite{Sleight_2018}, the scalar correlator in the free bosonic theory was shown to be a sum of single trace TCBs. In view of the higher spin symmetry, one expects to find a similar story for spinning correlators even though it was not shown explicitly. Finding a decomposition of a correlator into single trace twist conformal blocks also has significant consequences in the dual AdS bulk theories as demonstrated in \cite{Sleight_2018}. Let us now briefly review their argument.\\

Consider the connected part of the correlator $\langle O_1 O_1 O_1 O_1\rangle_{FB}$ in the $O(N)$ free bosonic theory with $O_1=\phi^i\phi^i$ where the scalar field $\phi^i$ is in the fundamental representation of $O(N)$. By the higher spin-CFT duality, this correlator is dual to a four-point scalar amplitude $\mathcal{A}_4$ in the AdS bulk which can further be written as a sum of exchange and contact diagrams as follows,
\\
\includegraphics[width=18 cm]{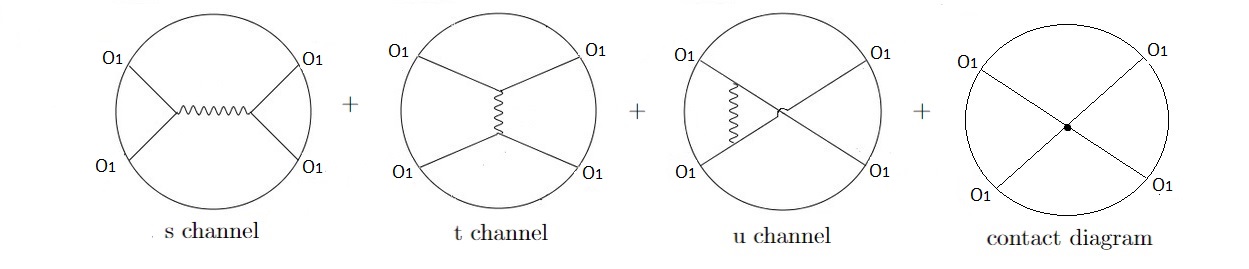}
\\
where the intermediate particle exchanges in s,t and u channels consist of a scalar particle with mass squared $m^2=-2$ and a tower of massless spin $s=2,4,6,\cdots$ gauge bosons where the $s=2$ particle is the graviton.
In the boundary perspective, bulk exchange amplitudes are represented by single trace TCBs, $\mathcal{H}^{(O_1 O_1|\tau|O_1 O_1)}$\footnote{This dictionary is true up to some double trace contributions, see \cite{Sleight_2018} for instance. We shall suppress these contributions as they do not affect the conclusions.}\footnote{The twist $\tau$ of an operator is defined as the difference between its scaling dimension $\Delta$ and spin $s$.}. 
Since all the single trace primary operators in this theory have $\tau=1$, the s channel exchange diagram in the boundary language is  equivalent to a single trace $\tau=1$ s channel TCB, and similarly for the t and u channels. The authors then showed that the full correlator is given by half the sum of the exchange diagrams \footnote{This half factor was first noticed in \cite{Diaz:2006nm}.}, which indicates that the bulk contact diagram is equivalent to minus half the sum of the exchange diagrams thereby revealing the non-locality of the bulk scalar amplitude at the four-point level.\\ 

With the scalar four-point correlator in the free bosonic theory and its bulk dual having been analyzed thoroughly, it is natural to investigate what happens for the spinning case, at higher point levels and also for the free fermionic theory to probe its bulk dual. Further, one can ask what happens if the higher spin symmetry is no longer exact but slightly broken. We investigate all these questions in this paper. Although the scalar TCBs have been developed very thoroughly, literature on spinning TCBs does not exist\footnote{At least to our knowledge.}. Thus, we shall fill this gap and present a way to systematically determine them. Further, based on our results we attempt to classify the set of CFTs whose correlators are completely given by single trace twist conformal blocks. We now provide a rough outline of the paper.
\subsection*{Outline}
In section \ref{sec:TCB}, we systematically determine both scalar and spinning twist conformal blocks, the latter of which we determine in two different ways. Then in section \ref{sec:FBFFtheories}, we probe the bulk locality of the theories with exact higher spin symmetry, that is, the free bosonic and free fermionic theories. In section \ref{sec:SBHSsec}, we do the same but for slightly broken higher spin theories, namely the critical bosonic and CS+matter theories. We then summarize our findings and discuss future directions in section \ref{sec:Discussion}. We also have several appendices that contain technical information and possible generalizations of our results. In appendix \ref{sec:FBFFCBQFoperators}, we provide the expressions for the conserved currents in the free bosonic and free fermionic theories. In appendix \ref{sec:weightshiftingspinraisingoperators}, we provide the explicit expressions for the weight shifting and spin raising operators that we used in the main text to obtain spinning twist conformal blocks from scalar seeds. In appendix \ref{sec:ddimensionalTCB}, we compute scalar twist conformal blocks at the four and five-point level in arbitrary dimensions. In appendix \ref{sec:detailsofsolutionstoTCB}, we explicitly provide the computational details for one of the spinning twist conformal blocks that we computed and also discuss the analysis of its parity odd counterpart. In appendix \ref{sec:HSE}, we show that it is not obvious from just the higher spin equations, that spinning correlators in the free bosonic theory should be given in terms of single trace twist conformal blocks just because the scalar one is. In appendix \ref{sec:misc}, we present a few degeneracy identities that we found useful while computing twist conformal blocks. In appendix \ref{sec:AdSContactdiagrams}, we discuss AdS contact diagrams and their structure. In appendix \ref{sec:OPE}, we present the general form of the OPE between arbitrary operators, both scalar and spinning. In appendix \ref{sec:dDisc}, we discuss the double discontinuity of correlators. In appendix \ref{sec: CBcorrelatorsappendix}, we present some details on the analysis of some critical bosonic theory correlators. Finally, in appendix \ref{TCBvsAmplitudes}, we compare twist conformal blocks to bulk amplitudes and also discuss the argument for the bulk locality for Euclidean AdS higher spin theories.

\section{Twist Conformal Blocks}\label{sec:TCB}
Twist conformal blocks, as we shall review below, re-sum the contribution to a correlator from all intermediate operators of a given twist. They are extremely useful in theories with a large twist degeneracy, such as higher spin theories. In this paper, we shall be concerned with such theories in three dimensions, and in this section, we shall systematically compute TCBs at the point of twist degeneracy, which is at twist equal to one. For a generalization of some of our results to $d$ dimensions, please refer to appendix \ref{sec:ddimensionalTCB}.

\subsection{Scalar Twist Conformal Blocks}\label{sec:scalarTCB}
Following \cite{Alday_2017}, we begin with a four-point function of identical scalars $O_{\Delta}$. The conformal symmetry tells us that it takes the following form:
\begin{align}\label{identicalscalars4pt}
    \langle O_{\Delta}(x_1)O_{\Delta}(x_2)O_{\Delta}(x_3)O_{\Delta}(x_4)\rangle=\frac{1}{x_{12}^{2\Delta}x_{34}^{2\Delta}}f(u,v),
\end{align}
where $u=\frac{x_{12}^2 x_{34}^2}{x_{13}^2 x_{24}^2}$ and $v=\frac{x_{14}^2 x_{23}^2}{x_{13}^2 x_{24}^2}$ are the conformally invariant cross ratios. Crossing symmetry implies that,
\begin{align}\label{fcrossingsym}
    f(u,v)=\bigg(\frac{u}{v}\bigg)^{\Delta}f(v,u).
\end{align}
Further, $f(u,v)$ admits an expansion in terms of conformal blocks,
\begin{align}
    f(u,v)=\sum_{\Delta,l}a_{\Delta,l}f^{(\Delta,l)}(u,v),
\end{align}
where $a_{\Delta,l}$ are the square of the OPE coefficients and $f^{(\Delta,l)}(u,v)$ are the corresponding conformal blocks.
We now introduce the twist $\tau=\Delta-l$ and write the above equation as,
\begin{align}\label{twist}
    f(u,v)=\sum_{\tau,l}a_{\tau,l}f^{(\tau,l)}(u,v).
\end{align}
We then substitute \eqref{twist} into \eqref{identicalscalars4pt} and carry out the sum over $l$ to obtain \footnote{In contrast to \cite{Alday_2017} but like in \cite{Sleight_2018}, we define our twist conformal blocks along with the pre-factor $\frac{1}{x_{12}^{2\Delta}x_{34}^{2\Delta}}$. We do this to make the generalization to the spinning case easier.},
\begin{align}
   \langle O_{\Delta}(x_1)O_{\Delta}(x_2)O_{\Delta}(x_3)O_{\Delta}(x_4)\rangle=\frac{1}{x_{12}^{2\Delta}x_{34}^{2\Delta}}\sum_{\tau,l}a_{\tau,l}f^{(\tau,l)}(u,v)=\sum_{\tau}\mathcal{H}^{(O_\Delta O_\Delta|\tau|O_\Delta O_\Delta)}.
\end{align}
$\mathcal{H}^{(O_\Delta O_\Delta|\tau|O_\Delta O_\Delta)}$ are known as twist conformal blocks (TCBs). $\mathcal{H}^{(O_\Delta O_\Delta|\tau|O_\Delta O_\Delta)}$ encodes the contribution from a given twist $\tau$, and all spins, to the correlator. The crossing symmetry equation \eqref{fcrossingsym} translates into the permutation symmetry under exchange of any of the $x_i$ for the $\mathcal{H}^{(O_\Delta O_\Delta|\tau|O_\Delta O_\Delta)}$. Consistency with the operator product expansion (OPE) also places restrictions on the allowed forms of the TCBs. For example, TCBs which receive contributions from the exchange of twist $\tau=1$ operators obey,
\begin{align}\label{OdeltaOPEanalysis}
    &\lim_{u\to 0,v\to 1}x_{12}^{2\Delta}x_{34}^{2\Delta}\mathcal{H}^{(O_\Delta O_\Delta|1|O_\Delta O_\Delta)}=\lambda_{O_{\Delta}O_{\Delta}\mathbf{I}}+\lambda_{O_{\Delta}O_{\Delta}O_1}\sqrt{u}+\cdots~,\notag\\
    &\lim_{u\to 1,v\to 0}x_{12}^{2\Delta}x_{34}^{2\Delta}\mathcal{H}^{(O_\Delta O_\Delta|1|O_\Delta O_\Delta)}=\lambda_{O_{\Delta}O_{\Delta}\mathbf{I}}\bigg(\frac{u}{v}\bigg)^{\Delta}+\lambda_{O_{\Delta}O_{\Delta}O_1}u^{\Delta}v^{\frac{1}{2}-\Delta}+\cdots~,\notag\\
    &\lim_{u\to \infty,v\to \infty}x_{12}^{2\Delta}x_{34}^{2\Delta}\mathcal{H}^{(O_\Delta O_\Delta|1|O_\Delta O_\Delta)}=\lambda_{O_{\Delta}O_{\Delta}\mathbf{I}}u^{\Delta}+\lambda_{O_{\Delta}O_{\Delta}O_1}u^{\Delta-\frac{1}{2}}+\cdots~.
\end{align}
Finally, the TCBs should also satisfy other consistency conditions such as the bound on Chaos \cite{Maldacena_2016}.\\
Following \cite{Sleight_2018} we now decompose TCBs of a given twist $\tau$ into a sum of s,t and u channel twist $\tau$ contributions\footnote{This is possible since we can do the OPE in the s,t or u channels. Also, in the bulk language, if we take the intermediate operators with twist $\tau$ to be single trace operators, then these correspond to s,t and u channel exchange diagrams with the appropriate particle exchanges. Thus, $\mathcal{H}^{(O_1O_1|\tau|O_1O_1)}$ can be thought of as a quantity representing the total bulk exchange amplitude. The overall factor of $\frac{1}{2}$ in \eqref{Htaufromstu} is purely conventional.},
\begin{align}\label{Htaufromstu}
  \mathcal{H}^{(O_\Delta O_\Delta|\tau|O_\Delta O_\Delta)}=\frac{1}{2}\bigg[ \mathcal{H}^{(O_\Delta O_\Delta|\tau|O_\Delta O_\Delta)}_s+\mathcal{H}^{(O_\Delta O_\Delta|\tau|O_\Delta O_\Delta)}_t+\mathcal{H}^{(O_\Delta O_\Delta|\tau|O_\Delta O_\Delta)}_u\bigg].
\end{align}
 Further, $ \mathcal{H}^{(O_\Delta O_\Delta|\tau|O_\Delta O_\Delta)}_s$ obeys a fourth-order partial differential equation, which we shall henceforth call Alday's s channel equation\footnote{Similarly, the t and u channel TCBs obey $\mathcal{A}_{14} \mathcal{H}^{(O_\Delta O_\Delta|\tau|O_\Delta O_\Delta)}_t=\lambda(\tau) \mathcal{H}^{(O_\Delta O_\Delta|\tau|O_\Delta O_\Delta)}_t$ and $\mathcal{A}_{13} \mathcal{H}^{(O_\Delta O_\Delta|\tau|O_\Delta O_\Delta)}_u=\lambda(\tau) \mathcal{H}^{(O_\Delta O_\Delta|\tau|O_\Delta O_\Delta)}_u$ respectively. }:
\begin{align}\label{schannelAldayequation}
    \mathcal{A}_{12}\mathcal{H}^{(O_\Delta O_\Delta|\tau|O_\Delta O_\Delta)}_s=\lambda(\tau) \mathcal{H}^{(O_\Delta O_\Delta|\tau|O_\Delta O_\Delta)}_s,
\end{align}
where,
\begin{align}\label{AldayOperator}
    &\mathcal{A}_{12}=-\frac{1}{2}\bigg(\mathcal{C}^{(12)}_4-(\mathcal{C}^{(12)}_2)^2-\frac{d(d-1)}{2}\mathcal{C}^{(12)}_2\bigg)-\frac{1}{4}(\mathcal{C}^{(12)}_2)^2+\frac{1}{2}\bigg(d^2-d(2\tau+3)+\tau^2+2\tau+2\bigg)\mathcal{C}^{(12)}_2,\notag\\
    &\lambda(\tau)=-\frac{\tau}{4}(d-\tau-2)(d-\tau)(2d-\tau-2).
\end{align}
     The superscript $(12)$ means that these operators act only on positions $1$ and $2$, for instance, $C_4^{(12)}=C_4^{(1)}+C_4^{(2)}$. $\mathcal{C}_2$ and $\mathcal{C}_4$ are the quadratic and quartic Casimirs of the Euclidean conformal group $SO(d+1,1)$ respectively. These Casimirs take the following simple forms in the $d+2$ dimensional embedding space\footnote{Please refer to say, \cite{Rychkov_2017} for details on the embedding space formalism.}:
\begin{align}\label{casimirs}
    &\mathcal{C}_2=-\frac{1}{2}M_{AB}M^{AB},\notag\\
    &\mathcal{C}_4=\frac{1}{2}M^{AB}M_{BC}M^{CD}M_{DA}.
\end{align}
where,
\begin{align}\label{scalarMAB}
    M_{AB}=X_A\frac{\partial}{\partial X^B}-X_B\frac{\partial}{\partial X^A},
\end{align}
is the generator of Lorentz transformations in the embedding space. \\

Armed with this knowledge, we now proceed to explicitly determine the scalar TCBs focussing our attention on $\tau=1$ twist conformal blocks\footnote{This is because the theories that we study in the subsequent sections have a large degeneracy in twist, containing infinitely many $\tau=1$ operators. The methods we develop however are easily generalized to study TCBs of arbitrary twist.}. For $d=3$ and $\tau=1$, the eigenvalue $\lambda(\tau)$ in \eqref{schannelAldayequation} vanishes and Alday's s channel equation thus reads\footnote{We suppress the $\tau=1$ superscript henceforth as our focus shall solely be on them.},
\begin{align}\label{3dAldayEquation}
    \mathcal{A}_{12} \mathcal{H}_s^{(O_\Delta O_\Delta|1|O_\Delta O_\Delta)}=0.
\end{align}
After obtaining the $t$ and $u$ channel TCBs by similar means we can then obtain the $\tau=1$ twist conformal block $ \mathcal{H}^{(O_\Delta O_\Delta|1|O_\Delta O_\Delta)}$ using \eqref{Htaufromstu}.

Our method to determine the identical scalar TCBs is given by the following steps:\\

\textcolor{blue}{Step 1:} Write an ansatz for $ \mathcal{H}^{(O_\Delta O_\Delta|1|O_\Delta O_\Delta)}$ that is conformally invariant and choose a half-integer polynomial form for the function of the cross ratios. Restrict this ansatz by taking into account its consistency with the OPE \eqref{OdeltaOPEanalysis} and permutation symmetry.\\ 
 
\textcolor{blue}{Step 2:} Apply Alday's s channel operator $\mathcal{A}_{12}$ \eqref{AldayOperator} on the ansatz for $ \mathcal{H}^{(O_\Delta O_\Delta|1|O_\Delta O_\Delta)}$. The part of the ansatz that satisfies \eqref{3dAldayEquation} is the s channel TCB $ \mathcal{H}_s^{(O_\Delta O_\Delta|1|O_\Delta O_\Delta)}$. Also, as is appropriate, impose $(1\leftrightarrow 2)$ and $(3\leftrightarrow 4)$ symmetry for $\mathcal{H}_s^{(O_\Delta O_\Delta|1|O_\Delta O_\Delta)}$.\\

\textcolor{blue}{Step 3}: Similarly, obtain the t and u channel TCBs and add all of them up as in \eqref{Htaufromstu} to obtain $\mathcal{H}^{(O_\Delta O_\Delta|1|O_\Delta O_\Delta)}$.\\

Let us now begin with the $\Delta=1$ case.

\subsection*{$\mathbf{\mathcal{H}^{(O_1 O_1|1|O_1 O_1)}}$}
We begin with the following half-integer polynomial ansatz,
\begin{align}\label{fuvansatz}
   \mathcal{H}^{(O_1 O_1|1|O_1 O_1)}=\frac{1}{x_{12}^2 x_{34}^2}\sum_{i,j\in\mathbb{Z}/2} c_{ij}u^{i}v^{j}.
\end{align}
 We then restrict the $c_{ij}$ further by demanding consistency with the OPE. The exchange of $O_1$ which is the leading contribution tells us that\footnote{Ignoring the contribution of the identity operator which contributes to the disconnected part of the correlator.},
\begin{align}\label{OPEdelta1}
    &\lim_{u\to 0,v\to 1}x_{12}^2 x_{34}^2 \mathcal{H}^{(O_1 O_1|1|O_1 O_1)}\sim \sqrt{u}+\cdots~,\notag\\
    &\lim_{u\to 1,v\to 0}x_{12}^2 x_{34}^2 \mathcal{H}^{(O_1 O_1|1|O_1 O_1)}\sim\sqrt{\frac{u}{v}}+\cdots~,\notag\\&
    \lim_{u\to\infty,v\to\infty}x_{12}^2  x_{34}^2 \mathcal{H}^{(O_1 O_1|1|O_1 O_1)}\sim \frac{u}{\sqrt{v}}+\cdots~.
\end{align}
This reduces our ansatz \eqref{fuvansatz} to the following expression\footnote{More precisely, this is the unique half integer polynomial expression consistent with \eqref{OPEdelta1}. },
\begin{align}\label{Huvupdated}
\mathcal{H}^{(O_1 O_1|1|O_1 O_1)}=\frac{1}{x_{12}^2 x_{34}^2}\bigg(c_1\sqrt{u}+c_2 \sqrt{\frac{u}{v}}+c_3\frac{u}{\sqrt{v}}\bigg).
\end{align}
Further, imposing permutation symmetry fixes $c_1=c_2=c_3$ thus yielding,
\begin{align}\label{Huvfixed}
   \mathcal{H}^{(O_1 O_1|1|O_1 O_1)}=\frac{c_1}{x_{12}^2 x_{34}^2}\bigg(\sqrt{u}+ \sqrt{\frac{u}{v}}+\frac{u}{\sqrt{v}}\bigg).
\end{align}
We see that to fix this TCB, we did not even have to solve Alday's equation \eqref{3dAldayEquation}. However, we shall now show that it can indeed be written as a sum of s,t and u channel TCBs. We do this by applying Alday's s channel operator $\mathcal{A}_{12}$ \eqref{AldayOperator} on \eqref{Huvupdated}. We find that for Alday's s channel equation \eqref{3dAldayEquation} to be satisfied we require $c_3=0$ whereas $c_1$ and $c_2$ can be arbitrary. However, since we are looking at TCBs of identical scalars, we must also impose a $(1\leftrightarrow 2)$ symmetry on the s channel TCB. This then enforces $c_2=c_1$. Ultimately, we are left with the following expression for the s channel TCB,
\begin{align}\label{HsDelta1}
    \mathcal{H}^{(O_1 O_1|1|O_1 O_1)}_s=\frac{c_1}{x_{12}^2 x_{34}^2}\bigg(\sqrt{u}+\sqrt{\frac{u}{v}}\bigg).
\end{align}
Similarly, we can identify the t and u channel TCBs but due to the permutation symmetry of the correlator, we can also obtain them by merely performing $(2\leftrightarrow 4)$ and $(2\leftrightarrow 3)$ exchanges respectively on the s channel TCB \eqref{HsDelta1}. The results are,
\begin{align}
    &\mathcal{H}^{(O_1 O_1|1|O_1 O_1)}_t=\frac{c_1}{x_{14}^2 x_{23}^2}\bigg(\sqrt{v}+\sqrt{\frac{v}{u}}\bigg),\notag\\
    &\mathcal{H}^{(O_1 O_1|1|O_1 O_1)}_u=\frac{c_1}{x_{13}^2 x_{24}^2}\bigg(\sqrt{\frac{1}{u}}+\sqrt{\frac{1}{v}}\bigg).
\end{align}
Finally, adding up the s,t and u channel TCBs like in \eqref{Htaufromstu} we obtain the following result,
\begin{align}\label{O1O1O1O1TCB}
&\mathcal{H}^{(O_1 O_1|1|O_1 O_1)}=\frac{1}{2}\bigg[\mathcal{H}^{(O_1 O_1|1|O_1 O_1)}_s+\mathcal{H}^{(O_1 O_1|1|O_1 O_1)}_t+\mathcal{H}^{(O_1 O_1|1|O_1 O_1)}_u\bigg]\notag\\&=\frac{c_1}{x_{12}^2 x_{34}^2}\bigg(\sqrt{u}+\sqrt{\frac{u}{v}}+\frac{u}{\sqrt{v}}\bigg).
\end{align}
which exactly matches with \eqref{Huvfixed}.
\subsection*{$\mathbf{\mathcal{H}^{(O_2 O_2|1|O_2 O_2)}}$}
Carrying out the algorithm for the $\Delta=2$ case outputs the following unique s channel TCB,
\begin{align}\label{O2O2O2O2sTCB}
    \mathcal{H}^{(O_2 O_2|1|O_2 O_2)}_s=\frac{c_2\sqrt{u}}{x_{12}^4 x_{34}^4 v^{3/2}}\bigg((1+\sqrt{v})v-(1+v^{5/2}-u(1+v^{3/2}))\bigg),
\end{align}
which solves the $\tau=1$ Alday's s channel equation \eqref{3dAldayEquation}. After obtaining the t and u channel TCBs and using \eqref{Htaufromstu} we get,
\begin{align}\label{O2stuTCB}
&\mathcal{H}^{(O_2 O_2|1|O_2 O_2)}=\frac{1}{2}\bigg[\mathcal{H}^{(O_2 O_2|1|O_2 O_2)}_s+\mathcal{H}^{(O_2 O_2|1|O_2 O_2)}_t+\mathcal{H}^{(O_2 O_2|1|O_2 O_2)}_u\bigg]\notag\\&=c_2\frac{\sqrt{u}}{x_{12}^4 x_{34}^4v^{3/2}}\bigg(-1-u^{5/2}+v-(v-1)v^{3/2}+u^{3/2}(1+v)+u(1+v^{3/2})\bigg).
\end{align}
It is interesting to note that this expression is consistent with the exchange of the stress tensor rather than a $\Delta=1$ scalar.\footnote{See section \ref{sec:FFtheory} for more discussion on this.}
\subsection*{$\mathbf{\mathcal{H}^{(O_\Delta O_\Delta|1|O_\Delta O_\Delta)}}$}
We now investigate $\tau=1$ TCBs for identical scalars with scaling dimension $\Delta$. We find that our results can be split into two cases, one of which is proportional to the $\Delta=1$ result \eqref{O1O1O1O1TCB} whilst the other is proportional to the $\Delta=2$ result \eqref{O2stuTCB}.\\

For $\Delta<2$, we obtain the following unique s channel TCB,
\begin{align}
    \mathcal{H}^{(O_\Delta O_\Delta|1|O_\Delta O_\Delta)}_s=\frac{c_{\Delta}}{x_{12}^{2\Delta}x_{34}^{2\Delta}}\bigg(\sqrt{u}+\sqrt{\frac{u}{v}}\bigg),
\end{align}
 and after obtaining the t and u channel TCBs by the appropriate exchanges, we obtain via \eqref{Htaufromstu},
 \begin{align}
     \mathcal{H}^{(O_\Delta O_\Delta|1|O_\Delta O_\Delta)}=\frac{c_{\Delta}}{x_{12}^{2\Delta}x_{34}^{2\Delta}}\bigg(\sqrt{u}+\sqrt{\frac{u}{v}}+\frac{u}{\sqrt{v}}\bigg),
 \end{align}
which is indeed proportional to the $\Delta=1$ expression \eqref{O1O1O1O1TCB}.

For $\Delta\ge 2$, we get the following unique s channel TCB:
\begin{align}
    \mathcal{H}^{(O_\Delta O_\Delta|1|O_\Delta O_\Delta)}=\frac{c_{\Delta}\sqrt{u}}{x_{12}^{2\Delta}x_{34}^{2\Delta} v^{3/2}}\bigg((1+\sqrt{v})v-(1+v^{5/2}-u(1+v^{3/2}))\bigg),
\end{align}
and yet again after obtaining the t and u channel TCBs and using \eqref{Htaufromstu} we obtain,
\begin{align}
    &\mathcal{H}^{(O_\Delta O_\Delta|1|O_\Delta O_\Delta)}=c_{\Delta}\frac{\sqrt{u}}{x_{12}^{2\Delta}x_{34}^{2\Delta}v^{3/2}}\bigg(-1-u^{5/2}+v-(v-1)v^{3/2}+u^{3/2}(1+v)+u(1+v^{3/2})\bigg),
\end{align}
which is indeed proportional to the $\Delta=2$ TCB \eqref{O2stuTCB}.
\subsection{Spinning Twist Conformal Blocks}\label{sec:spinningTCB}
As far as we are aware, only scalar TCBs have been considered so far in the literature \cite{Alday_2017,Alday1_2017}. The spinning case has not been considered yet. An arbitrary spinning CFT four-point function can be written as \cite{Costa_2011},
\begin{align}
    \langle J_{s_1}(x_1,z_1)J_{s_2}(x_2,z_2)J_{s_3}(x_3,z_3)J_{s_4}(x_4,z_4)\rangle=\frac{(\frac{x_{24}}{x_{14}})^{\sigma_1-\sigma_2}(\frac{x_{14}}{x_{13}})^{\sigma_3-\sigma_4} }{x_{12}^{\sigma_1+\sigma_2}x_{34}^{\sigma_3+\sigma_4}}\sum_{k}f_k(u,v)\mathbf{Q}^{(k)}(\{x_i,z_i\}),
\end{align}
where the $z_i$ are auxiliary null polarization vectors and $\sigma_i=\Delta_i+s_i$. The $\mathbf{Q}^{(k)}$ are conformally invariant tensor structures that serve as building blocks to spinning correlators. One can now expand $f_k(u,v)$ in terms of spinning conformal blocks, thus obtaining,
\begin{align}
     \langle J_{s_1}(x_1,z_1)J_{s_2}(x_2,z_2)J_{s_3}(x_3,z_3)J_{s_4}(x_4,z_4)\rangle=\frac{(\frac{x_{24}}{x_{14}})^{\sigma_1-\sigma_2}(\frac{x_{14}}{x_{13}})^{\sigma_3-\sigma_4} }{x_{12}^{\sigma_1+\sigma_2}x_{34}^{\sigma_3+\sigma_4}}\sum_{k}\sum_{\Delta,l}f^{\Delta,l}_k(u,v)\mathbf{Q}^{(k)}(\{x_i,z_i\}).
\end{align}
We then eliminate $\Delta$ in favour of $\tau$ using $\Delta=\tau+l$ and perform the sum over $l$ to obtain,
\begin{align}\label{spinningTCB}
    \langle J_{s_1}(x_1,z_1)J_{s_2}(x_2,z_2)J_{s_3}(x_3,z_3)J_{s_4}(x_4,z_4)\rangle=\sum_{\tau} \mathcal{H}^{\langle J_{s_1}J_{s_2}J_{s_3}J_{s_4}\rangle_{(\tau)}},
\end{align}
where the TCBs are defined via,
\begin{align}
    \mathcal{H}^{\langle J_{s_1}J_{s_2}J_{s_3}J_{s_4}\rangle_{(\tau)}}=\frac{(\frac{x_{24}}{x_{14}})^{\sigma_1-\sigma_2}(\frac{x_{14}}{x_{13}})^{\sigma_3-\sigma_4} }{x_{12}^{\sigma_1+\sigma_2}x_{34}^{\sigma_3+\sigma_4}}\sum_{k}\sum_{l}f^{\tau,l}_k(u,v)\mathbf{Q}^{(k)}(\{x_i,z_i\}).
\end{align}
The spinning $\mathcal{H}^{\langle J_{s_1}J_{s_2}J_{s_3}J_{s_4}\rangle_{(\tau)}}$ are constrained by the OPE and also permutation symmetry where applicable.
We then write $\mathcal{H}^{\langle J_{s_1}J_{s_2}J_{s_3}J_{s_4}\rangle_{(\tau)}}$ as a sum of s,t and u channel twist conformal blocks,
\begin{align}\label{HfromstuSpinning}
    \mathcal{H}^{\langle J_{s_1}J_{s_2}J_{s_3}J_{s_4}\rangle_{(\tau)}}=\frac{1}{2}\bigg(\mathcal{H}^{(J_{s_1}J_{s_2}|\tau|J_{s_3}J_{s_4})}_s+\mathcal{H}^{(J_{s_1} J_{s_4}|\tau|J_{s_2} J_{s_3})}_t+\mathcal{H}^{(J_{s_1}J_{s_3}|\tau|J_{s_2} J_{s_4})}_u\bigg),
\end{align}
which are in general, independent quantities. For instance, if we want TCBs for identical spinning operators, all with spin $s$ and scaling dimension $\Delta$, then $\mathcal{H}^{(J_s J_s|\tau|J_s J_s)}_t$ and $\mathcal{H}^{(J_s J_s|\tau|J_s J_s)}_u$ are obtained via $(2\leftrightarrow 4)$ and $(2\leftrightarrow 3)$ exchanges from $\mathcal{H}^{(J_s J_s|\tau|J_s J_s)}_s$ much like in the identical scalar case. However, if we want TCBs where not all operators are identical, say, those involving two operators with spin $s_1$ and scaling dimension $\Delta_1$ and two other operators with spin $s_2$ and scaling dimension $\Delta_2$, then we have $\mathcal{H}^{(J_{s_1}J_{s_2}|\tau|J_{s_1}J_{s_2})}_u=\mathcal{H}^{(J_{s_1}J_{s_2}|\tau|J_{s_1}J_{s_2})}_t(3\leftrightarrow 4)$ but $\mathcal{H}^{(J_{s_1}J_{s_1}|\tau|J_{s_2}J_{s_2})}_s$ is completely independent of $\mathcal{H}^{(J_{s_1}J_{s_2}|\tau|J_{s_1}J_{s_2})}_t$ and $\mathcal{H}^{(J_{s_1}J_{s_2}|\tau|J_{s_1}J_{s_2})}_u$.

Alday's s channel equation for the spinning TCBs reads\footnote{The t and u channel equations are $\mathcal{A}_{14}\mathcal{H}^{(J_{s_1}J_{s_4}|\tau|J_{s_2}J_{s_3})}_t=\lambda(\tau)\mathcal{H}^{(J_{s_1}J_{s_4}|\tau|J_{s_2}J_{s_3})}_t$ and $\mathcal{A}_{13}\mathcal{H}^{(J_{s_1}J_{s_3}|\tau|J_{s_2}J_{s_4})}_u=\lambda(\tau)\mathcal{H}^{(J_{s_1}J_{s_3}|\tau|J_{s_2}J_{s_4})}_u$ respectively.},
\begin{align}\label{schannelspinningAldayEq}
    \mathcal{A}_{12}\mathcal{H}^{(J_{s_1}J_{s_2}|\tau|J_{s_3}J_{s_4})}_s=\lambda(\tau)\mathcal{H}^{(J_{s_1}J_{s_2}|\tau|J_{s_3}J_{s_4})}_s,
\end{align}

where $\mathcal{A}_{12}$ and $\lambda(\tau)$ are the same as in \eqref{AldayOperator} with the Casimir's given as in \eqref{casimirs} but with the generator of Lorentz transformations now given by,
\begin{align}\label{MABspinning}
     M_{AB}=X_A\frac{\partial}{\partial X^B}-X_B\frac{\partial}{\partial X^A}+Z_A\frac{\partial}{\partial Z^B}-Z_B\frac{\partial}{\partial Z^A},
\end{align}
due to the non-trivial transformation properties of spinning operators in contrast to the scalar case \eqref{scalarMAB}.
 $Z_A$ are the embedding space counterparts to the $z_i$ which are the  auxiliary null vectors. We now present two distinct methods to obtain spinning TCBs.
\subsubsection{Solving the twist eigenvalue equation}
 As with the case of identical scalars, we shall be focused on determining single trace $\tau=1$ TCBs. For $d=3$ and $\tau=1$, Alday's s channel equation \eqref{schannelspinningAldayEq} reads,
\begin{align}\label{3dAldayEquationSpinning}
    \mathcal{A}_{12}\mathcal{H}^{(J_{s_1}J_{s_2}|\tau|J_{s_3}J_{s_4})}_s=0.
\end{align}
 Our algorithm for computing the single trace $\tau=1$ spinning TCBs is the following:\\
 
 \textcolor{blue}{Step 1:} Write a conformally invariant ansatz for the TCB and choose half-integer polynomial ansätze for the functions of the cross ratios taking into account consistency with the OPE.\\ 
 
 \textcolor{blue}{Step 2:}  Apply Alday's s channel operator $\mathcal{A}_{12}$ \eqref{AldayOperator} on the ansatz for $\mathcal{H}^{\langle J_{s_1}J_{s_2}J_{s_3}J_{s_4}\rangle_{\tau=1}}$. The part of the ansatz that satisfies \eqref{3dAldayEquationSpinning} is the s channel TCB $\mathcal{H}^{(J_{s_1}J_{s_2}|\tau|J_{s_3}J_{s_4})}_s$.\\
 
 \textcolor{blue}{Step 3:} Similarly, obtain the t and u channel TCBs and add them up as in \eqref{HfromstuSpinning} to obtain $\mathcal{H}^{\langle J_{s_1}J_{s_2}J_{s_3}J_{s_4}\rangle_{\tau=1}}$.\\
 
  \textcolor{blue}{Step 4:} For TCBs involving conserved currents, demand the conservation of $\mathcal{H}^{\langle J_{s_1}J_{s_2}J_{s_3}J_{s_4}\rangle_{\tau=1}}$ at non-coincident operator insertions.\\
  
  The output is then a $\tau=1$ spinning TCB.\\

  {\underline{Note}: We solve Alday's equation \eqref{3dAldayEquationSpinning} in the $5$ dimensional embedding space and in the end convert our answers back into the $3$ dimensional real space\footnote{See for instance, \cite{Rychkov_2017} for more details on the embedding space formalism}. This is essential for making progress as the ansätze for spinning TCBs get exponentially complicated for increasing spin. We provide explicit details of some computations in appendix \ref{sec:detailsofsolutionstoTCB}. \\

  We shall now present several examples of spinning TCBs.
Further, we construct both parity even and parity odd twist conformal blocks. We begin with the parity even case.
\subsection*{\underline{Parity-even Spinning Twist Conformal Blocks}}
The parity even building blocks at the four-point level are the $V_{i,jk}$ and $H_{ij}$ whose expressions in the $5$ dimensional embedding space are \cite{Costa_2011},
\begin{align}
    &V_{i,jk}=\frac{(Z_i\cdot X_j)(X_i\cdot X_k)-(Z_i\cdot X_k)(X_i\cdot X_j)}{(X_j\cdot X_k)},\notag\\
    &H_{ij}=-2\bigg((Z_i\cdot Z_j)(X_i\cdot X_j)-(Z_i\cdot X_j)(Z_j\cdot X_i)\bigg).
\end{align}
  \subsubsection*{$\mathbf{\mathcal{H}^{\langle TO_1O_1O_1\rangle_{\tau=1}}}$}
  The most general ansatz for this TCB is,
\begin{align}
   \mathcal{H}^{\langle TO_1O_1O_1\rangle_{\tau=1}}=\frac{x_{34}^2}{x_{12}^2 x_{13}^4 x_{14}^4}\bigg(f(u,v)V_{1,23}^2+g(u,v)V_{1,24}^2+V_{1,34}^2 h(u,v)\bigg).
\end{align}
Crossing symmetry demands that $g(u,v)=f(\frac{u}{v},\frac{1}{v})$ and $h(u,v)=\frac{v}{u}f(v,u)$ which gives,
\begin{align}\label{TOOOansatz1}
     \mathcal{H}^{\langle TO_1O_1O_1\rangle_{\tau=1}}=\frac{x_{34}^2}{x_{12}^2 x_{13}^4 x_{14}^4}\bigg(f(u,v)V_{1,23}^2+f(\frac{u}{v},\frac{1}{v})V_{1,24}^2+V_{1,34}^2 \frac{v}{u}f(v,u)\bigg).
\end{align}
Next, we demand an ansatz of the form \eqref{fuvansatz} for $f(u,v)$,
\begin{align}\label{TOOOansatz1fuv}
    &f(u,v)=\sum_{i,j\in\mathbb{Z}/2}f_{ij}u^i v^j.
\end{align}
where the coefficients $f_{ij}$ are restricted via the crossing symmetry equation, $f(u,v)=f(\frac{u}{v},\frac{1}{v})$. Finally, consistency with the OPE (please refer to appendix \ref{sec:OPE} for more details) where we consider a $\Delta=1$ scalar exchange in each channel tells us that $f(u,v)$ takes the unique form of $\frac{v^2}{u^{3/2}}$. This yields the following expression: 
\begin{align}\label{TO1O1O1TCB}
  \mathcal{H}^{\langle TO_1O_1O_1\rangle_{\tau=1}}=\frac{c_1 x_{34}^2}{x_{12}^2 x_{13}^4 x_{14}^4}\bigg(\frac{v^2}{u^{{3/2}}}V_{1,23}^2+\frac{u}{\sqrt{v}}V_{1,34}^2+\frac{1}{u^{3/2}v^{1/2}}V_{1,24}^2\bigg).
\end{align}
We will now show that this result can also be written as a sum of s,t and u channel TCBs. By following steps similar to that of the identical scalar case \eqref{HsDelta1}, we obtain the s channel TCB,
\begin{align}\label{TO1O1O1sTCB}
    \mathcal{H}^{(TO_1|1|O_1O_1)}_s=\frac{c_1x_{34}^2}{x_{12}^2 x_{13}^4 x_{14}^4}\bigg(\frac{v^2}{u^{{3/2}}}V_{1,23}^2+\frac{1}{u^{3/2}v^{1/2}}V_{1,24}^2\bigg).
\end{align}
 One can check that it solves the $\tau=1$ Alday's s channel equation \eqref{3dAldayEquationSpinning}.
Similarly, we obtain the $t$ and $u$ channel TCBs and add all of them up as in \eqref{HfromstuSpinning} to obtain,
\begin{align}
    &\mathcal{H}^{\langle TO_1O_1O_1\rangle_{\tau=1}}=\frac{1}{2}\bigg(\mathcal{H}^{(TO_1|1|O_1O_1)}_s+\mathcal{H}^{(TO_1|1|O_1O_1)}_t+\mathcal{H}^{(TO_1|1|O_1O_1)}_u\bigg)\notag\\&=\frac{c_1 x_{34}^2}{x_{12}^2 x_{13}^4 x_{14}^4}\bigg(\frac{v^2}{u^{{3/2}}}V_{1,23}^2+\frac{u}{\sqrt{v}}V_{1,34}^2+\frac{1}{u^{3/2}v^{1/2}}V_{1,24}^2\bigg).
\end{align}
which exactly coincides with \eqref{TO1O1O1TCB}.
\subsection*{$\mathbf{\mathcal{H}^{\langle J_4 O_1O_1O_1\rangle_{(\tau=1)}}}$}
We begin with the following ansatz,
\begin{align}
   \mathcal{H}^{\langle J_4 O_1O_1O_1\rangle_{(\tau=1)}}=\frac{x_{34}^6}{x_{12}^2x_{13}^8 x_{14}^8}\bigg(f_1(u,v) V_{1,23}^4+ f_2(u,v)V_{1,24}^4+f_3(u,v)V_{1,34}^4\bigg).
\end{align}
Imposing the permutation symmetry and demanding OPE consistency (please refer to appendix \ref{sec:OPE} for the relevant OPE formulae) yields the following solution,
\begin{align}\label{J4O1O1O1TCB}
    &\mathcal{H}^{\langle J_4 O_1O_1O_1\rangle_{(\tau=1)}}=c_1\frac{ x_{34}^6}{x_{12}^2x_{13}^8 x_{14}^8}\bigg(\frac{v^4}{u^{7/2}}V_{1,23}^4+\frac{1}{u^{7/2}v^{1/2}}V_{1,24}^4+\frac{u}{v^{1/2}}V_{1,34}^4\bigg).
\end{align}
We then identified the s channel TCB that solves Alday's s channel equation \eqref{3dAldayEquationSpinning},
\begin{align}\label{J4O1O1O1sTCB}
    \mathcal{H}^{(J_4O_1|1|O_1O_1)}_s=c_1\frac{x_{34}^6}{x_{12}^2x_{13}^8 x_{14}^8}\bigg(\frac{v^4}{u^{{7/2}}}V_{1,23}^4+\frac{1}{u^{7/2}v^{1/2}}V_{1,24}^4\bigg),
\end{align}
and after obtaining the t and u channel TCBs by $(2\leftrightarrow 4)$ and $(2\leftrightarrow 3)$ exchanges on \eqref{J4O1O1O1sTCB} and using \eqref{HfromstuSpinning} we get,
\begin{align}
    &\mathcal{H}^{\langle J_4 O_1O_1O_1\rangle_{(\tau=1)}}=\frac{1}{2}\bigg(\mathcal{H}^{(J_4O_1|1|O_1O_1)}_s+\mathcal{H}^{(J_4O_1|1|O_1O_1)}_t+\mathcal{H}^{(J_4O_1|1|O_1O_1)}_u\bigg)\notag\\&=c_1\frac{ x_{34}^6}{x_{12}^2x_{13}^8 x_{14}^8}\bigg(\frac{v^4}{u^{7/2}}V_{1,23}^4+\frac{1}{u^{7/2}v^{1/2}}V_{1,24}^4+\frac{u}{v^{1/2}}V_{1,34}^4\bigg).
\end{align}
which is identical to \eqref{J4O1O1O1TCB}.
\subsection*{$\mathbf{\mathcal{H}^{\langle J_s O_1O_1O_1\rangle_{(\tau=1)}}}$}
So far, We have obtained the spinning s channel $\tau=1$ TCBs for $\langle TO_1O_1O_1\rangle$ in \eqref{TO1O1O1sTCB} and $\langle J_4O_1O_1O_1\rangle$ in  \eqref{J4O1O1O1sTCB}.
Based on the pattern, we guess the following form for $\mathcal{H}^{(J_s O_1|1|O_1O_1)}_s$:
\begin{align}\label{JsO1O1O1sTCB}
   \mathcal{H}^{(J_s O_1|1|O_1O_1)}_s=c_1\frac{ x_{34}^{2(s-1)}}{x_{12}^2 x_{13}^{2s}x_{14}^{2s}}\bigg(\frac{v^\textbf{s}}{u^{\frac{(2\textbf{s}-1)}{2}}}V_{1,23}^\textbf{s}+\frac{1}{u^{\frac{(2\textbf{s}-1)}{2}}v^{1/2}}V_{1,24}^\textbf{s}\bigg).
\end{align}
Remarkably, the result satisfies Alday's equation \eqref{3dAldayEquationSpinning} for any even spin \footnote{Correlators of the form $\langle J_s O_1O_1O_1\rangle$ for odd $s$ can be non-zero only if we give the scalar operators a non-abelian charge so we stick to the even $s$ case.}. By obtaining the $t$ and $u$ channel expressions by permutations, we get the following beautiful form for the TCB,
\begin{align}\label{JsO1O1O1TCB}
       &\mathcal{H}^{\langle J_s O_1O_1O_1\rangle_{(\tau=1)}}=\frac{1}{2}\bigg(\mathcal{H}^{(J_s O_1|1|O_1O_1)}_s+\mathcal{H}^{(J_s O_1|1|O_1O_1)}_t+\mathcal{H}^{(J_s O_1|1|O_1O_1)}_u\bigg)\notag\\&=c_1\frac{(1+(-1)^s)}{2}\frac{x_{34}^{2(s-1)}u^{\frac{1}{2}-s}}{x_{12}^2x_{13}^{2s}x_{14}^{2s}v^{1/2}}\bigg(v^{\frac{1}{2}+\textbf{s}}V_{1,23}^\textbf{s}+V_{1,24}^\textbf{s}+u^{\frac{1}{2}+\textbf{s}}V_{1,34}^\textbf{s}\bigg)\notag\\
       &c_1\frac{x_{34}^{2(s-1)}u^{\frac{1}{2}-s}}{x_{12}^2x_{13}^{2s}x_{14}^{2s}v^{1/2}}\bigg(v^{\frac{1}{2}+\textbf{s}}V_{1,23}^\textbf{s}+V_{1,24}^\textbf{s}+u^{\frac{1}{2}+\textbf{s}}V_{1,34}^\textbf{s}\bigg).
\end{align}
 Crossing symmetry is manifest and consistency with the OPE (please refer to appendix \ref{sec:OPE} for the relevant formulae) can also be checked. Conservation of the spin $s$ current is also satisfied at non-coincident points. \\

 We now move on to a TCB that involves two spinning operators.
\subsection*{$\mathbf{\mathcal{H}^{\langle JJO_1O_1\rangle_{(\tau=1)}}}$}
The most general ansatz for this TCB is \cite{Rychkov_2017},
\begin{align}
    &\mathcal{H}^{\langle JJO_1O_1\rangle_{(\tau=1)}}=\frac{x_{34}^4}{x_{13}^6 x_{24}^6}\bigg(W_1 W_2 f_1(u,v)+\Bar{W}_1\Bar{W}_2 f_2(u,v)+(W_1\Bar{W}_2-\Bar{W}_1 W_2)f_3(u,v)+H_{12}f_4(u,v)\bigg),
\end{align}
where,
\begin{align}
    &W_1=V_{1,23}+V_{1,24}, \Bar{W}_1=V_{1,23}-V_{1,24},\notag\\
    &W_2=V_{2,13}+V_{2,14}, \Bar{W}_2=V_{2,13}-V_{2,14}.
\end{align}
Crossing symmetry implies that,
\begin{align}
    &f_i(u,v)=\frac{1}{v^3}f_i(\frac{u}{v},\frac{1}{v}),i=1,2,4,\notag\\
    &f_3(u,v)=-\frac{1}{v^3}f_3(\frac{u}{v},\frac{1}{v}).
\end{align}
We then take half integer polynomial ansätze for each of the $f_i(u,v)$,
\begin{align}
    f_i(u,v)=\sum_{\alpha,\beta\in\mathbb{Z}/2}c_i^{(\alpha,\beta)}u^
    \alpha v^\beta.
\end{align}
which we then restrict using OPE consistency (please refer to appendix \ref{sec:OPE} for the relevant formulae).
There are two independent expressions that can be constructed out of $\tau=1$ twist conformal blocks that we can form, consistent with the $(1\leftrightarrow 2)$ and $(3\leftrightarrow 4)$ permutation symmetries, which are the s channel TCB and the sum of the t and u channel TCBs respectively.

The first of these can be obtained by solving Alday's equation in the s channel \eqref{3dAldayEquationSpinning} which yields,
\small
\begin{align}\label{JJO1O1sTCB}
    &\mathcal{H}^{(J J|1|O_1O_1)}_s=\frac{ x_{34}^4}{x_{13}^6 x_{24}^6 u^{5/2}v^{3/2}}\bigg((W_1 W_2-\Bar{W}_1\Bar{W}_2)(1+v^{5/2})+(W_1 \Bar{W}_2-\Bar{W}_1 W_2)(1-v^{5/2})+4 H_{12}v(1+\sqrt{v})\bigg).
\end{align}
\normalsize
Similarly, solving Alday's equation in the t channel yields,
\begin{align}\label{JJO1O1tTCB}
   \mathcal{H}^{(J O_1|1|J O_1)}_t=\frac{ x_{34}^4}{4 x_{13}^6 x_{24}^6}\bigg(W_1 W_2 h_1(u,v)+\Bar{W}_1\Bar{W}_2 h_2(u,v)+(W_1\Bar{W}_2-\Bar{W}_1 W_2 h_3(u,v)+H_{12}h_4(u,v)\bigg),
\end{align}
where,
\begin{align}
    &h_1(u,v)=\frac{-2+\sqrt{u}(v-1)^2}{2 u^{5/2}v^{3/2}},h_2(u,v)=\frac{2-\sqrt{u}(v+1)^2}{2 u^{5/2}v^{3/2}},\notag\\
    &h_3(u,v)=-\frac{2+\sqrt{u}(v^2-1)}{2u^{5/2}v^{3/2}},h_4(u,v)=-\frac{4}{u^{5/2}\sqrt{v}}.
\end{align}
and $\mathcal{H}^{(J O_1|1|J O_1)}_u=\mathcal{H}^{(J O_1|1|J O_1)}_t(3\leftrightarrow 4)$. Thus, we obtain the following $\tau=1$ TCB,
\begin{align}\label{JJO1O1stu}
     \mathcal{H}^{\langle JJO_1O_1\rangle_{(\tau=1)}}=c_1\mathcal{H}^{(J J|1|O_1 O_1)}_s+c_2\bigg(\mathcal{H}^{(J O_1|1|J O_1)}_t+\mathcal{H}^{(J O_1|1|J O_1)}_u\bigg),
\end{align}
which is a two parameter ($c_1$ and $c_2$) family of solutions.
For details on this computation please refer to appendix \ref{sec:detailsofsolutionstoTCB}.
\subsection*{$\mathbf{\mathcal{H}^{\langle JJO_2O_2\rangle_{(\tau=1)}}}$}
The most general ansatz for this TCB is,
\begin{align}
   \mathcal{H}^{\langle JJO_2O_2\rangle_{(\tau=1)}}=\frac{x_{34}^2}{x_{13}^6 x_{24}^6}\bigg(W_1 W_2 f_1(u,v)+\Bar{W}_1\Bar{W}_2 f_2(u,v)+(W_1\Bar{W}_2-\Bar{W}_1 W_2)f_3(u,v)+H_{12}f_4(u,v)\bigg).
\end{align}
Just like the $\Delta=1$ case, we impose crossing symmetry, take half-integer polynomial ansätze for the $f_i(u,v)$, take into account consistency with the OPE and then we solve Alday's equation \eqref{3dAldayEquationSpinning} in the s,t and u channels to identify the respective TCBs.
The result we obtained for the s channel $\tau=1$ TCB is,
\small
\begin{align}\label{JJO2O2s}
    \mathcal{H}^{(JJ|1|O_2O_2)}_s=\frac{ x_{34}^2}{x_{13}^6 x_{24}^6 }\bigg((W_1 W_2 g_1(u,v)+\Bar{W}_1\Bar{W}_2 g_2(u,v)+(W_1\Bar{W}_2-\Bar{W}_1 W_2)g_3(u,v)+H_{12}g_4(u,v)\bigg),
\end{align}
where,
\begin{align}
    &g_1(u,v)=\frac{-1+v+v^{3/2}-v^{5/2}}{u^{5/2}v^{3/2}}, g_2(u,v)=\frac{(v-1)(v^{3/2}-1)}{u^{5/2}v^{3/2}},\notag\\
    &g_3(u,v)=\frac{(v+1)(v^{3/2}-1)}{u^{5/2}v^{3/2}},g_4(u,v)=\frac{2(-1+u+v+(1+u)v^{3/2}-v^{5/2})}{u^{5/2}v^{3/2}}.
\end{align}
\normalsize
 whilst for the t channel TCB we got,
\begin{align}\label{JJO2O2t}
    &\mathcal{H}^{(JO_2|1|J O_2)}_t=\frac{x_{34}^2}{x_{13}^6 x_{24}^6}\bigg(W_1 W_2 h_1(u,v)+\Bar{W}_1\Bar{W}_2 h_2(u,v)+(W_1\Bar{W}_2-\Bar{W}_1 W_2)h_3(u,v)+H_{12}h_4(u,v)\bigg),
\end{align}
where,
\begin{align}
    &h_1(u,v)=\frac{2(v-1)+u^{3/2}(v+1)}{u^{5/2}v^{3/2}},h_2(u,v)=-\frac{2(v-1)+u^{3/2}(v+1)}{u^{5/2}v^{3/2}},\notag\\
    &h_3(u,v)=-\frac{u^{3/2}(v-1)+2(v+1)}{u^{5/2}v^{3/2}},h_4(u,v)=\frac{2(2u+u^{3/2}(-1+u-v)+2(v-1))}{u^{5/2}v^{3/2}}.
\end{align}
The u channel TCB can then be determined using $\mathcal{H}^{(JO_2|1|J O_2)}u=\mathcal{H}^{(JO_2|1|J O_2)}_t(3\leftrightarrow 4)$. We then obtain the following $\tau=1$ TCB,
\begin{align}
    \mathcal{H}^{\langle JJO_2O_2\rangle_{(\tau=1)}}= c_1\mathcal{H}^{(JJ|1|O_2 O_2)}_s+c_2\bigg( \mathcal{H}^{(JO_2|1|J O_2)}_t+ \mathcal{H}^{(JO_2|1|J O_2)}_u\bigg).
\end{align}
 which is a two-parameter family ($c_1$ and $c_2$) of $\tau=1$ TCBs \footnote{One can also carry out this analysis for other correlators such as ones with more spin. However, for our purposes, the current computations suffice.}.
\subsection*{\underline{Parity-odd Twist Conformal Blocks}}
The parity odd building blocks at the four-point level are the $S_{i,jk}$ and $Y_{ij,k}$ whose expressions in the $5$ dimensional embedding space are given by,
\begin{align}
    &S_{i,jkl}=\frac{1}{(X_j\cdot X_k)^{1/2}(X_k\cdot X_l)^{1/2}(X_l\cdot X_j)^{1/2}}\epsilon^{Z_i X_i X_j X_k X_l},\notag\\
    &Y_{ij,m}=\frac{(X_i\cdot X_j)^{1/2}}{(X_i\cdot X_m)^{1/2}(X_j\cdot X_m)^{1/2}}\epsilon^{Z_i Z_j X_i X_j X_m},
\end{align}
where we have used the shorthand,
\begin{align}
    \epsilon^{ABCDE}X_{A}Y_{B}Z_{C}U_{D}V_{E}:=\epsilon^{X Y Z U V}.
\end{align}
\subsubsection*{${\mathbf{\mathcal{H}^{\langle TO_2O_2O_2\rangle_{\tau=1,\text{odd}}}}}$}
This parity odd TCB is constructed out of the $V_{i,jk}$ and the parity odd structure $S_{1,234}$.
 The most general ansatz is given by,
\begin{align}
    &\mathcal{H}^{\langle TO_2O_2O_2\rangle_{\tau=1,\text{odd}}}=\frac{x_{23}x_{34}}{x_{13}^6 x_{12}^2 x_{14}^2 x_{24}^3}S_{1,234}\bigg(V_{1,23}f_1(u,v)-V_{1,24}f_2(u,v)+V_{1,34}f_3(u,v)\bigg).
\end{align}
Our methods to solve Alday's equation \eqref{3dAldayEquation} are agnostic to the parity of the correlator and hence we proceed through with it which yields the following unique form for the TCB,
\begin{align}\label{TO2O2O2oddTCB}
    \mathcal{H}^{\langle TO_2O_2O_2\rangle_{\tau=1,\text{odd}}}=\frac{1}{2}\bigg(\mathcal{H}_{s,\text{odd}}^{(TO_2|1|O_2O_2)}+\mathcal{H}_{t,\text{odd}}^{(TO_2|1|O_2O_2)}+\mathcal{H}_{u,\text{odd}}^{(TO_2|1|O_2O_2)}\bigg),
\end{align}
where,
\begin{align}\label{TO2O2O2odds}
   \mathcal{H}_{s,\text{odd}}^{(TO_2|1|O_2O_2)}=\frac{c_1 x_{23}^{3}x_{34}^{3}}{x_{13}^{10} x_{24}^{7}}\frac{1}{(uv)^{5/2}}S_{1,234}\bigg(V_{1,23} v^{5/2}-V_{1,24}\bigg),
\end{align}
is a $\tau=1$ s channel TCB that solves Alday's equation \eqref{schannelspinningAldayEq}.\\

Let us now generalize this to situation where the spinning operator has arbitrary spin.
\subsection*{$\mathbf{\mathcal{H}^{\langle J_s O_2|1|O_2O_2\rangle_{\tau=1,\text{odd}}}}$}
By solving Alday's equation \eqref{3dAldayEquation}, imposing OPE consistency and crossing symmetry we obtained the following $\tau=1$ TCB,
\begin{align}
     \mathcal{H}^{\langle J_sO_2O_2O_2\rangle_{\tau=1,\text{odd}}}=\frac{1}{2}\bigg(\mathcal{H}_{s,\text{odd}}^{(J_sO_2|1|O_2O_2)}+\mathcal{H}_{t,\text{odd}}^{(J_sO_2|1|O_2O_2)}+\mathcal{H}_{u,\text{odd}}^{(J_s O_2|1|O_2O_2)}\bigg),
\end{align}
where\footnote{We also notice a very interesting relationship between $\mathcal{H}_{s,\text{odd}}^{(J_sO_2|1|O_2O_2)}$ \eqref{JsOOOodds} and $\mathcal{H}_{s,\text{even}}^{(J_sO_1|1|O_1O_1)}$ \eqref{JsO1O1O1sTCB}. Upon the replacements $S_{1,234}V_{1,23}^{s-1}\to V_{1,23}^s$ and $S_{1,234}V_{1,24}^{s-1}\to -V_{1,24}^s$ we have,
\begin{align}
    \mathcal{H}_{s,\text{odd}}^{(J_sO_2|1|O_2O_2)}\to\frac{1}{x_{23}x_{34}x_{24}} \mathcal{H}_{s,\text{even}}^{(J_sO_1|1|O_1O_1)}.\notag
\end{align}},
\begin{align}\label{JsOOOodds}
    \mathcal{H}_{s,\text{odd}}^{(J_sO_2|1|O_2O_2)}=\frac{x_{23}^{2s-1}x_{34}^{2s-1}}{x_{13}^{4s+2}x_{24}^{2s+3}}\frac{1}{(uv)^{s+\frac{1}{2}}}S_{1,234}\bigg(v^{s+\frac{1}{2}}V_{1,23}^{s-1}-V_{1,24}^{s-1}\bigg),
\end{align}
which for $s=2$ does indeed reproduce \eqref{TO2O2O2odds}.\\
 \subsubsection{Using weight-shifting and spin-raising operators}
 Due to the proliferation of tensor structures for sufficient spin, solving Alday's equation \eqref{schannelspinningAldayEq} for the spinning case is not straightforward. The degeneracy that accompanies the growing number of tensor structures becomes hard to tackle as well, see appendix \ref{sec:misc} for a few degeneracy formulae. Thus, we seek alternative methods to obtain them.
One alternative is to employ the technology of weight-shifting and spin-raising operators and act with them appropriately on scalar seed TCBs to produce the spinning TCBs. Usually, these operators work for a particular exchange but despite the fact that a TCB involves infinitely many exchanges, we find that we are indeed able to reproduce some of the results that we obtained via solving Alday's equation \eqref{schannelspinningAldayEq} by acting with these operators channel by channel. Thus, we see that this provides us with a useful and powerful tool. For the convenience of the reader, we provide the actions of the weight-shifting and spin-raising operators that we employ in this work.
\begin{align}\label{spinraisops}
    &\mathcal{D}_{ij}: (\Delta_i,s_i,\Delta_j,s_j)\to (\Delta_i,s_i+1,\Delta_j-1,s_j),\notag\\
    &\mathcal{H}_{ij}: (\Delta_i,s_i,\Delta_j,s_j)\to (\Delta_i-1,s_i+1,\Delta_j-1,s_j+1),\notag\\
    &\mathcal{W}^{(n)}_{ij}: (\Delta_i,s_i,\Delta_j,s_j)\to (\Delta_i-n,s_i,\Delta_j-n,s_j).
\end{align}
Importantly, Alday's operator commutes with these particular weight-shifting and spin-raising operators,
\begin{align}
    &[\mathcal{A}_{ij},\mathcal{D}_{ij}]=0,\notag\\
    &[\mathcal{A}_{ij},\mathcal{H}_{ij}]=0,\notag\\
    &[\mathcal{A}_{ij},\mathcal{W}^{(n)}_{ij}]=0.
\end{align}
This implies that the action of these operators map solutions of Alday's equation to other solutions of Alday's equation, a fact that we will exploit to produce spinning TCBs from scalar TCBs. We also notice that the spinning TCBs that we produce via this method automatically satisfy all our consistency conditions such as consistency with the OPE. Also, for the explicit form of these weight-shifting and spin-raising operators, please refer to appendix \ref{sec:weightshiftingspinraisingoperators}.
  \subsection*{$\mathbf{\mathcal{H}^{\langle TO_1O_1O_1\rangle_{\tau=1}}}$}
We can obtain this spinning TCB using weight-shifting and spin-raising operators. For instance, starting with the $\Delta=1$ s channel scalar TCB $\mathcal{H}_s^{(O_1O_1|1|O_1O_1)}$ \eqref{HsDelta1} we construct,
\begin{align}
    &\frac{4}{3}\mathcal{D}_{12}\mathcal{D}_{12}\mathcal{W}_{12}^{(-2)}\mathcal{H}_s^{(O_1O_1|1|O_1O_1)}=\frac{4c_1}{3}\mathcal{D}_{12}\mathcal{D}_{12}\mathcal{W}_{12}^{(-2)}\frac{1}{x_{12}^2 x_{34}^2}\bigg(\sqrt{u}+\sqrt{\frac{u}{v}}\bigg)\notag\\&=\frac{c_1x_{34}^2}{x_{12}^2 x_{13}^4 x_{14}^4}\bigg(\frac{v^2}{u^{{3/2}}}V_{1,23}^2+\frac{1}{u^{3/2}v^{1/2}}V_{1,24}^2\bigg).
\end{align}
which is identical to the expression that we obtained via solving Alday's equation \eqref{TO1O1O1sTCB}. We can then obtain the t and u channel TCBs via the appropriate permutations to obtain $\mathcal{H}^{\langle TO_1O_1O_1\rangle_{\tau=1}}$ exactly as in \eqref{TO1O1O1TCB}.
\subsection*{$\mathbf{\mathcal{H}^{\langle J_4 O_1 O_1 O_1\rangle_{\tau=1}}}$}
We can obtain the s channel TCB as follows:
\begin{align}
    \mathcal{H}_s^{(J_4 O_1|1|O_1 O_1)}=\frac{16 c_1}{105}\mathcal{D}_{12}^4\mathcal{W}_{12}^{(-4)}\mathcal{H}_s^{(O_1O_1|1|O_1O_1)}=c_1\frac{x_{34}^6}{x_{12}^2x_{13}^8 x_{14}^8}\bigg(\frac{v^4}{u^{{7/2}}}V_{1,23}^4+\frac{1}{u^{7/2}v^{1/2}}V_{1,24}^4\bigg),
\end{align}
where $\mathcal{H}_s^{(O_1O_1|1|O_1O_1)}$ is given in \eqref{HsDelta1}. We see that this is identical to the result that we obtained via solving Alday's equation \eqref{J4O1O1O1sTCB}. We then obtain the t and u channel TCBs and form $\mathcal{H}^{\langle J_4 O_1 O_1 O_1\rangle_{\tau=1}}$ as in \eqref{J4O1O1O1TCB}.
\subsection*{$\mathbf{\mathcal{H}^{\langle J_s O_1 O_1 O_1\rangle}_{\tau=1}}$}
One can obtain this s channel TCB as follows:
\begin{align}
    \mathcal{H}^{(J_sO_1|1|O_1O_1)}_s\propto\mathcal{D}_{12}^s\mathcal{W}_{12}^{(-s)}\mathcal{H}^{(O_1O_1|1|O_1O_1)}_s\propto \frac{c_1 x_{34}^{2(s-1)}}{x_{12}^2 x_{13}^{2s}x_{14}^{2s}}\bigg(\frac{v^\textbf{s}}{u^{\frac{(2\textbf{s}-1)}{2}}}V_{1,23}^\textbf{s}+\frac{1}{u^{\frac{(2\textbf{s}-1)}{2}}v^{1/2}}V_{1,24}^\textbf{s}\bigg),
\end{align}
with $\mathcal{H}^{(O_1O_1|1|O_1O_1)}_s$ given in \eqref{HsDelta1}. This coincides exactly with the result in \eqref{JsO1O1O1sTCB} that we guessed earlier. Thus, similarly obtaining the t and u channel TCBs, we obtain $\mathcal{H}^{\langle J_s O_1 O_1 O_1\rangle}_{\tau=1}$ just like in \eqref{JsO1O1O1TCB}. We see through this example that spin-raising and weight-shifting operators provide us with a systematic way to construct spinning twist conformal blocks of arbitrary spin. 
\subsection*{$\mathbf{\mathcal{H}^{\langle TO_2O_2O_2\rangle_{\tau=1,\text{odd}}}}$}
We can obtain the s channel expression for $\mathcal{H}^{\langle TO_2O_2O_2\rangle_{\tau=1,\text{odd}}}$ by acting with appropriate spin raising and weight shifting operators on the $\Delta=1$ s channel scalar TCB \eqref{HsDelta1}:
\begin{align}
    &\mathcal{H}_{s,\text{odd}}^{(TO_2|1|O_2O_2)}=\mathcal{D}_{12}\tilde{\mathcal{D}}_{12}\mathcal{W}_{12}^{(-2)}\mathcal{W}_{34}^{(-1)}\mathcal{H}^{(O_1O_1|1|O_1O_1)}\notag\\&\propto \frac{x_{23}x_{34}}{x_{13}^6 x_{12}^2 x_{14}^2 x_{24}^3}S_{1,234}\bigg(V_{1,23}\frac{v}{u^{3/2}}-V_{1,24}\frac{1}{(uv)^{3/2}}\bigg),
\end{align}
which is precisely what we found by solving Alday's equation for the s channel answer\eqref{TO2O2O2odds}. We then obtain the the t and u channel expressions similarly and obtain $\mathcal{H}^{\langle TO_2O_2O_2\rangle_{\tau=1,\text{odd}}}$ as in \eqref{TO2O2O2oddTCB}. The explicit form of the operator $\tilde{D_{12}}$ is provided in appendix \ref{sec:weightshiftingspinraisingoperators}.
\subsection*{$\mathbf{\mathcal{H}^{\langle JJO_1O_1\rangle_{\tau=1}}}$}
We construct the following quantity that we will choose as our seed\footnote{The seeds that we take to produce spinning TCBs do not themselves necessarily have to be TCBs, that is, they do not have to satisfy all the consistency criteria mentioned in section \ref{sec:scalarTCB}. The only criteria there that they have to satisfy is that they are solutions to Alday's equation, for instance,  \eqref{3dAldayEquation} if one wants to produce a s channel TCB. The quantity \eqref{seedForJJOO} we have chosen as our seed is consistent with the OPE but not the $(1\leftrightarrow 4)$ permutation symmetry that is expected of a t channel TCB but nevertheless we require it to produce the t channel TCB for $\langle JJO_1O_1\rangle$.}
\begin{align}\label{seedForJJOO}
    F_t(x_1,x_2,x_3,x_4)=\frac{a}{4x_{14}^2 x_{23}^2}\bigg(\sqrt{v}-2 \sqrt{\frac{v}{u}}\bigg)
\end{align}
where $a$ is a constant. It solves Alday's TCB equation in the t channel\footnote{The t channel equation is obtained via a $(2\leftrightarrow 4)$ exchange from the s channel equation \eqref{3dAldayEquation}.},
\begin{align}
    \mathcal{A}_{14}F_t=0.
\end{align}
We now perform the following actions on our seed,
\begin{align}
&\mathcal{D}_{23}\mathcal{D}_{14}\mathcal{W}_{14}^{(-1)}\mathcal{W}_{23}^{(-1)} F_t(x_1,x_2,x_3,x_4)\notag\\&=\frac{ x_{34}^4}{4 x_{13}^6 x_{24}^6}\bigg(W_1 W_2 (\frac{-2+\sqrt{u}(v-1)^2}{2 u^{5/2}v^{3/2}})+\Bar{W}_1\Bar{W}_2 (\frac{2-\sqrt{u}(v+1)^2}{2 u^{5/2}v^{3/2}})+(W_1\Bar{W}_2-\Bar{W}_1 W_2) (-\frac{2+\sqrt{u}(v^2-1)}{2u^{5/2}v^{3/2}})\notag\\&+H_{12}(-\frac{4}{u^{5/2}\sqrt{v}})\bigg)\propto \mathcal{H}_t^{(JO_1|1|JO_1)},
\end{align}
where $\mathcal{H}_t^{(JO_1|1|JO_1)}$ is given in \eqref{JJO1O1tTCB}. We obtain the u channel TCB via a $(3\leftrightarrow 4)$ exchange and add it with the t channel expression a valid TCB. However, we were unable to obtain the s channel TCB \eqref{JJO1O1sTCB} by using weight shifting and spin raising operators.\\

Now, armed with the machinery of spinning TCBs, we return to investigate whether the higher spin and slightly broken higher spin theory correlators can be written in terms of the single trace TCBs.
 \section{Theories with Exact Higher Spin Symmetry}\label{sec:FBFFtheories}
In three dimensions, examples of CFTs with higher spin symmetry include the free bosonic and free fermionic theories. In this section, we shall decompose correlators in these theories in terms of the single trace TCBs that we computed in the previous section. Based on this decomposition, we shall discuss the issue of bulk non-locality for both these theories. \\

Note: We will suppress the overall factor of $N$ and overall numerical factors in the correlators that we consider for notational simplicity.
\subsection{Free Bosonic Theory}\label{sec:FBtheory}
Consider a theory of massless free bosons in the fundamental representation of $SU(N)$,
\begin{align}
    S=\int d^3 x \partial_\mu \bar{\phi}^i\partial_\mu \phi^i.
\end{align}
This theory, apart from being a conformal field theory(CFT) also has an infinite number of conserved currents with all positive integer spins. The $SU(N)$ singlet sector of this theory has been conjectured to have a higher spin bulk dual. The spectrum of single trace primary operators in this theory consists of a $\Delta=1$ scalar $O_1=\bar{\phi}\phi(x)$ and conserved currents $J_{s}$ for $s=1,2,\cdots$ schematically of the form $\bar{\phi}\partial_{\mu_1}\cdots \partial_{\mu_s}\phi+\cdots$. All these operators have twist $\tau=1$. For explicit expressions of these currents please refer to appendix \ref{sec:FBFFCBQFoperators}.

\subsubsection{Scalar Correlators}
\subsection*{Scalar Four Point Function}
We obtain via Wick contractions,
\begin{align}\label{FBOOOO}
    \langle O_1 O_1 O_1 O_1\rangle_{FB}=\frac{2}{x_{12}^2 x_{34}^2}\bigg(\sqrt{u}+\sqrt{\frac{u}{v}}+\frac{u}{\sqrt{v}}\bigg).
\end{align}
Using the TCB that we obtained earlier \eqref{O1O1O1O1TCB} we see that,
\begin{align}\label{O1O1O1O1FBasTCB}
    \langle O_1 O_1 O_1 O_1\rangle_{FB}=\frac{1}{2}\bigg(\mathcal{H}_s^{(O_1O_1|1|O_1 O_1)}+\mathcal{H}_t^{(O_1O_1|1|O_1 O_1)}+\mathcal{H}_u^{(O_1O_1|1|O_1 O_1)}\bigg),
\end{align}
a fact that was previously noticed in \cite{Sleight_2018}.\\

Now, as is done usually in CFT, $\langle O_1 O_1O_1O_1\rangle_{FB}$ can be decomposed in any channel (s,t or u) into single trace and double trace contributions. Indeed, \eqref{FBOOOO} can be written as\footnote{In this example and all those to follow, we shall decompose correlators into s channel single trace and double trace contributions. It is easy to perform an analogous analysis in the t and u channels as well.},
\begin{align}\label{O1O1O1O1stdt}
    \langle O_1O_1O_1O_1\rangle_{FB}=\mathcal{H}_s^{\text{single-trace}}+\mathcal{H}_s^{\text{double-trace}},
\end{align}
where $\mathcal{H}_s^{\text{single-trace}}$ is the s channel $\tau=1$ TCB $\mathcal{H}_s^{(O_1O_1|1|O_1 O_1)}$ \eqref{O1O1O1O1TCB},
\begin{align}
    &\mathcal{H}_s^{\text{single-trace}}=\mathcal{H}_s^{(O_1O_1|1|O_1 O_1)},
\end{align}
and,
\begin{align}
    \mathcal{H}_s^{\text{double-trace}}=\frac{2}{x_{12}^2 x_{34}^2}\bigg(\frac{u}{\sqrt{v}}\bigg).
\end{align}
One can then check that the double discontinuity (please refer to appendix \ref{sec:dDisc} for the definition and details) of the full correlator is purely due to the single trace contributions,
\begin{align}
&\mathbf{dDisc_s}(\mathcal{H}_s^{(O_1O_1|1|O_1 O_1)})=\mathbf{dDisc_s}(\langle O_1O_1O_1O_1\rangle_{FB})\ne 0,\notag\\&
    \mathbf{dDisc_s}(\mathcal{H}_s^{\text{double-trace}})=0,
\end{align}
which is consistent with equations \eqref{scalartau1dDisc1} and \eqref{scalartaudtdDisc1} in appendix \ref{sec:dDisc}.
\subsection*{Scalar Five Point Function}
The most general parity even five-point function of scalars with scaling dimension $\Delta$ can be written as,
\begin{align}\label{fptscalar}
    \langle O_{\Delta}(x_1) O_{\Delta}(x_2) O_{\Delta}(x_3) O_{\Delta}(x_4) O_{\Delta}(x_5)\rangle=\frac{1}{(x_{12}x_{13}x_{14}x_{15}x_{23}x_{24}x_{25}x_{34}x_{35}x_{45})^{\frac{\Delta}{2}}}F(u_1,v_1,u_2,v_2,w),
\end{align}
where the cross ratios are given by,
\begin{align}
    &u_1=\frac{x_{12}^2x_{34}^2}{x_{13}^2 x_{24}^2},v_1=\frac{x_{14}^2 x_{23}^2}{x_{13}^2 x_{24}^2}, u_2=\frac{x_{23}^2x_{45}^2}{x_{24}^2 x_{35}^2},v_2=\frac{x_{25}^2 x_{34}^2}{x_{24}^2 x_{35}^2},w=\frac{x_{15}^2 x_{23}^2 x_{34}^2}{x_{13}^2 x_{24}^2 x_{35}^2}.
\end{align}
We now compute the free bosonic correlator via Wick contractions. Writing it in the form of \eqref{fptscalar}, we obtain,
\begin{align}
    &\langle O_1 O_1 O_1 O_1 O_1\rangle_{FB}=\frac{2}{(x_{12}x_{13}x_{14}x_{15}x_{23}x_{24}x_{25}x_{34}x_{35}x_{45})^{\frac{1}{2}}(u_1 u_2 v_1 v_2 w^2)^{\frac{1}{4}}}\bigg((\sqrt{u_2}+\sqrt{v}_2)w\notag\\&+\sqrt{v_1}(w+\sqrt{v_2}(1+\sqrt{u_2}+w))+\sqrt{u_1}(\sqrt{v_1 v_2}+w+\sqrt{u_2}(1+\sqrt{v_1}+\sqrt{v_2}+w))\bigg).
\end{align}
Consider now the following s channel $\tau=1$ TCB \footnote{The fact that this is a $\tau=1$ s channel TCB can be verified by checking that it solves Alday's s channel equation \eqref{3dAldayEquation}. We leave a more systematic analysis of five-point correlators to a future work. },
\begin{align}\label{5ptTCB}
   &\mathcal{H}_s^{(O_1O_1|1|O_1O_1O_1)} =\frac{2}{(x_{12}x_{13}x_{14}x_{15}x_{23}x_{24}x_{25}x_{34}x_{35}x_{45})^{\frac{1}{2}}}\bigg(\frac{(\sqrt{u_2}+\sqrt{v_2})w+\sqrt{v_1}(w+\sqrt{v_2}(1+\sqrt{u_2}+w))}{(u_1 u_2 v_1 v_2 w^2)^{\frac{1}{4}}}\bigg).
\end{align}
The u,t and v channel twist conformal blocks are constructed as follows:
\begin{align}\label{utvchannels}
 &\mathcal{H}_u^{(O_1O_1|1|O_1O_1O_1)}=\mathcal{H}_s^{(O_1O_1|1|O_1O_1O_1)}(2\leftrightarrow 3),\notag\\
&\mathcal{H}_t^{(O_1O_1|1|O_1O_1O_1)}=\mathcal{H}_s^{(O_1O_1|1|O_1O_1O_1)}(2\leftrightarrow 4),\notag\\
&\mathcal{H}_v^{(O_1O_1|1|O_1O_1O_1)}=\mathcal{H}_s^{(O_1O_1|1|O_1O_1O_1)}(2\leftrightarrow5).
\end{align}
Using these expressions, we see that,
\begin{align}\label{OOOOOTCB}
    \langle O_1O_1O_1O_1O_1\rangle_{FB}=\frac{1}{2}\bigg(\mathcal{H}_s^{(O_1O_1|1|O_1O_1O_1)}+\mathcal{H}_u^{(O_1O_1|1|O_1O_1O_1)}+\mathcal{H}_t^{(O_1O_1|1|O_1O_1O_1)}+\mathcal{H}_v^{(O_1O_1|1|O_1O_1O_1)}\bigg).
\end{align}
thus decomposing a five-point function into a sum of single trace TCBs.\\

Let us now turn our attention to the case of spinning correlators in the free bosonic theory. One might think that due to the higher spin symmetry, the spinning correlators should be given by $\tau=1$ TCBs since the scalar correlator is. However, this is not so obvious as we show in appendix \ref{sec:HSE}. 
\subsubsection{Spinning Correlators}
\subsection*{$\mathbf{\langle T O_1 O_1 O_1\rangle_{FB}}$}
Performing the Wick contractions, we obtain the following expression for the correlator:
\begin{align}\label{FBTOOO}
    \langle TO_1O_1O_1\rangle_{FB}=\frac{x_{34}^2}{x_{12}^2 x_{13}^4 x_{14}^4}\bigg(\frac{v^2}{u^{{3/2}}}V_{1,23}^2+\frac{u}{\sqrt{v}}V_{1,34}^2+\frac{1}{u^{3/2}v^{1/2}}V_{1,24}^2\bigg).
\end{align}
We can see by inspection that we can decompose this answer as a sum of $\tau=1$ TCBs we obtained earlier \eqref{TO1O1O1TCB}.
Thus we have, very much analogous to the scalar case,
\begin{align}\label{TO1O1O1asTCB}
    \langle TO_1O_1O_1\rangle_{FB}=\frac{1}{2}\bigg(\mathcal{H}^{(TO_1|1|O_1O_1)}_s+\mathcal{H}^{(TO_1|1|O_1O_1)}_t+\mathcal{H}^{(TO_1|1|O_1O_1)}_u\bigg).
\end{align}
Also, just like the scalar case \eqref{O1O1O1O1stdt}, we can decompose this correlator into single-trace and double-trace contributions,
\begin{align}
    \langle TO_1O_1O_1\rangle_{FB}=\mathcal{H}_s^{\text{single-trace}}+\mathcal{H}_s^{\text{double-trace}},
\end{align}
where $\mathcal{H}_s^{\text{single-trace}}$ is the $\tau=1$ TCB \eqref{TO1O1O1TCB},
\begin{align}
    \mathcal{H}_s^{\text{single-trace}}=\mathcal{H}^{(TO_1|1|O_1O_1)}_s,
\end{align}
and,
\begin{align}
    \mathcal{H}_s^{\text{double-trace}}=\frac{x_{34}^2}{x_{12}^2 x_{13}^4 x_{14}^4}\bigg(\frac{u}{\sqrt{v}}V_{1,34}^2\bigg).
\end{align}
One can then see that the double discontinuity of the correlator is entirely due to the single trace TCB,
\begin{align}
    &\mathbf{dDisc_s}(\mathcal{H}_s^{(TO_1|1|O_1O_1)})=\mathbf{dDisc_s}(\langle TO_1O_1O_1\rangle_{FB}),\notag\\
    &\mathbf{dDisc_s}(\mathcal{H}_s^{\text{double-trace}})=0.
\end{align}
which is consistent with our expectations (please refer to equations \eqref{TOOOrtau1dDisc1} and \eqref{OOdoubletraceTOOO}).
\subsection*{$\mathbf{\langle J_4 O_1 O_1 O_1\rangle}$}
We obtain via Wick contractions,
\begin{align}\label{FBJ4OOO}
    \langle J_4 O_1O_1O_1\rangle_{FB}=\frac{x_{34}^6}{x_{12}^2x_{13}^8 x_{14}^8}\bigg(\frac{v^4}{u^{7/2}}V_{1,23}^4+\frac{1}{u^{7/2}v^{1/2}}V_{1,24}^4+\frac{u}{v^{1/2}}V_{1,34}^4\bigg).
\end{align}
Just like $\langle TO_1O_1O_1\rangle_{FB}$ \eqref{TO1O1O1asTCB}, this correlator too can be written in terms of a TCB that we have already computed earlier,  \eqref{J4O1O1O1sTCB} in this case. We then see that,
\begin{align}\label{J4O1O1O1asTCB}
    \langle J_4 O_1 O_1 O_1\rangle_{FB}=\frac{1}{2}\bigg(\mathcal{H}^{(J_4O_1|1|O_1O_1)}_s+\mathcal{H}^{(J_4O_1|1|O_1O_1)}_t+\mathcal{H}^{(J_4O_1|1|O_1O_1)}_u\bigg),
\end{align}
thereby decomposing a higher spin correlator in terms of single trace TCBs.\\

As with the previous examples, we can decompose this correlator into single-trace and double-trace contributions,
\begin{align}
    \langle J_4O_1O_1O_1\rangle_{FB}=\mathcal{H}_s^{\text{single-trace}}+\mathcal{H}_s^{\text{double-trace}},
\end{align}
where $\mathcal{H}_s^{\text{single-trace}}$ is the $\tau=1$ TCB \eqref{J4O1O1O1sTCB} which accounts for the full double discontinuity of the correlator and $\mathcal{H}^{\text{double-trace}}$ is the  difference between the full correlator \eqref{FBJ4OOO} and $\mathcal{H}_s^{\text{single-trace}}$.
\begin{align}
    &\mathcal{H}_s^{\text{single-trace}}=\mathcal{H}^{(J_4O_1|1|O_1O_1)}_s,\notag\\
    &\mathcal{H}_s^{\text{double-trace}}=\frac{x_{34}^6}{x_{12}^2x_{13}^8 x_{14}^8}\bigg(\frac{u}{v^{1/2}}V_{1,34}^4\bigg).
\end{align}
\subsection*{$\mathbf{\langle J J O_1 O_1\rangle}$}
Computing this correlator via Wick contractions we find,
\begin{align}
    &\langle J(x_1,z_1)J(x_2,z_2)O_1(x_3)O_1(x_4)\rangle_{FB}\notag\\&=\frac{x_{34}^4}{4x_{13}^6 x_{24}^6}\bigg(W_1 W_2 f_1(u,v)+\Bar{W}_1\Bar{W}_2 f_2(u,v)+(W_1\Bar{W}_2-\Bar{W}_1 W_2)f_3(u,v)+H_{12}f_4(u,v)\bigg),
\end{align}
where,
\begin{align}
    &f_1(u,v)=\frac{-1+\sqrt{u}(v-1)^2-v^{5/2}}{(u^{5/2}v^{3/2})},f_2(u,v)=\frac{1+v^{5/2}-\sqrt{u}(v+1)^2}{u^{5/2}v^{3/2}}\notag\\
    &f_3(u,v)=\frac{-1+v^{5/2}-\sqrt{u}(v^2-1)}{u^{5/2}v^{3/2}},f_4(u,v)=-\frac{4(1+\sqrt{v})}{u^{5/2}\sqrt{v}}.
\end{align}
We see by inspection that this result can be written in terms of the $t$ and $u$ channel $\tau=1$ twist conformal blocks that we computed independently earlier, please refer to \eqref{JJO1O1tTCB} and the equations just below it. We see that,
\begin{align}\label{JJO1O1asTCB}
   \langle JJO_1O_1\rangle_{FB}=\frac{1}{2}\bigg(\mathcal{H}^{(J O_1|1|J O_1)}_t+\mathcal{H}^{(J O_1|1|J O_1)}_u\bigg).
\end{align}
We notice the contrast to the previous examples where the correlators were given by a sum of s,t and u channel TCBs. Here, just the t and u channel TCBs suffice to reproduce the correlator. The fact that the s channel TCB \eqref{JJO1O1sTCB} is absent in the above expression can be understood as follows. As usual, we write the correlator in the s channel as a sum of single trace and double trace contributions,
\begin{align}
    \langle JJO_1O_1\rangle_{FB}=\mathcal{H}_s^{\text{single-trace}}+\mathcal{H}_s^{\text{double-trace}}.
\end{align}
The first term in this equation is the $\tau=1$ TCB \eqref{JJO1O1sTCB} (which accounts for the full double discontinuity of the correlator) while the second term accounts for the double-trace operator exchanges.
\begin{align}
    &\mathcal{H}_s^{\text{single-trace}}=\mathcal{H}^{(J J|1|O_1 O_1)}_s,\notag\\&
     \mathcal{H}_s^{\text{double-trace}}=\frac{x_{34}^4}{2 u^2 v^{3/2}x_{13}^6 x_{24}^6}(v V_{1,23}-V_{1,24})(-V_{2,13}+v V_{2,14}).
\end{align}
Thus, it is clear that just the s channel $\tau=1$ TCB cannot reproduce the entire correlator, one needs to add to it all the s channel double trace contributions. The t and u channel $\tau=1$ TCBs are needed if the entire correlator is to be expressed purely in terms of $\tau=1$ TCBs. Due to the permutation symmetry of the correlator, the t and u channel TCBs can only appear as a sum. Thus, \eqref{JJO1O1stu} is the most general $\tau=1$ TCB contribution that can give rise to a valid correlator and for the FB case, we find that just the sum of the t and u channel contributions suffices to reproduce the full correlator \eqref{JJO1O1asTCB}.
\subsection*{$\mathbf{\langle J J J J\rangle}$}
We compute this correlator via Wick contractions and find,
\small
\begin{align}\label{FBJJJJ}
    &\langle J(x_1,z_1)J(x_2,z_2)J(x_3,z_3)J(x_4,z_4)\rangle_{FB}=\frac{1}{x_{13}^6 x_{24}^6(uv)^{5/2}}\Bigg(H_{12}\bigg(v^2(1+\sqrt{v})H_{34}+v V_{3,42}V_{4,31}+v^{7/2}V_{3,41}V_{4,32}\bigg)\notag\\
    &+H_{13}uv\bigg( H_{24}u(\sqrt{u}+\sqrt{v})v+\sqrt{u}V_{2,13}(V_{4,31}-v V_{4,32})+v V_{2,14}(-\sqrt{u}V_{4,31}+(\sqrt{u}+\sqrt{v})v V_{4,32})\bigg)\notag\\
    &+H_{14}\bigg(u^{3/2}(H_{23}\sqrt{u}(1+\sqrt{u})+v V_{2,14}(v V_{3,41}-V_{3,42}))+u V_{2,13}(V_{3,42}+\sqrt{u}(-v V_{3,41}+V_{3,42}))\bigg)\notag\\
    &+H_{23}\bigg(u^{3/2}v V_{1,23}(-V_{4,31}+v V_{4,32})+u V_{1,24}((1+\sqrt{u})V_{4,31}-\sqrt{u}v V_{4,32})\bigg)\notag\\
    &+H_{24}uv\bigg(\sqrt{u}V_{1,24}(-v V_{3,41}+V_{3,42})+v V_{1,23}((\sqrt{u}+\sqrt{v})v V_{3,41}-\sqrt{u}V_{3,42})\bigg)+H_{34}\bigg( v V_{1,24}V_{2,13}+v^{7/2}V_{1,23}V_{2,14}\bigg)\notag\\&+V_{1,23}V_{2,13}\sqrt{u}v(v V_{3,41}-V_{3,42})(V_{4,31}-v V_{4,32})+V_{1,23}V_{2,14}v^2\bigg(\sqrt{u}V_{3,42}(V_{4,31}-v V_{4,32})\notag\\&+v V_{3,41}(-\sqrt{u}V_{4,31}+(\sqrt{u}+\sqrt{v})v V_{4,32})+V_{1,24}V_{2,13}+V_{1,24}V_{2,14}\sqrt{u}v(-vV_{3,41}+V_{3,42})(-V_{4,31}+v V_{4,32})\notag\\
    &+(V_{3,42}+\sqrt{u}(-v V_{3,41}+V_{3,42}))V_{4,31}+\sqrt{u}v(v V_{3,41}-V_{3,42})V_{4,32}\bigg).
\end{align}
\normalsize
We find that this correlator can be written as a sum of $s,t$ and $u$ channel $\tau=1$ TCBs. The s channel TCB is given by\footnote{Unlike the previous examples, we did not present a systematic analysis of twist conformal blocks for $\langle JJJJ\rangle$ due to the technical difficulty of the rising number of linearly independent tensor structures.},
\begin{align}\label{JJJJsTCB}
    &\mathcal{H}_s^{(JJ|1|JJ)}=\frac{2}{x_{13}^6 x_{24}^6 u^{5/2}v^{5/2}}\bigg( H_{14}H_{23}u^2+ H_{12}H_{34}v^2+ H_{12}H_{34}v^{5/2}+H_{13}H_{24}u^2 v^{5/2}+ H_{14} V_{2,13}V_{3,42}u\notag\\
    &+H_{12}V_{3,42}V_{4,31}v+ H_{23}V_{1,24}V_{4,31}u+V_{1,24}V_{2,13}H_{34} v+V_{1,24}V_{2,13}V_{3,42}V_{4,31}+ H_{13}V_{2,14}V_{4,32}u v^{7/2}\notag\\
    &+ H_{12}V_{3,41}V_{4,32}v^{7/2}+ H_{24}V_{1,23}V_{3,41} u v^{7/2}+H_{34}V_{1,23}V_{2,14}v^{7/2}+V_{1,23}V_{2,14}V_{3,41}V_{4,32}v^{9/2}\bigg),
\end{align}
and the $t$ and $u$ channel expressions are obtained by $(2\leftrightarrow 4)$ and $(2\leftrightarrow 3)$ exchanges respectively. We see that,
\begin{align}\label{JJJJasTCB}
    \langle JJJJ\rangle_{FB}=\frac{1}{2}\bigg(\mathcal{H}_s^{(JJ|1|JJ)}+\mathcal{H}_t^{(JJ|1|JJ)}+\mathcal{H}_u^{(JJ|1|JJ)}\bigg).
\end{align}
Thus, we have decomposed a correlator with all spinning insertions into a sum of single trace TCBs thus indicating that all correlators in the FB theory will have a similar decomposition as well.\\

Also, as with the other examples, we can write \eqref{FBJJJJ} as a sum of single trace and double trace contributions,
\begin{align}
    \langle JJJJ\rangle_{FB}=\mathcal{H}_s^{\text{single-trace}}+\mathcal{H}_s^{\text{double-trace}},
\end{align}
where $\mathcal{H}_s^{\text{single-trace}}$ is exactly the $\tau=1$ TCB \eqref{JJJJsTCB} and $\mathcal{H}_s^{\text{double-trace}}$ is the difference between the correlator \eqref{FBJJJJ} and $\mathcal{H}_s^{\text{single-trace}}$, 
\small
\begin{align}
    &\mathcal{H}_s^{\text{single-trace}}=\mathcal{H}_s^{(JJ|1|JJ)},\notag\\
    &\mathcal{H}_s^{\text{double-trace}}=\frac{1}{x_{13}^6 x_{24}^6 u^2 v^{5/2}}\Bigg( u\bigg(u(H_{14}H_{23}+H_{13}H_{24}v^2)+V_{2,13}(-H_{14}v V_{3,41}+H_{14}V_{3,42}+H_{13}v(V_{4,31}-v V_{4,32}))\notag\\
&+v V_{2,14}(H_{14}v V_{3,41}-H_{14}V_{3,42}+H_{13}v(-V_{4,31}+v V_{4,32}))\bigg)+v V_{1,23}\bigg(H_{23}u(-V_{4,31}+v V_{4,32})+v V_{3,41}(H_{24}u v\notag\\&
+V_{2,13}(V_{4,31}-v V_{4,32})+v V_{2,14}(-V_{4,31}+v V_{4,32}))-V_{3,42}(H_{24}u v+V_{2,13}(V_{4,31}-v V_{4,32})+v V_{2,14}(-V_{4,31}+v V_{4,32}))\bigg)\notag\\
&+V_{1,24}\bigg(H_{23}u(V_{4,31}-v V_{4,32})-v V_{3,41}(H_{24}uv+V_{2,13}(V_{4,31}-v V_{4,32})+v V_{2,14}(-V_{4,31}+v V_{4,32}))+V_{3,42}(H_{24}uv\notag\\
&+V_{2,13}(V_{4,31}-v V_{4,32})+v V_{2,14}(-V_{4,13}+v V_{4,32}))\bigg)\Bigg).
\end{align}
\normalsize
using which one can check that the double discontinuity of the full correlator comes entirely due to $\mathcal{H}_s^{\text{single-trace}}$.
\subsubsection{Bulk Locality}
We shall now investigate the locality of the bulk dual of the free bosonic theory following \cite{Sleight_2018}. Just as in that paper, our criteria for the non-locality of the bulk amplitudes is if the bulk contact diagram contains a piece proportional to the exchange diagrams\footnote{There exist more sophisticated criteria for bulk locality between the string scale and AdS radius \cite{Maldacena:2015iua} based on the singularity structure of the correlators. However, in this paper, we shall be primarily investigating the bulk locality in the language of \cite{Sleight_2018}}.
\subsubsection*{Four point scalar amplitude}
By the higher spin-CFT duality, this correlator is dual to a four-point scalar amplitude $\mathcal{A}_4$ in the AdS bulk which can be written as,
\begin{align}\label{bulkboundarydictionary}
\langle O_1 O_1 O_1 O_1\rangle_{FB}=\mathcal{A}_4=\mathcal{A}_{\text{exchange}}+\mathcal{A}_{\text{contact}},
\end{align}
where,
\begin{align}\label{ONHSS}
\mathcal{A}_{\text{exchange}}=\mathcal{A}_{s}+\mathcal{A}_t+\mathcal{A}_u.
\end{align}
By the bulk-boundary dictionary, we have\footnote{The $\cdots$ indicate some double trace contributions which correspond to local contact terms in the bulk, see \cite{Sleight_2018} for instance. We shall henceforth suppress the $\cdots$ henceforth since it does not affect our conclusions about the bulk locality.},
\begin{align}
    &\mathcal{A}_s=\mathcal{H}_s^{(O_1O_1|1|O_1O_1)}+\cdots,\mathcal{A}_t=\mathcal{H}_t^{(O_1O_1|1|O_1O_1)
    }+\cdots,\mathcal{A}_u=\mathcal{H}_u^{(O_1O_1|1|O_1O_1)}+\cdots.
\end{align}
Thus we see that the $\langle O_1O_1O_1O_1\rangle_{FB}$ correlator \eqref{O1O1O1O1FBasTCB} can be written as,
\begin{align}\label{OOOOstu}
    &\langle O_1 O_1 O_1 O_1\rangle_{FB}=\frac{1}{2}\bigg(\mathcal{A}_s+\mathcal{A}_t+\mathcal{A}_u\bigg)=\frac{1}{2}\mathcal{A}_{\text{exchange}},
\end{align}
which together with \eqref{bulkboundarydictionary} implies that,
\begin{align}
    \mathcal{A}_{\text{contact}}=\langle O_1O_1O_1O_1\rangle_{FB}-\mathcal{A}_{\text{exchange}}=-\frac{1}{2}\mathcal{A}_{\text{exchange}},
\end{align}
thereby revealing the non-locality of the bulk scalar amplitude at the four-point level as noticed in \cite{Sleight_2018}. Let us now see what the story is at the five-point level.
\subsubsection*{Five point scalar amplitude}
In the bulk language, we have,
\begin{align}\label{5ptbulk}
\langle O_1O_1O_1O_1O_1\rangle_{FB}=\mathcal{A}_5=\mathcal{A}_{\text{ exchange-1}}+\mathcal{A}_{\text{ exchange-2}}+\mathcal{A}_{\text{contact}},
\end{align}
where,
\begin{align}\label{exchange1}
    \mathcal{A}_{\text{exchange-1}}=\mathcal{A}_s+\mathcal{A}_t+\mathcal{A}_u+\mathcal{A}_v,
\end{align}
and by the dictionary that relates bulk exchange amplitudes to boundary single trace TCBs,
\begin{align}\label{stuv}
    &\mathcal{A}_s=\mathcal{H}_s^{(O_1O_1|1|O_1O_1O_1)},\mathcal{A}_t=\mathcal{H}_t^{(O_1O_1|1|O_1O_1O_1)},\mathcal{A}_u=\mathcal{H}_u^{(O_1O_1|1|O_1O_1O_1)},\mathcal{A}_v=\mathcal{H}_v^{(O_1O_1|1|O_1O_1O_1)}.
\end{align}
$\mathcal{A}_{exchange-2}$ represent the other exchange diagrams in the bulk and $\mathcal{A}_{\text{contact}}$ is the five point contact diagram.

Thus we see using \eqref{OOOOOTCB}, \eqref{5ptbulk}, \eqref{exchange1} and \eqref{stuv} that,
\begin{align}
    \langle O_1O_1O_1O_1O_1\rangle_{FB}=\mathcal{A}_5=\frac{1}{2}\mathcal{A}_{\text{exchange-1}}.
\end{align}
Using these equations we find,
\begin{align}
    &\mathcal{A}_{\text{exchange-2}}+\mathcal{A}_{\text{contact}}=-\frac{1}{2}\bigg(\mathcal{H}_s^{(O_1O_1|1|O_1O_1O_1)}+\mathcal{H}_u^{(O_1O_1|1|O_1O_1O_1)}+\mathcal{H}_t^{(O_1O_1|1|O_1O_1O_1)}+\mathcal{H}_v^{(O_1O_1|1|O_1O_1O_1)}\bigg)\notag\\&=-\frac{1}{2}\mathcal{A}_{\text{exchange-1}}.
\end{align}
Thus, we see that the singularity structure of the bulk diagrams indicates the presence of non-locality since the contact diagram and exchange diagrams involving quartic vertices ($\mathcal{A}_{\text{exchange-2}}$) contain singularities that should only appear in $\mathcal{A}_{\text{exchange-1}}$. 
\subsubsection*{Four point spinning amplitudes}
\subsubsection*{$\mathbf{\langle TO_1O_1O_1\rangle}$}
 In the bulk language, this correlator is given by a one graviton-three scalar amplitude,
\begin{align}
    \langle TO_1O_1O_1\rangle_{FB}=\mathcal{A}_4=\mathcal{A}_{\text{exchange}}+\mathcal{A}_{\text{contact}}=\mathcal{A}_s+\mathcal{A}_t+\mathcal{A}_u+\mathcal{A}_{\text{contact}}.
\end{align}
Using the dictionary between bulk exchange amplitudes and single trace twist conformal blocks we have,
\begin{align}
    &\mathcal{A}_s=\mathcal{H}^{(TO_1|1|O_1O_1)}_s,\mathcal{A}_t=\mathcal{H}^{(TO_1|1|O_1O_1)}_t,\mathcal{A}_u=\mathcal{H}^{(TO_1|1|O_1O_1)}_u,
\end{align}
which using \eqref{TO1O1O1asTCB} implies that,
\begin{align}
    \mathcal{A}_{\text{contact}}=-\frac{1}{2}\bigg(\mathcal{H}^{(TO_1|1|O_1O_1)}_s+\mathcal{H}^{(TO_1|1|O_1O_1)}_t+\mathcal{H}^{(TO_1|1|O_1O_1)}_u\bigg)=-\frac{1}{2}\mathcal{A}_{\text{exchange}},
\end{align}
thus showing a non-locality at the level of spinning four-point functions.
\subsubsection*{$\mathbf{\langle J_4O_1O_1O_1\rangle}$}
By the same analysis as for $\langle TO_1O_1O_1\rangle_{FB}$ we find using \eqref{J4O1O1O1asTCB} and the bulk-boundary dictionary that the bulk contact diagram is proportional to the exchange diagrams,
\begin{align}
    \mathcal{A}_{\text{contact}}=-\frac{1}{2}\bigg(\mathcal{H}^{(J_4O_1|1|O_1O_1)}_s+\mathcal{H}^{(J_4O_1|1|O_1O_1)}_t+\mathcal{H}^{(J_4O_1|1|O_1O_1)}_u\bigg)=-\frac{1}{2}\mathcal{A}_{\text{exchange}}.
\end{align}
\subsubsection*{$\mathbf{\langle JJO_1O_1\rangle}$}
In the bulk language, our result \eqref{JJO1O1asTCB} implies that,
\begin{align}\label{JJOOnonlocal}
    \mathcal{A}_s+\mathcal{A}_{\text{contact}}=-\frac{1}{2}\bigg( \mathcal{A}_t+ \mathcal{A}_u\bigg)\implies \mathcal{A}_{\text{contact}}=  -\frac{1}{2}\bigg(2 \mathcal{A}_s+\mathcal{A}_t+ \mathcal{A}_u\bigg).
\end{align}
 This equation, although different from the previous examples in that the s channel exchange constitutes twice as much of the contact diagram compared to the t and u channels, is still yet another indicator of the non-locality of the bulk theory.
\subsubsection*{$\mathbf{\langle JJJJ\rangle}$}
By the same arguments as that of the previous examples we find using \eqref{JJJJasTCB} that the bulk contact diagram is proportional to the exchanges:
 \begin{align}
     \mathcal{A}_{\text{contact}}=-\frac{1}{2}\mathcal{A}_{\text{exchange}}.
 \end{align}
 \subsection*{Summary}
We have found at the five-point level that the scalar correlator can be written in terms of $\tau=1$ TCBs. We also analyzed several spinning correlators in this theory and found that they too can be decomposed into $\tau=1$ TCBs. Based on this pattern, we expect every correlator in this theory to have a similar decomposition and hence establish that the dual bulk theory is non-local\footnote{All n point correlators in the free bosonic and free fermionic theories were computed in \cite{Didenko:2013bj}. To check whether they are all made up of $\tau=1$ TCBs, one first needs to compute the spinning TCBs which for general spins becomes increasingly complicated as we have seen explicitly for \eqref{JJJJasTCB}.}.
\subsection{Free Fermionic Theory}\label{sec:FFtheory}
In \cite{Sleight_2018}, the scalar four-point function in the free bosonic theory was considered, and in subsection \ref{sec:FBtheory}, we have analyzed the scalar five-point function and several spinning correlators. In this subsection, We shall now investigate another important theory, that is, the free fermionic theory.\\

Consider a theory of $SU(N)$ massless free Dirac fermions in three dimensions:
\begin{align}
    S=\int d^3 x~i \bar{\psi}^a(\slashed{\partial})\psi^a.
\end{align}
This theory, apart from being a CFT also contains conserved currents of all positive integer spins. It's spectrum of primary operators contains a scalar $O_2(x)=\bar{\psi}^a\psi^a(x)$ with scaling dimension $\Delta=2$ and a tower of conserved currents $J_{s},s=1,2,3,4\cdots,$ all twist $\tau=1$. For explicit expressions for these currents please refer to appendix \ref{sec:FBFFCBQFoperators}.
\subsubsection{Scalar Correlators}
From Wick contractions we obtain,
\small
\begin{align}\label{O2O2O2O2FF}
    \langle O_2O_2O_2O_2\rangle=\frac{1}{2x_{12}^4 x_{34}^4}\frac{\sqrt{u}}{ v^{3/2}}(-1-u^{5/2}+v-(-1+v)v^{3/2}+u^{3/2}(1+v)+u(1+v^{3/2})\bigg).
\end{align}
\normalsize
Just as in the case of the free bosonic scalar four-point function, we can decompose this correlator into a sum of s,t and u channel TCBs. Using the $\tau=1$ TCB that we obtained earlier \eqref{O2O2O2O2sTCB} we find that,
\begin{align}\label{FFO2O2O2O2asTCB}
     \langle O_2O_2O_2O_2\rangle_{FF}=\frac{1}{2}\bigg( \mathcal{H}_s^{(O_2O_2|1|O_2O_2)}+\mathcal{H}_t^{(O_2O_2|1|O_2O_2)}+\mathcal{H}_u^{(O_2O_2|1|O_2O_2)}\bigg).
\end{align}
It is very interesting to note that our steps to determine the TCB outputted the free fermionic result which has no $O_1$ exchange despite us not providing that information while computing the TCB.\\

Now, just as we did for the FB correlators, we can decompose FF theory correlators into sums of single-trace and double-trace contributions based on their contribution to the double discontinuity of the correlator. We have,
\begin{align}\label{O2O2O2O2stdt}
    \langle O_2O_2O_2O_2\rangle_{FF}=\mathcal{H}_s^{\text{single-trace}}+\mathcal{H}_s^{\text{double-trace}},
\end{align}
where $\mathcal{H}_s^{\text{single-trace}}$ is exactly the $\tau=1$ TCB \eqref{O2O2O2O2sTCB} which accounts for the full double discontinuity of the correlator and $\mathcal{H}_s^{\text{double-trace}}$ is the difference between the full correlator \eqref{O2O2O2O2FF} and $\mathcal{H}_s^{\text{single-trace}}$ which has zero double discontinuity.
\begin{align}
    &\mathcal{H}_s^{\text{single-trace}}=\mathcal{H}_s^{(O_2O_2|1|O_2O_2)},\notag\\
&\mathcal{H}_s^{\text{double-trace}}=\frac{1}{2 x_{12}^4 x_{34}^4}\frac{u^2(1-u+v)}{v^{3/2}}.
\end{align}
\subsubsection{Spinning Correlators}
\subsection*{$\mathbf{\langle T O_2 O_2 O_2\rangle}$}
 From Wick contractions we obtain,
\begin{align}\label{FFTO2O2O2}
    \langle TO_2O_2O_2\rangle_{FF}=\frac{x_{23}x_{34}}{x_{13}^6 x_{12}^2 x_{14}^2 x_{24}^3}S_{1,234}\bigg(V_{1,23}\frac{v}{u^{3/2}}-V_{1,24}\frac{1}{(uv)^{3/2}}+V_{1,34}\frac{u}{v^{3/2}}\bigg).
\end{align}
We see that this expression can also be decomposed into a sum of s,t and u channel parity odd $\tau=1$ TCBs that we obtained earlier \eqref{TO2O2O2odds}. We see that,
\begin{align}\label{FFTO2O2O2asTCB}
    \langle TO_2O_2O_2\rangle=\frac{1}{2}\bigg(\mathcal{H}^{(TO_2|1|O_2O_2)}_{s,\text{odd}}+\mathcal{H}^{(TO_2|1|O_2O_2)}_{t,\text{odd}}+\mathcal{H}^{(TO_2|1|O_2O_2)}_{u,\text{odd}}\bigg),
\end{align}
thus decomposing a parity odd correlator in terms of single trace TCBs.\\

Again, like the scalar case \eqref{O2O2O2O2stdt}, we can write this as a sum of single trace and double trace contributions based on the fact that the double discontinuity of the full correlator entirely comes from the former.
\begin{align}
    \langle TO_2O_2O_2\rangle_{FF}=\mathcal{H}_s^{\text{single-trace}}+\mathcal{H}_s^{\text{double-trace}},
\end{align}
where $\mathcal{H}_s^{\text{single-trace}}$ is the s channel $\tau=1$ TCB \eqref{TO2O2O2odds} whereas $\mathcal{H}_s^{\text{double-trace}}$ is the difference between the full correlator \eqref{FFTO2O2O2} $\mathcal{H}_s^{\text{single-trace}}$ and has zero double discontinuity.
\begin{align}
   &\mathcal{H}_s^{\text{single-trace}}=\mathcal{H}^{(TO_2|1|O_2O_2)}_{s,\text{odd}},\notag\\
   &\mathcal{H}_s^{\text{double-trace}}=\frac{x_{23}x_{34}}{x_{13}^6 x_{12}^2 x_{14}^2 x_{24}^3}S_{1,234}\bigg(V_{1,34}\frac{u}{v^{3/2}}\bigg).
\end{align}
\subsection*{$\mathbf{\langle J J O_2 O_2 \rangle}$}
We obtain via Wick contractions,
\begin{align}\label{JJO2O2FF}
    &\langle JJO_2O_2\rangle_{FF}=\frac{2 x_{34}^2}{x_{13}^6 x_{24}^6 u^{5/2}v^{3/2}}\bigg(W_1 W_2 f_1(u,v)+\Bar{W}_1\Bar{W}_2 f_2(u,v)+(W_1 \Bar{W}_2-\Bar{W}_1 W_2)f_3(u,v)+H_{12}f_4(u,v)\bigg).
\end{align}
where,
\begin{align}
    &f_1(u,v)=\frac{1}{2}\bigg(-1+v-(v-1)v^{3/2}+u^{3/2}(1+v)\bigg)\notag\\&f_2(u,v)=\frac{1}{2}\bigg(1-v+(v-1)v^{3/2}-u^{3/2}(1+v)\bigg)\notag\\
    &f_3(u,v)=\frac{1}{2}\bigg(-1-u^{3/2}(v-1)-v+v^{3/2}(1+v)\bigg)\notag\\&f_4(u,v)=-1+u^{5/2}+v-(v-1)v^{3/2}-u^{3/2}(1+v)+u(1+v^{3/2}).
\end{align}
We see that much like the free bosonic $\langle JJO_1O_1\rangle$ that this correlator too can be written purely in terms of the $\tau=1$ t and u channel TCBs that we obtained earlier \eqref{JJO2O2t}. We see that,
\begin{align}\label{FFJJO2O2asTCB}
    \langle JJO_2O_2\rangle_{FF}=\frac{1}{2}\bigg(\mathcal{H}^{(JO_2|1|JO_2)}_t+\mathcal{H}^{(JO_2|1|JO_2)}_u\bigg),
\end{align}
which is very much analogous to the free bosonic result for $\langle JJO_1O_1\rangle $\eqref{JJO1O1asTCB}. Again, we can decompose this correlator in the s channel into a sum of single-trace and double-trace contributions,
\begin{align}
    \langle JJO_2O_2\rangle_{FF}=\mathcal{H}_s^{\text{single-trace}}+\mathcal{H}_s^{\text{double-trace}},
\end{align}
where the former is given by the $\tau=1$ s channel TCB \eqref{JJO2O2s} and accounts for the entire double discontinuity of the correlator whereas the latter has zero double discontinuity and is given by the difference between the full correlator and $\mathcal{H}_s^{\text{single-trace}}$.
\small
\begin{align}
    &\mathcal{H}_s^{\text{single-trace}}=\mathcal{H}_s^{(JJ|1|O_2O_2))},\notag\\&
    \mathcal{H}_s^{\text{double-trace}}=\frac{x_{34}^2}{x_{13}^6 x_{24}^6}\bigg(2 H_{12}(-1+u-v)+(1+v)W_1 W_2+(v-1)\Bar{W}_1 W_2+W_1\Bar{W}_2-(v W_1+(1+v)\Bar{W}_1)\Bar{W}_2\bigg).
\end{align}
\normalsize
\subsubsection{Bulk Locality}
We shall now probe the locality of the bulk dual of the free fermionic theory.
\subsubsection*{Four point scalar amplitude}
In the bulk language, we have,
\begin{align}
    \langle O_2O_2O_2O_2\rangle=\mathcal{A}_4=\mathcal{A}_{\text{exchange}}+\mathcal{A}_{\text{contact}}=\mathcal{A}_s+\mathcal{A}_t+\mathcal{A}_u+\mathcal{A}_{\text{contact}}.
\end{align}
Using the fact that,
\begin{align}
    &\mathcal{A}_s= \mathcal{H}_s^{(O_2O_2|1|O_2O_2)},\mathcal{A}_t= \mathcal{H}_t^{(O_2O_2|1|O_2O_2)},\mathcal{A}_u=\mathcal{H}_u^{(O_2O_2|1|O_2O_2)},
\end{align}
we see using \eqref{FFO2O2O2O2asTCB} that,
\begin{align}
    \mathcal{A}_{\text{contact}}=-\frac{1}{2}\mathcal{A}_{\text{exchange}},
\end{align}
thus showing that the dual bulk theory exhibits non-locality at the level of scalar four-point functions.
\subsubsection*{Four point spinning amplitudes}
\subsubsection*{$\mathbf{\langle TO_2O_2O_2\rangle}$}
By the usual arguments and using \eqref{FFTO2O2O2asTCB} we find that the bulk contact diagram is proportional to the exchange diagrams,
\begin{align}
\mathcal{A}_{\text{contact}}=-\frac{1}{2}\bigg(\mathcal{H}^{(TO_2|1|O_2O_2)}_{s,\text{odd}}+\mathcal{H}^{(TO_2|1|O_2O_2)}_{t,\text{odd}}+\mathcal{H}^{(TO_2|1|O_2O_2)}_{u,\text{odd}}\bigg)=-\frac{1}{2}\mathcal{A}_{\text{exchange}}.
\end{align}
\subsubsection*{$\mathbf{\langle JJO_2O_2\rangle}$}
Using \eqref{FFJJO2O2asTCB} and the usual bulk-boundary dictionary we see that,

\begin{align}
    \mathcal{A}_{s}+\mathcal{A}_{\text{contact}}=-\frac{1}{2}\bigg(  \mathcal{A}_{t}+  \mathcal{A}_{u}\bigg)\implies \mathcal{A}_{\text{contact}}=-\frac{1}{2}\bigg( 2  \mathcal{A}_{s}+ \mathcal{A}_{t}+  \mathcal{A}_{u}\bigg).
\end{align}
Thus we see that this equation is just like its free bosonic theory bulk dual counterpart \eqref{JJOOnonlocal} and is manifestly non-local.
\subsection*{Summary}
Based on the examples considered, much like what we concluded for the free bosonic theory, we expect all correlators in this theory to be written in terms of single trace TCBs and hence conclude that the bulk dual of the free fermionic theory is non-local.
\section{Large $N$ Slightly Broken Higher Spin Theories}\label{sec:SBHSsec}
Large $N$ Slightly broken higher spin theories are a class of conformal field theories that possess an approximate higher spin symmetry, that is they possess a tower of currents $J_s$ which for $s>2$ are weakly non-conserved, that is, their non-conservation starts at $\order{\frac{1}{N}}$. The critical bosonic theory and Chern-Simons theories coupled to matter in the fundamental representation are examples of such theories. In this section we first analyze the structure of four-point correlators in the critical bosonic theory and then based on the results of \cite{Jain:2022ajd} we see the implications of our results for Chern Simons matter theories.\\

\subsection{Critical Bosonic Theory}\label{sec:CBtheory}
The critical bosonic theory is the theory that we obtain by flowing to the Wilson-Fisher fixed point of the following action:
\begin{align}
    S=\int d^3 x\bigg(\frac{1}{2}\partial_\mu \bar{\phi}^a\partial^\mu \phi^a+\frac{1}{2}m^2\phi^a\phi^a+\frac{\lambda_4}{N}(\phi^a\phi^a)^2\bigg).
\end{align}
At large $N$ and large $\lambda_4$, keeping $\kappa=\frac{\lambda_4}{N}$ fixed, the spectrum of this theory consists of a $\Delta=2+\order{\frac{1}{N}}$ scalar $O_2$, conserved spin one and spin two currents $J$ and $T$ and weakly non conserved currents $J_{s}$ with $s=3,4,\cdots$ with twist $\tau=1+\order{\frac{1}{N}}$. 
\subsubsection{Scalar Correlators}
We have,
\begin{align}
    \langle O_2(x_1) O_2(x_2) O_2(x_3) O_2(x_4)\rangle_{CB}=\int\frac{d^3 x_5 d^3 x_6 d^3 x_7 d^3 x_8}{x_{15}^4x_{26}^4x_{37}^4 x_{48}^4}\langle O_1(x_5)O_1(x_6)O_1(x_7)O_1(x_8)\rangle_{FB}.
\end{align}
Upon performing the integration \cite{Li:2019twz} we see that,
\begin{align}\label{CBO2O2O2O2asTCB}
    \langle O_2(x_1) O_2(x_2) O_2(x_3) O_2(x_4)\rangle_{CB}=\langle O_2(x_1) O_2(x_2) O_2(x_3) O_2(x_4)\rangle_{FF}.
\end{align}
We have already shown that the free fermionic scalar four-point function can be written in terms of $\tau=1$ TCBs \eqref{FFO2O2O2O2asTCB}. Thus, by the above equality, the same holds true for the critical bosonic scalar four-point function.
\subsubsection{Spinning Correlators}
\subsection*{$\mathbf{\langle T O_2 O_2 O_2\rangle}$}
We have\footnote{The contact terms represent the contribution to this correlator coming due to the $O_2$ conformal block. However since $\langle O_2(y_1)O_2(y_2)O_2(y_3)\rangle_{CB}\propto \delta^3(y_1-y_2)\delta^3(y_2-y_3)$ is a contact term, we can ignore this term while working at separated points. In fact, for any  $\langle J_sO_2O_2O_2\rangle_{CB}$, the contribution of the $O_2$ conformal block is a contact term and hence we can ignore it.},
\begin{align}\label{CBintTOOO}
     \langle T(x_1,z_1) O_2(x_2) O_2(x_3) O_2(x_4)\rangle_{CB}=\int\frac{d^3 x_5 d^3 x_6 d^3 x_7 }{x_{25}^4x_{36}^4x_{47}^4 }\langle T(x_1,z_1)O_1(x_5)O_1(x_6)O_1(x_7)\rangle_{FB}+\text{contact terms}.
\end{align}
We obtain \cite{Li:2019twz, Silva_2021},
\begin{align}\label{TO2O2O2CBcorrelator}
    &\langle TO_2O_2O_2\rangle_{CB}=\frac{x_{24}^3}{x_{12}^7x_{14}^3x_{34}^4}\bigg(\sqrt{u v}(1+u-v)(1+u+3v)V_{1,23}^2\notag\\&+\frac{\sqrt{u}}{v^{3/2}}(-1+u+v)(3+u+v)V_{1,24}^2+\frac{u^{5/2}}{v^{3/2}}(1-u+v)(1+3u+v)V_{1,34}^2\bigg).
\end{align}
We now write this correlator in the following form:
\begin{align}
    \langle TO_2O_2O_2\rangle_{CB}=\frac{1}{2}\bigg(\langle TO_2O_2O_2\rangle_s+\langle TO_2O_2O_2\rangle_t+\langle TO_2O_2O_2\rangle_u\bigg),
\end{align}
where we identify the s channel expression as,
\begin{align}
   &\langle TO_2O_2O_2\rangle_{s}=\frac{x_{24}^3}{x_{12}^7x_{14}^3x_{34}^4}\bigg(\sqrt{u v}(1+u-v)(1+u+3v)V_{1,23}^2+\frac{\sqrt{u}}{v^{3/2}}(-1+u+v)(3+u+v)V_{1,24}^2\bigg).
\end{align}
Plugging this into the TCB eigenvalue equation \eqref{3dAldayEquationSpinning} shows that it is not $\tau=1$ TCB. However, it can be written as a TCB plus something extra. First, we attempt to form a $\tau=1$ TCB by following the algorithm in section \ref{sec:spinningTCB}. We then find the following expression that solves Alday's s channel equation \eqref{3dAldayEquationSpinning},
\begin{align}\label{TO2O2O2CBtcb}
    &\mathcal{H}^{(TO_2|1|O_2O_2)}_s=\frac{x_{24}^3}{x_{12}^7 x_{14}^3 x_{34}^4}\bigg(\sqrt{u}v^{3/2}(1+u-v)V_{1,23}^2+\frac{\sqrt{u}}{v^{3/2}}(-1+u+v)V_{1,24}^2\bigg),
\end{align}
and by obtaining the t and u channel solutions by $(2\leftrightarrow 4)$ and $(2\leftrightarrow 3)$ exchanges on $\mathcal{H}^{(TO_2|1|O_2O_2)}_s$ and adding all of them up we get,
\begin{align}\label{TOOOCBtcb}
   &\mathcal{H}^{(TO_2|1|O_2O_2)}=\frac{1}{2}\bigg(\mathcal{H}^{(TO_2|1|O_2O_2)}_s+\mathcal{H}^{(TO_2|1|O_2O_2)}_t+\mathcal{H}^{(TO_2|1|O_2O_2)}_u\bigg)\notag\\&=\frac{x_{24}^3}{x_{12}^7 x_{14}^3 x_{34}^4}\bigg(\sqrt{u}v^{3/2}(1+u-v)V_{1,23}^2+\frac{\sqrt{u}}{v^{3/2}}(-1+u+v)V_{1,24}^2+\frac{u^{7/2}}{v^{3/2}}(1-u+v)V_{1,34}^2\bigg).
\end{align}
Comparing with \eqref{TO2O2O2CBcorrelator}, we see that $\langle TO_2O_2O_2\rangle_{CB}$ can be written as a sum of $\mathcal{H}^{(TO_2|1|O_2O_2)}$ and an extra piece whose expression is given by,
\begin{align}\label{extrapieceTOOOCB}
    &\langle TO_2O_2O_2\rangle_{Extra}=\frac{x_{24}^3}{x_{12}^7x_{14}^3 x_{34}^4}\bigg(\sqrt{u v}(1+u)(1+u-v)V_{1,23}^2\notag\\
    &+\frac{\sqrt{u}}{v^{3/2}}(-1+u+v)(u+v)V_{1,24}^2-\frac{u^{5/2}}{v^{3/2}}(-1+u-v)(1+v)V_{1,34}^2\bigg).
\end{align}
We see that,
\begin{align}\label{TO2O2O2CBtcbplusextra}
    \langle TO_2O_2O_2\rangle_{CB}=\mathcal{H}^{(TO_2|1|O_2O_2)}+\langle TO_2O_2O_2\rangle_{Extra}
\end{align}
thus showing that the full correlator is not equivalent to a $\tau=1$ TCB.\\

There is also another way to obtain this extra piece starting from the TCB $\mathcal{H}^{(TO_2|1|O_2O_2)}$. We notice that $\mathcal{H}^{(TO_2|1|O_2O_2)}$ is not conserved with respect to the stress tensor at non-coincident points. We then write an ansatz for an expression to add to $\mathcal{H}^{(TO_2|1|O_2O_2)}$ to make it conserved. This expression turns out to be the same as the extra piece \eqref{extrapieceTOOOCB} which when added to $\mathcal{H}^{(TO_2|1|O_2O_2)}$ reproduces the CB theory correlator \eqref{TO2O2O2CBtcbplusextra}. Perhaps, this method could be an algorithm to construct CB theory correlators with conserved currents. \\

Now, as we did for all the free theory correlators, let us write the CB correlator as a sum of single-trace and double-trace contributions.
\begin{align}
    \langle TO_2O_2O_2\rangle_{CB}=\mathcal{H}_s^{\text{single-trace}}+\mathcal{H}_s^{\text{double-trace}}.
\end{align}
However, in contrast to the earlier cases, these contributions cannot be distinguished by their contribution to the double discontinuity of the full correlator as both contributions actually are zero as we can see from equations \eqref{TOOOrtau1dDisc1} and \eqref{OOdoubletraceTOOO} in appendix \ref{sec:dDisc}.
\begin{align}\label{TO2O2O2CBdDisc}
    \mathbf{dDisc_s}(\langle TO_2O_2O_2\rangle_{CB})=\mathbf{dDisc}(\mathcal{H}_s^{\text{single-trace}})=\mathbf{dDisc}(\mathcal{H}_s^{\text{double-trace}})=0.
\end{align}
However, we have independently calculated the single trace contribution (\eqref{TO2O2O2CBtcb} for the s channel $\tau=1$ TCB) and hence the difference between the full correlator \eqref{TO2O2O2CBcorrelator} and the s channel $\tau=1$ TCB \eqref{TO2O2O2CBtcb} is precisely the s channel double trace contribution.

\begin{align}
&\mathcal{H}_s^{\text{double-trace}}=\frac{x_{24}^3}{x_{12}^7 x_{14}^3 x_{34}^4}\frac{\sqrt{u}}{v^{3/2}}\bigg(V_{1,23}^2 (1+u)(1+u-v)v^2+V_{1,24}^2(u+v)(-1+u+v)\notag\\&~~~~~~~~~~~~~~~~~~~~~~~~~~~~~~~~~~~~~-V_{1,34}^2(1+3u+v)(-1+u-v)u^2\bigg).
\end{align}

\subsection*{$\mathbf{\langle J_4 O_2 O_2 O_2\rangle}$}
This correlator is given by
\begin{align}
    \langle J_4(x_1,z_1) O_2(x_2) O_2(x_3) O_2(x_4)\rangle_{CB}=\int\frac{d^3 x_5 d^3 x_6 d^3 x_7 }{x_{25}^4x_{36}^4x_{47}^4 }\langle J_4(x_1,z_1)O_1(x_5)O_1(x_6)O_1(x_7)\rangle_{FB}+\text{contact terms}.
\end{align}
This equals \cite{Silva_2021},
\begin{align}\label{J4OOOCB}
    &\langle J_4 OOO\rangle_{CB}=\frac{x_{34}^3}{x_{12}^4 x_{13}^7 x_{14}^7}\bigg(V_{1,23}^4 F_1(u,v)+V_{1,24}^4(\frac{v}{u})^{\frac{3}{2}}F_1(\frac{v}{u},\frac{1}{u})+V_{1,34}^4(\frac{v}{u})^{3/2}F_1(u,v)\notag\\&+V_{1,23}^2 V_{1,24}^2F_2(u,v)+V_{1,23}^2V_{1,34}^2\frac{1}{u^{3/2}}F_2(\frac{1}{u},\frac{v}{u})+V_{1,24}^2V_{1,34}^2(\frac{v}{u})^{\frac{3}{2}}F_2(v,u)\bigg),
\end{align}
where,
\begin{align}\label{F1F2}
   &F_1(u,v)=\frac{1}{u^3 (\sqrt{u}+\sqrt{v}+1)^3}v^{5/2}(15 u^{7/2}+45 u^3 (\sqrt{v}+1)+45 \sqrt{u} (\sqrt{v}+1)^2+15 (\sqrt{v}+1)^3\notag\\&
   +(u^{3/2} (8 v+33 \sqrt{v}))+u^{5/2} (45 v+90 \sqrt{v}+41)+3 u^2 (5 v+15 \sqrt{v}+11) \sqrt{v}+u (-8 v+33 \sqrt{v}+41))\notag\\
   &\text{and}\notag\\
   &F_2(u,v)=-\frac{1}{u^3 (\sqrt{u}+\sqrt{v}+1)^3}\sqrt{v}(4 u^{5/2} (17 v+42 \sqrt{v}+17)+28 u^{7/2}+84 u^3 (\sqrt{v}+1)\notag\\
   &-4 u^2 (8 v^{3/2}-9 v-9 \sqrt{v}+8)+9 \sqrt{u} (7 v^2-4 v+7) (\sqrt{v}+1)^2+3 (7 v^2-4 v+7) (\sqrt{v}+1)^3\notag\\
   &u^{3/2} (132 v^{3/2}+51 v^2+112 v+132 \sqrt{v}+51)+7 u (-20 v^{3/2}+5 v^{5/2}-3 v^2-20 v-3 \sqrt{v}+5)).
\end{align}
We can now write this answer as follows,
\begin{align}\label{J4O2O2O2CBnotTCB}
    \langle J_4 O_2O_2O_2\rangle_{CB}=\frac{1}{2}\bigg(\langle J_4 O_2 O_2 O_2\rangle_s+\langle J_4 O_2 O_2 O_2\rangle_t+\langle J_4 O_2 O_2 O_2\rangle_u\bigg),
\end{align}
where,
\small
\begin{align}\label{J4O2O2O2s}
    \langle J_4 O_2 O_2 O_2\rangle_s=\frac{x_{34}^3}{x_{12}^4 x_{13}^7 x_{14}^7}\bigg(V_{1,23}^4 F_1(u,v)+V_{1,24}^4(\frac{v}{u})^{\frac{3}{2}}F_1(\frac{v}{u},\frac{1}{u})+V_{1,23}^2V_{1,34}^2\frac{1}{u^{3/2}}F_2(\frac{1}{u},\frac{v}{u})+V_{1,24}^2V_{1,34}^2(\frac{v}{u})^{\frac{3}{2}}F_2(v,u)\bigg),
\end{align}
\normalsize
with $F_1(u,v)$ and $F_2(u,v)$ the same as in \eqref{F1F2}. Notice that this answer has a pole of the form $\frac{1}{(1+\sqrt{u}+\sqrt{v})^3}$ which is a bulk point singularity \cite{Silva_2021}. However, this does not indicate that the dual bulk theory is local since one would expect a more singular pole for local bulk contact diagrams \cite{Silva_2021}. Now, plugging \eqref{J4O2O2O2s} into Alday's s channel equation \eqref{3dAldayEquation} shows that it is not a s channel $\tau=1$ TCB\footnote{One can attempt to modify the algorithm of section \ref{sec:spinningTCB} by changing the fourth step from demanding conservation to instead requiring the correct non-conservation of $J_4$. However, we saw that this does not yield a valid TCB.}. Further, just like for $\langle TO_2O_2O_2\rangle_{CB}$ \eqref{TO2O2O2CBdDisc}, this correlator has zero double discontinuity as both the single trace and double trace operators do not contribute to the double discontinuity as can be checked using the formulae in appendix \ref{sec:dDisc}.
\subsection*{$\mathbf{\langle J_{s_1}O_2 O_2 O_2\rangle}$}
Correlators of this form up to $s_1=14$ were explicitly computed in \cite{Silva_2021}. Based on the examples considered above, we expect that these correlators are not entirely determined in terms of single-trace TCBs. For an abstract analysis at the integrand level, please refer to appendix \ref{sec: CBcorrelatorsappendix} where we show that these correlators are not single trace $\tau=1$ TCBs.

\subsection*{$\mathbf{\langle J J O_2 O_2 \rangle}$}
Based on analysis at the integrand level (please refer to appendix \ref{sec: CBcorrelatorsappendix}), we see that this correlator cannot be entirely written in terms of $\tau=1$ TCBs.
\subsection*{$\mathbf{\langle J_{s_1}J_{s_2}J_{s_3}O_2\rangle}$}
We have,
\small
\begin{align}\label{Js1Js2Js3O2CB}
     &\langle J_{s_1}(x_1)J_{s_2}(x_2)J_{s_3}(x_3)O_2(x_4)\rangle_{CB}=\int \frac{d^3 x_5}{x_{45}^4}\langle J_{s_1}(x_1)J_{s_2}(x_2)J_{s_3}(x_3)O_1(x_5)\rangle_{FB}\notag\\&-\frac{1}{N}\bigg[ \int \frac{d^3 x_5 d^3 x_6}{x_{56}^2}\langle J_{s_1}(x_1)O_2(x_4)O_2(x_5)\rangle_{CB}\langle O_2(x_6) J_{s_2}(x_2)J_{s_3}(x_3)\rangle_{CB}\notag\\
     &+(1\leftrightarrow 2)+(1\leftrightarrow 3)\bigg].
\end{align}
\normalsize
We re-write the correlator as a sum of various channel contributions\footnote{Based on the results of section \ref{sec:FBtheory} we use, $\langle J_{s_1}J_{s_2}J_{s_3}O_1\rangle_{FB}=\frac{1}{2}\bigg(\mathcal{H}_s^{(J_{s_1}J_{s_2}|1|J_{s_3}O_1)}+\mathcal{H}_t^{(J_{s_1}O_1|1|J_{s_2}J_{s_3})}+\mathcal{H}_u^{(J_{s_1}J_{s_3}|1|J_{s_2}O_1)}\bigg)$}. We have,
\small
\begin{align}\label{JJJOCBstu}
     &\langle J_{s_1}(x_1)J_{s_2}(x_2)J_{s_3}(x_3)O_2(x_4)\rangle_{CB}\notag\\&=\Bigg[\int \frac{d^3 x_5}{2x_{45}^4}\mathcal{H}_s^{(J_{s_1}J_{s_2}|1|J_{s_3}O_1)}(x_1,z_1,x_2,z_2,x_3,z_3,x_5)-\frac{1}{N}\int\frac{d^3 x_5 d^3 x_6}{x_{56}^2}\langle J_{s_1}(x_1)J_{s_2}(x_2)O_2(x_5)\rangle\langle O_2(x_6)J_{s_3}(x_3)O_2(x_4)\rangle\notag\\&+\int \frac{d^3 x_5}{2x_{45}^4}\mathcal{H}_s^{(J_{s_1}O_1|1|J_{s_2}J_{s_3})}(x_1,z_1,x_2,z_2,x_3,z_3,x_5)-\frac{1}{N}\int\frac{d^3 x_5 d^3 x_6}{x_{56}^2}\langle J_{s_1}(x_1)O_2(x_4)O_2(x_5)\rangle\langle O_2(x_6)J_{s_2}(x_2)J_{s_3}(x_3)\rangle\notag\\&+\int \frac{d^3 x_5}{2x_{45}^4}\mathcal{H}_s^{(J_{s_1}J_{s_3}|1|J_{s_2}O_1)}(x_1,z_1,x_2,z_2,x_3,z_3,x_5)-\frac{1}{N}\int\frac{d^3 x_5 d^3 x_6}{x_{56}^2}\langle J_{s_1}(x_1)J_{s_3}(x_2)O_2(x_5)\rangle\langle O_2(x_6)J_{s_2}(x_3)O_2(x_4)\rangle\Bigg]\notag\\&=\frac{1}{2}\bigg[\langle J_{s_1}J_{s_2}J_{s_3}O_2\rangle_{CB,s}+\langle J_{s_1}J_{s_2}J_{s_3}O_2\rangle_{CB,t}+\langle J_{s_1}J_{s_2}J_{s_3}O_2\rangle_{CB,u}\bigg],
\end{align}
\normalsize
where the the s,t and u channel expressions represent the second, third and fourth lines of \eqref{JJJOCBstu} respectively. For instance, consider the s channel expression,
\small
\begin{align}\label{JJJOCBeq1}
   &\langle J_{s_1}J_{s_2}J_{s_3}O_2\rangle_{CB,s}\notag\\&=\int \frac{d^3 x_5}{2x_{45}^4}\mathcal{H}_s^{(J_{s_1}J_{s_2}|1|J_{s_3}O_1)}(x_1,z_1,x_2,z_2,x_3,z_3,x_5)-\frac{1}{N}\int\frac{d^3 x_5 d^3 x_6}{x_{56}^2}\langle J_{s_1}(x_1)J_{s_2}(x_2)O_2(x_5)\rangle\langle O_2(x_6)J_{s_3}(x_3)O_2(x_4)\rangle.
\end{align}
The first term in \eqref{JJJOCBeq1} can easily be checked to be a $\tau=1$ TCB by applying Alday's s channel operator $\mathcal{A}_{12}$. The second term on the other hand is the $O_2$ conformal block and by acting with $\mathcal{A}_{12}$ we see that it is a $\tau=2$ TCB as it should be. Similarly, we find the $u$ channel expression to also be comprised of a $\tau=1$ TCB and a $\tau=2$ $O_2$ conformal block by acting with $\mathcal{A}_{13}$. As for the t channel expression, if we act on it  with $\mathcal{A}_{23}$\footnote{We use the fact that acting on a t channel expression with $\mathcal{A}_{23}$ is equivalent to acting on it with $\mathcal{A}_{14}$.}, it is clear that it too is a  sum of a $\tau=1$ TCB and a $\tau=2$ $O_2$ conformal block.\\

To summarize, correlators in the CB theory with three spinning operators and one scalar operator are indeed entirely given by $\tau=1$ and $\tau=2$ twist conformal blocks. For instance the first term in \eqref{JJJOCBeq1} (The s channel expression) is a $\tau=1$ TCB whilst the second is a $\tau=2$ TCB.
\normalsize
\subsection*{$\mathbf{\langle J_{s_1}J_{s_2}J_{s_3}J_{s_4}\rangle}$}
We have,
\small
\begin{align}\label{Js1Js2Js3Js4CB}
     &\langle J_{s_1}(x_1)J_{s_2}(x_2)J_{s_3}(x_3)J_{s_4}(x_4)\rangle_{CB}=\langle J_{s_1}(x_1)J_{s_2}(x_2)J_{s_3}(x_3)J_{s_4}(x_4)\rangle_{FB}\notag\\&-\frac{1}{N}\bigg[ \int \frac{d^3 x_5 d^3 x_6}{x_{56}^2}\langle J_{s_1}(x_1)J_{s_2}(x_2)O_2(x_5)\rangle_{CB}\langle O_2(x_6) J_{s_3}(x_3)J_{s_4}(x_4)\rangle_{CB}+(2\leftrightarrow 3)+(2\leftrightarrow 4)\bigg].
\end{align}
Based on our discussion in section \ref{sec:FBtheory}, the first term in \eqref{Js1Js2Js3Js4CB} is a $\tau=1$ TCB. The term in the second line of \eqref{Js1Js2Js3Js4CB} is the contribution coming due to the $O_2$ exchange and can easily be checked to be a $\tau=2$ TCB.\\

Thus correlators in the CB theory without scalar insertions are indeed completely given by single-trace TCBs.
\subsubsection{Bulk Locality}
\subsubsection*{Scalar four-point amplitude}
We have already shown the bulk non-locality for the free fermionic scalar four-point function \eqref{FFO2O2O2O2asTCB} and hence by \eqref{CBO2O2O2O2asTCB} we conclude that the same story holds for the critical bosonic theory scalar four-point function. The story for the spinning correlators is however, quite different.
\subsubsection*{Spinning four-point amplitudes}
\subsubsection*{$\mathbf{\langle TO_2O_2O_2\rangle}$}
By the bulk-boundary dictionary, we have,
\begin{align}
   \langle TO_2O_2O_2\rangle=\mathcal{A}_{\text{exchange}}+\mathcal{A}_{\text{contact}}.
\end{align}
Using the fact that $\mathcal{A}_{\text{exchange}}$ is given by $\mathcal{H}^{\langle TO_2O_2O_2\rangle_{\tau=1}}$ and using \eqref{TO2O2O2CBtcbplusextra} we see that,
\begin{align}
    \mathcal{A}_{\text{contact}}=-\frac{1}{2}\mathcal{H}^{(TO_2|1|O_2O_2)}+\frac{1}{2}\langle TO_2O_2O_2\rangle_{Extra}=-\frac{1}{2}\mathcal{A}_{\text{exchange}}+\mathcal{A}_{\text{extra}},
\end{align}
where the extra piece is not a $\tau=1$ TCB and hence is not accounted for by bulk exchange diagrams. Further, in the CFT language, we have accounted for the single trace contributions via the $\tau=1$ TCB and hence the extra piece is purely a double trace contribution which has zero double discontinuity in every channel. In \cite{Turiaci:2018nua}, the possible "extra" pieces of a correlator after taking into account the contribution of the single trace operators were determined to be in correspondence with AdS contact diagrams. Hence, by a similar analysis, it may be possible to show that this extra piece is indeed a sum of AdS contact diagrams involving one graviton and 3 scalars. However, AdS contact diagrams in general are complicated $D$ functions (See appendix \ref{sec:AdSContactdiagrams}) and hence, one might require infinitely many contact diagrams to reproduce an expression as simple as \eqref{extrapieceTOOOCB}. Thus, the fact that the extra piece is not given by the exchanges implies that it is a local term in the sense of bulk locality criteria put forth in \cite{Sleight_2018}. 
\subsubsection*{$\mathbf{\langle J_s O_2O_2O_2\rangle}$ and $\mathbf{\langle J_{s_1}J_{s_2}O_2O_2\rangle}$}
As we have seen explicitly for $s=4$ in \eqref{J4O2O2O2CBnotTCB}, and abstractly for the general case in appendix \ref{sec: CBcorrelatorsappendix}, correlators of the form $\langle J_s O_2O_2O_2\rangle$ are not entirely made out of single trace $\tau=1$ TCBs. Thus we conclude that the associated bulk amplitudes are not entirely given by $\tau=1$ TCBs and hence the bulk contact diagram is not proportional to the sum of the exchanges.\\

As for $\langle J_{s_1}J_{s_2}O_2O_2\rangle$, as we have seen for the $s_1=s_2=1$ case in \eqref{JJOOCBastu}, this correlator is not entirely made up of single trace TCBs. Therefore, much like the $\langle J_{s}O_2O_2O_2\rangle$, we conclude that the contact diagram in the dual bulk amplitude need not be proportional to a sum of exchange diagrams.
\subsubsection*{$\mathbf{\langle J_{s_1}J_{s_2}J_{s_3}O_2\rangle}$ and $\mathbf{\langle J_{s_1}J_{s_2}J_{s_3}J_{s_4}\rangle}$}
As we have discussed just below \eqref{JJJOCBstu}, correlators of the form $\langle J_{s_1}J_{s_2}J_{s_3}O_2\rangle$ are indeed given by a sum of s,t and u channel single trace $\tau=1$ and $\tau=2$ TCBs. Hence, the contact diagram in the dual bulk amplitude is proportional to the exchange diagrams and hence exhibits non-locality.\\

As for $\langle J_{s_1}J_{s_2}J_{s_3}J_{s_4}\rangle$, we have seen that \eqref{Js1Js2Js3Js4CB} is also entirely given by single trace TCBs. Therefore, the associated bulk contact diagram is indeed proportional to the sum of the exchange diagrams and hence also exhibits non-locality.
\subsection*{Summary}
It is very clear from this analysis that the correlators in the CB theory are quite different from those in the FB and FF theories. Several types of CB correlators are given by a sum of single trace TCBs and an extra piece that is not a single trace TCB. Thus by the analysis of \cite{Sleight_2018}, the  bulk contact diagram is given by a sum of exchanges and an extra piece which in some sense looks like the result of local contact interactions. However as discussed in \cite{Silva_2021, Maldacena:2015iua}, the contact diagrams in a local theory should have a more singular pole when $\sqrt{u}+\sqrt{v}=-1$ than the exchange diagrams \footnote{More precisely, the bulk point singularity of contact diagrams should be enhanced as compared to that of an individual conformal block.}. For instance, $\langle J_4 O_2O_2O_2\rangle$, given in \eqref{J4OOOCB}, has a $\frac{1}{(1+\sqrt{u}+\sqrt{v})^3}$ pole in contrast to the $\frac{1}{(1+\sqrt{u}+\sqrt{v})^{10}}$ required for a local contact interaction. Therefore, even though the contact diagram is not the same as the exchange diagrams, the bulk theory still suffers from non-locality.
\subsection{Chern-Simons Matter theories}\label{sec:CStheories}
In this subsection, we shall focus on a Chern-Simons theory coupled to matter, known as the quasi-fermionic theory. The quasi-fermionic theory refers to two theories that are dual to each other, that is, a Chern-Simons gauge field coupled to a fermion and a Chern-Simons gauge field coupled to a critical boson. The action description of the $SU(N_f)$ fermionic theory is,
\begin{align}
    S_{FF,CS}=\int d^3 x\bigg(\bar{\psi}\slashed{D}\psi+i\epsilon^{\mu\nu\rho}\frac{\kappa_f}{4\pi}\Tr{A_\mu\partial_\nu A_\rho-\frac{2i}{3}A_\mu A_\nu A_\rho}\bigg).
\end{align}
As usual, we take the large $N_f\to\infty $ limit. We also take the limit $\kappa_f\to \infty$ keeping the t'Hooft coupling $\lambda_f=\frac{N_f}{\kappa_f}$ fixed. The spectrum of this theory includes a parity odd $\Delta=2+\order{\frac{1}{N_f}}$ scalar $O_2$, conserved spin $1$ and spin $2$ currents $J$ and $T$ and weakly non-conserved currents $J_s,s=3,4,\cdots$ with $\Delta_s=s+1+\order{\frac{1}{N_f}}$.\\

The $SU(N_b)$ critical bosonic theory on the other hand is the theory at the Wilson-Fisher fixed point of the following action:
\begin{align}
    S_{CB,CS}=\int d^3 x\bigg(D_\mu \bar{\phi}D_\mu \phi+\frac{\lambda_4}{N_b}(\bar{\phi}\phi)^2+i\epsilon^{\mu\nu\rho}\frac{\kappa_b}{4\pi}\Tr{A_\mu\partial_\nu A_\rho-\frac{2i}{3}A_\mu A_\nu A_\rho}\bigg).
\end{align}
We take the limits $N_b\to\infty,\lambda_4\to \infty$ and $\kappa_b\to \infty$ keeping fixed $\frac{\lambda_4}{N_b}$ and $\lambda_b=\frac{N_b}{\kappa_b}$.  The spectrum of this theory includes a parity even $\Delta=2+\order{\frac{1}{N_b}}$ scalar $O_2$, conserved spin $1$ and spin $2$ currents $J$ and $T$ and weakly non-conserved currents $J_s,s=3,4,\cdots$ with $\Delta=s+1+\order{\frac{1}{N_b}}$.\\

The quasi-fermionic theory is a theory that interpolates between these two theories. For convenience we define $\Tilde{N}=2N_f\frac{\sin\pi\lambda_f}{\pi \lambda_f}$ and $\Tilde{\lambda}=\tan(\frac{\pi\lambda_f}{2})$ and choose a normalization where $0\le \lambda_f\le 1$. We also define the \textit{epsilon transform} which reads,
\begin{align}\label{epsilontransform}
    \langle \epsilon\cdot J_s^{(\mu_1\cdots \mu_s)}(x)\cdots\rangle=\epsilon^{\sigma\alpha(\mu_1}\int\frac{d^3 y}{|x-y|^2}\partial_{y}^{\sigma}\langle J_s^{\alpha\mu_2\cdots\mu_s)}(y)\cdots\rangle.
\end{align}
The epsilon transform commutes with conformal invariance as was shown in \cite{Caron_Huot_2021}. Further, in momentum space, the epsilon transform becomes much simpler than \eqref{epsilontransform},
\begin{align}\label{eptmomspace}
    \langle \epsilon\cdot J_s^{(\mu_1\cdots \mu_s)}(p)\cdots\rangle=\epsilon^{\sigma\alpha(\mu_1}\frac{p^{\sigma}}{p}\langle J_s^{\alpha\mu_2\cdots\mu_s)}(p)\cdots\rangle.
\end{align}
Even more simplification occurs in spinor helicity variables where the epsilon transform of a correlator, depending on the helicity, is just equal to $\pm i$ times the original correlator \cite{Jain:2022ajd}. This is an important fact for the following reason: if a correlator can be written in terms of single trace TCBs, then so can its epsilon transform even though it may not be obvious given the position space relation \eqref{epsilontransform}.\\

The n-point correlation functions in this theory were systematically obtained by solving the higher spin equations in \cite{Jain:2022ajd} and here, since our interest is in four-point correlators, we shall direct our focus towards them, beginning with the scalar case. 
\subsubsection{Scalar Correlators}
We have,
\begin{align}
    \langle O_2O_2O_2O_2\rangle_{QF}=\tilde{N}(1+\tilde{\lambda}^2)^2\langle O_2O_2O_2O_2\rangle_{FF}.
\end{align}
We know that the RHS is a $\tau=1$ twist conformal block \eqref{FFO2O2O2O2asTCB} and hence we can immediately conclude that the QF correlator is also a $\tau=1$ TCB.
\subsubsection{Spinning Correlators}
\subsubsection*{$\mathbf{\langle J_s O_2 O_2 O_2\rangle_{QF}}$}
We have,
\begin{align}\label{JOOOQF}
    \langle J_s O_2 O_2 O_2\rangle_{QF}=\Tilde{N}(1+\tilde{\lambda}^2)\bigg(\langle J_sOOO\rangle_{FF}+\tilde{\lambda}\langle J_s OOO\rangle_{CB}\bigg).
\end{align}
As we have discussed in subsections \ref{sec:FFtheory} and \ref{sec:CBtheory} respectively, the FF correlator is entirely given by a sum of single trace $\tau=1$ TCBs while the CB correlator is not. Therefore we expect that at generic $\tilde{\lambda}$, the QF correlator is a sum of a $\tau=1$ TCB and an extra piece that is not a $\tau=1$ TCB.\\
\subsubsection*{$\mathbf{\langle J_{s_1}J_{s_2}O_2O_2\rangle_{QF}}$}
We have,
\begin{align}\label{JJOOQF}
    \langle J_{s_1}J_{s_2}O_2O_2\rangle_{QF}=\Tilde{N}\bigg(\langle J_{s_1}J_{s_2}O_2O_2\rangle_{FF}+\tilde{\lambda}\langle \epsilon\cdot J_{s_1}J_{s_2}O_2O_2\rangle_{FF-CB}+\tilde{\lambda}^2\langle J_{s_1}J_{s_2}O_2O_2\rangle_{CB}\bigg).
\end{align}
Based on our discussion in subsection \ref{sec:FFtheory}, the FF term is indeed given by a sum of single-trace TCBs.  The CB correlator, based on our discussion in section \ref{sec:CBtheory}, is not entirely made up of single-trace TCBs. The fact whether the parity odd epsilon transform contribution is given by single-trace TCBs as discussed below \eqref{eptmomspace} is determined based on whether the correlator which is being epsilon transformed is a single-trace TCB or not. In this case, the FF correlator is a TCB but the CB correlator is not. Therefore, $\langle \epsilon\cdot J_{s_1}J_{s_2}O_2O_2\rangle_{FF-CB}$ is not entirely given by single-trace TCBs.  Therefore, these correlators are not entirely determined in terms of single-trace TCBs.
\subsubsection*{$\mathbf{\langle J_{s_1}J_{s_2}J_{s_3}O_2\rangle_{QF}}$}
We have,
\begin{align}\label{JJJOQF}
    \langle J_{s_1}J_{s_2}J_{s_3}O_2\rangle_{QF}=\tilde{N}\bigg(\langle J_{s_1}J_{s_2}J_{s_3}O_2\rangle_{FF}+\tilde{\lambda}\langle J_{s_1}J_{s_2}J_{s_3}O_2\rangle_{CB}\bigg).
\end{align}
As we have seen in sections \ref{sec:FFtheory} and \ref{sec:CBtheory}, both the FF and CB correlators are entirely given in terms of single trace TCBs and hence by the above formula, so is the QF correlator.
\subsubsection*{$\mathbf{\langle J_{s_1}J_{s_2}J_{s_3}J_{s_4}\rangle_{QF}}$}
We have,
\begin{align}\label{Js1234QFtcb}
    \langle J_{s_1}J_{s_2}J_{s_3}J_{s_4}\rangle_{QF}=\frac{\tilde{N}}{(1+\tilde{\lambda}^2)}\bigg(\langle J_{s_1}J_{s_2}J_{s_3}J_{s_4}\rangle_{FF}+\tilde{\lambda}\langle \epsilon\cdot J_{s_1}J_{s_2}J_{s_3}J_{s_4}\rangle_{FF-CB}+\tilde{\lambda}^2\langle J_{s_1}J_{s_2}J_{s_3}J_{s_4}\rangle_{CB}\bigg).
\end{align}
Based on our discussion in subsection \ref{sec:FFtheory} and section \ref{sec:CBtheory}, the first and third terms in the above equation are indeed given by single trace TCBs. As for the parity odd epsilon transform term in the middle, it is also given by single trace TCBs since the FF and CB correlators are, as can be understood based on our discussion after \eqref{eptmomspace}.
\subsubsection{Bulk Locality}
\subsubsection*{Scalar four point amplitudes}
Since the scalar four-point function is proportional to the free fermionic expression, the associated bulk amplitudes are equivalent up to an overall factor and hence the bulk contact diagram is indeed proportional to the exchange diagrams, thus exhibiting non-locality.
\subsubsection*{Spinning four point amplitudes}
For correlators with a no scalar insertions and one scalar insertion, $\langle J_{s_1}J_{s_2}J_{s_3}J_{s_4}\rangle$ and $\langle J_{s_1}J_{s_2}J_{s_3}O_2\rangle$, we have seen that they can entirely be written in terms of single trace TCBs (please refer to \eqref{Js1234QFtcb} and \eqref{JJJOQF} respectively) and hence just like in the free bosonic and free fermionic theories, the associated bulk theory has a non-local quartic interaction.
However, correlators of the form $\langle J_{s_1}O_2O_2O_2\rangle$ (\eqref{JOOOQF}) and $\langle J_{s_1}J_{s_2}O_2O_2\rangle$  (\eqref{JJOOQF}) cannot be entirely decomposed into a sum of single trace TCBs. Thus, in the bulk language, we have for arbitrary spins $s_1$ and $s_2$ \footnote{We have a similar formula for $\langle J_{s_1}O_2O_2O_2\rangle$ as well.},
\begin{align}
    \langle J_{s_1}J_{s_2}O_2O_2\rangle_{CB}=\frac{1}{2}\mathcal{H}+\frac{1}{2}\mathcal{J}=\mathcal{A}_4=\mathcal{A}_{\text{exchange}}+\mathcal{A}_{\text{contact}}=\mathcal{H}+\mathcal{A}_{\text{contact}},
\end{align}
where $\mathcal{H}$ represents the single trace twist conformal block contributions and $\mathcal{J}$ represents the part of the correlator that is not a single trace TCB. 
We see that,
\begin{align}
\mathcal{A}_{\text{contact}}=\frac{1}{2}\bigg(\mathcal{J}-\mathcal{H}\bigg).
\end{align}
We see that the contact diagram is not given by the exchange diagrams fully. There is a part of the contact diagram which is local in the sense of \cite{Sleight_2018}.
\subsection*{Summary}
Scalar correlators in this theory are given entirely by single trace $\tau=1$ TCBs. Correlators that have three or two scalar operators are in general not completely determined by single-trace TCBs. Correlators with just one scalar operator or no scalar operators, on the other hand, are entirely given by the single trace TCBs. Thus, there exists the possibility of finding a sub-sector of the theory that has a local bulk dual. 

\section{Discussion}\label{sec:Discussion}
\subsection*{Summary}
In this paper, we have systematically obtained TCBs (both scalar and spinning) by explicitly solving the TCB eigenvalue equation and also by acting with weight shifting and spin raising operators on scalar seeds in section \ref{sec:TCB}. Using these results, we then showed that scalar and spinning correlators in the free bosonic (section \ref{sec:FBtheory}) and free fermionic theories (section \ref{sec:FFtheory}) can be written purely in terms of single trace TCBs. We then investigated the bulk-locality of the free bosonic and free fermionic theories at the level of spinning four-point functions where we find that the non-locality found at the scalar four-point level in \cite{Sleight_2018} still persists. We also investigated the bulk-locality at the level of the five-point function of scalars in the free bosonic theory and found a non-locality in the bulk quintic vertex as well. After analyzing the bulk locality in these exact higher spin theories, we went on to investigate the bulk locality of theories where the higher spin symmetry is slightly broken. Beginning with the critical bosonic theory in section \ref{sec:CBtheory}, we found that correlators of the form $\langle J_s O_2O_2O_2\rangle$ and $\langle J_{s_1}J_{s_2}O_2O_2\rangle$ are not completely determined in terms of single trace TCBs and hence the bulk contact diagram is not completely given by the exchanges as was the case with the FB and FF theory correlators. We then used our results for the FF and CB theory correlators to probe the bulk-locality of the quasi-fermionic theory in section \ref{sec:CStheories}. We found at the four-point level that correlators with one or two scalar insertions in this theory cannot be written purely in terms of single trace TCBs, rather they can be written as a sum of single trace TCBs and quantities that are not single trace TCBs thus indicating the possibility of finding a sub-sector of this theory that has a local bulk dual. 
\subsection*{Future directions}
\subsubsection*{A Classification of CFTs via TCBs}
In \cite{Giombi:2011rz}, three-point functions 
involving (higher spin) conserved currents and scalar operators in three dimensions were computed.    
 At the level of three points, one can use the conformal symmetry to fix all possible allowed structures that can appear in a correlator. They found that the three-point function of spinning conserved currents is given by,
\begin{align}
    \langle J_{s_1}J_{s_2}J_{s_3}\rangle=a\langle J_{s_1}J_{s_2}J_{s_3}\rangle_{FB}+b\langle J_{s_1}J_{s_2}J_{s_3}\rangle_{FF}+c\langle J_{s_1}J_{s_2}J_{s_3}\rangle_{\text{odd}}.
\end{align}
$\langle J_{s_1}J_{s_2}J_{s_3}\rangle_{FB}$ and $\langle J_{s_1}J_{s_2}J_{s_3}\rangle_{FF}$ are the parity preserving structures from the free bosonic and free fermionic theory and $\langle J_{s_1}J_{s_2}J_{s_3}\rangle_{\text{odd}}$ is a parity-violating structure that exists only when the triangle inequality, $s_i+s_j\ge s_k\forall i,j,k\in\{1,2,3\}$ is obeyed\footnote{For correlators involving scalar operators, the number of allowed structures is even lesser than the case of all non zero spins.}. Further, it was found that when the higher spin symmetry is slightly broken, the odd structure can exist even when the triangle inequality is broken.\\

We now ask the following question: is it possible to make such a classification at the level of four-point functions? The first hurdle is that the four-point function is fixed only up to functions of the cross ratios, unlike their three-point counterpart. Thus, we need additional assumptions to make progress. One such route is to make the assumption that the correlator is given by a sum of single trace twist conformal blocks\footnote{For correlators of single trace operators, these single trace twist conformal blocks are equivalent to higher spin blocks \cite{Alday_2017}.}. Thus, along with the other natural assumptions such as OPE consistency, crossing symmetry and conservation where applicable as we discussed in greater detail in section \ref{sec:TCB}, we obtained the  single trace $\tau=1$ TCBs as shown in Table \ref{table:1}. As expected, our results are able to reproduce the free bosonic and free fermionic answers. However, when we get to $\langle JJO_\Delta O_\Delta\rangle$ for $\Delta=1$ and $\Delta=2$, we obtained a two-parameter family of solutions with the free bosonic and free fermionic correlators being a single point in the respective parameter spaces. Thus, it would be interesting to check if they can arise from some other theory. Our suspicion is that they could be realized in the critical fermionic and critical bosonic theories.
\begin{table}[h!]
  \begin{center}
    \begin{tabular}{|c|l|}
    \hline    
      \textbf{TCB} & \textbf{Solution} \\
      \hline
    $\mathcal{H}^{(O_{\Delta}O_{\Delta}|1|O_{\Delta}O_{\Delta})}$ & Unique\\
      \hline
    $\mathcal{H}^{\langle J_sO_1O_1O_1\rangle}$ & Unique\\
      \hline
    $\mathcal{H}^{\langle J_s O_2O_2O_2\rangle_{\text{odd}}}$ & Unique\\
      \hline
    $\mathcal{H}^{\langle JJO_1O_1\rangle_{\tau=1}}$ & Two Parameter Family\\
      \hline
    $\mathcal{H}^{\langle JJO_2O_2\rangle_{\tau=1}}$ & Two Parameter Family\\
      \hline
    \end{tabular}
    \caption{Number of $\tau=1$ TCBs}
    \label{table:1}
  \end{center}
\end{table}
\subsubsection*{An algorithm to compute SBHS theory correlators}
For the theories with exactly conserved higher spin symmetry such as free bosonic or free fermionic theories, the principle to compute correlation functions is to simply compute single trace TCBs. One can ask how this principle changes when we consider theories with slightly broken higher spin symmetry.  
 One might expect that TCB or slight generalization of TCB should be good enough to compute but as we have explicitly seen in section \ref{sec:SBHSsec}, in general, the four-point functions in the SBHS theories are extensively different from their exactly conserved counterparts.  In the critical bosonic theory where the higher spin symmetry is slightly broken, the scalar correlator, $\langle O_2O_2O_2O_2\rangle_{CB}$ is still a TCB. However $\langle TO_2O_2O_2\rangle_{CB}$ is not. Rather, it is a sum of a TCB and an extra piece. However, this extra piece can be determined by demanding the conservation of the correlator since the TCB by itself is not conserved with respect to the stress tensor. For other correlators, it is not very clear what the algorithm is to compute correlators. Given the fact that the breaking of higher spin symmetry in this theory is quite controlled, it would be really interesting if we can pin down an algorithm that will give us an independent way of computing correlation functions in this theory and also other SBHS theories such as CS theories with matter.
\subsubsection*{Chiral Higher Spin Theory}
Chiral higher spin theory \cite{Skvortsov:2022wzo,Sharapov:2022awp,Sharapov:2022wpz} in flat space in the light cone gauge contains only cubic vertices. However, in AdS, it can contain vertices of all orders.
Further, the Chiral higher spin theory in AdS is a conjectured local bulk dual to a sub-sector of CS matter theories. We believe that our analysis in this paper should be helpful in this pursuit and we hope to come back to this problem in the future.

\subsubsection*{Momentum space twist conformal blocks}
The analysis of CFT in momentum space leads to many simplifications and reveals structures that were hidden in position space, such as relations between parity odd and parity even correlators \cite{Jain:2021vrv,Jain:2021wyn,Jain4_2021}. For the SBHS theories that are of interest to us, many drastic simplifications also occur in momentum space, especially in the formulae that relate interacting theory correlators to the free theory ones. For instance, the Legendre transformation that relates $\langle TO_2O_2O_2\rangle_{CB}$ to $\langle TO_1O_1O_1\rangle_{FB}$ (please see equation \eqref{CBintTOOO}) which is a complicated integral transform becomes a simple multiplicative transform in momentum space.
\begin{align}\label{easyLT}
    \langle TO_2O_2O_2\rangle_{CB}=p_2p_3p_4\langle TO_1O_1O_1\rangle_{FB}+\text{contact terms}.
\end{align}
The epsilon transformations of free fermionic and critical bosonic theory correlators  that are present in quasi fermionic theory correlators \eqref{epsilontransform} also become much more simple in momentum space \eqref{eptmomspace}.
\begin{align}\label{easyEPT}
    \epsilon^{\sigma\alpha(\mu_1}\int\frac{d^3 y}{|x-y|^2}\partial_{y}^{\sigma}\langle J_s^{\alpha\mu_2\cdots\mu_s)}(y)\cdots\rangle\to \epsilon^{\sigma\alpha(\mu_1}\frac{p^{\sigma}}{p}\langle J_s^{\alpha\mu_2\cdots\mu_s)}(p)\cdots\rangle.
\end{align}
Further, if we go to the helicity basis, there is no distinction between parity even and parity odd correlators other than overall factors of $\pm i$ depending on the helicity of the operators in a  correlator. This is an important fact for the following reason: if a correlator can be written in terms of single trace TCBs, then so can its epsilon transform even though it may not be obvious given the position space relation ( the LHS of \eqref{easyEPT}). Thus, correlators in CS+matter theories such as $\langle JJO_2O_2\rangle_{QF}$ and $\langle JJJJ\rangle_{QF}$  (please refer to equations \eqref{JJOOQF} and \eqref{Js1234QFtcb} respectively) become much simpler to analyze in spinor helicity variables.\\

Thus, we see that it is natural to develop the formalism presented in this paper in momentum space and perhaps even spinor helicity variables. Due to the simplicity of expressions such as \eqref{easyLT} and \eqref{easyEPT}, one has a greater handle on correlators in the SBHS theories. It would be interesting to approach the classification problem discussed earlier from the momentum space perspective. As far as the formalism of momentum space CFT at the four-point level goes, in the past few years, a simplex representation for scalar four-point functions has been developed \cite{Bzowski_2020,Bzowski_2021,Caloro_2023}. Using our method of producing spinning TCBs channel by channel from scalar TCB seeds using spin raising and weight shifting operators, it is possible to derive simplex representations for the spinning correlators as well. 

\subsubsection*{Finite $N$ corrections}
In this paper, we investigated correlators in the critical bosonic theory and CS theory with matter (the quasi fermionic theory) at the leading order in $N$\footnote{More precisely, we investigated the connected part of the correlators at leading order in $N$.}. In \cite{Aharony:2018npf}, the authors used the leading order results for the scalar correlators to systematically determine the subleading corrections in $\order{1/N}$ using large spin perturbation theory. The idea of large spin perturbation theory (LSPT) \cite{Alday_2017,Alday1_2017,van_Loon_2018,Henriksson:2018myn,Alday:2017zzv,Aharony:2018npf}  is to use the crossing equation and inversion formula to derive perturbative results in conformal field theories.  The results of this procedure yields a set of CFT data (OPE coefficients and anomalous dimensions) or an explicit expression for the correlator.\\

Thus, an interesting direction is to extend this study to compute  $\order{1/N}$ corrections to spinning correlators such as $\langle TO_2O_2O_2\rangle$ in the critical bosonic theory or the CS matter theories which will allow us to compute say, corrections to the anomalous dimensions of several double trace operators. However, there are hurdles such as the fact that $\langle TO_2O_2O_2\rangle_{CB}$ has zero double discontinuity. The usual process of LSPT however involves a step where the double discontinuities are matched on either side of the crossing symmetry equations. Thus we expect the generalization to using LSPT for spinning correlators to be more difficult than the application to the scalar case but nevertheless, it should be possible and we hope to come back to this problem in the future.

\section*{Acknowledgements}
We would like to thank A.Gadde, A.Mehta, S.Minwalla, E.Skvortsov and J.Yoon for valuable discussions. We would also like to thank  O.Aharony and E.Skvortsov for comments on an earlier version of the draft. 
S.J would also like to thank organizers of higher spin $2022$, Pohang, organized by APCTP for providing a stimulating environment.
S.J. would also like to thank the participants of gravity beyond Riemannian paradigm, Jeju, APCTP for their valuable comments. 
The work of S.J is supported by the Ramanujan Fellowship.  
We acknowledge our debt to the people of India for their steady support of research in basic sciences.

\appendix
\section{Primary single trace operators in the higher spin theories }\label{sec:FBFFCBQFoperators}
In this appendix, we provide the expressions for the spin s currents in the $U(N)$ free bosonic and free fermionic theories that we used in sections \ref{sec:FBtheory} and \ref{sec:FFtheory} to compute the correlators via Wick contractions. 
In the free bosonic theory the currents are given by,
\begin{align}
    J_s^{FB}(x,z)=\sum_{k=0}^{s}\frac{(-1)^{(k+s)}}{s!}\frac{(2s)!}{(2k)!(2s-2k)!}(z\cdot \partial)^k\Bar{\phi}(z\cdot \partial)^{s-k}\phi.
\end{align}
while in the free fermionic theory they are,
\begin{align}
    J_s^{FF}=\sum_{k=0}^{s-1}\frac{(-1)^{k+s+1}}{(2s)!}\frac{(2s)!}{(2k+1)!(2s-2k-1)!}(z\cdot \partial)^k \Bar{\psi}(z\cdot \sigma)(z\cdot \partial)^{s-k-1}\psi.
\end{align}
where we have suppressed the trace over the $SU(N)$ indices. For details on the derivation of these expressions, please refer to \cite{Giombi_2017}.
\section{Weight shifting and spin raising operators}\label{sec:weightshiftingspinraisingoperators}
Weight shifting and spin raising operators 
\cite{Karateev_2018} provide us with a tool to bootstrap spinning correlators from scalar seeds \cite{Baumann_2020}. In this paper the expressions for the parity preserving spin raising and weight shifting operators that we used are,
\begin{align}
    &\mathcal{D}_{ij}=(\Delta_i+S_i)Z_i\cdot X_j+(X_i\cdot X_j)Z_i\cdot\frac{\partial}{\partial X_i}\notag\\
    &\mathcal{H}_{ij}=-2((Z_i\cdot Z_j)(X_i\cdot X_j)-(Z_i\cdot X_j)(Z_j\cdot X_i))\notag\\
    &\mathcal{W}_{ij}^{n}=(X_i\cdot X_j)^n,
\end{align}
with their action given in \eqref{spinraisops}. 

We also employed the following parity odd operator while constructing $\langle TO_2O_2O_2\rangle_{\text{odd}}$ in subsection \ref{sec:spinningTCB}:
\begin{align}
      &\mathcal{\Tilde{D}}_{ij}=\epsilon^{Z_i, X_i, X_j, \frac{\partial}{\partial X_i},\frac{\partial}{\partial X_j}},\notag\\
\end{align}
where we have used the shorthand,
\begin{align}
\epsilon^{ABCDE}X_{A}Y_{B}Z_{C}U_{D}V_{E}:=\epsilon^{X Y Z U V}.
\end{align}
The action of this operator is to raise the spin at $i$ by $1$ while flipping the parity of the correlator.
\section{Twist conformal blocks in $d$ dimensions}\label{sec:ddimensionalTCB}
In this appendix, we present results for scalar TCBs in $d$ dimensions involving exchanges of $\tau=d-2$ operators. For $\Delta\ge d-1$ we find the following solution to Alday's s channel equation \eqref{schannelAldayequation},
\begin{align}\label{Hsddim}
   \mathcal{H}_s^{(O_{d-1}O_{d-1}|d-2|O_{d-1}O_{d-1})}=\frac{a}{x_{12}^{d+1}x_{34}^{d+1}}u^{\frac{d-2}{2}}v^{-\frac{d}{2}}\bigg((u-1)+ v+v^{\frac{d}{2}}(1+(u-v))\bigg), a\in\mathbb{R},
\end{align}
which for $d=3$ reduces to \eqref{O2O2O2O2sTCB}.  We then see that the free fermionic correlator can be reproduced by adding the s,t and u channel $\tau=1$ expressions,
\begin{align}
    &\langle O_{d-1} O_{d-1}O_{d-1}O_{d-1}\rangle_{FF}=\frac{1}{2 u^{d/2} v^{d/2}x_{13}^{2(d-1)}x_{24}^{2(d-1)}}\bigg(-1+u-u^{1+\frac{d}{2}}+u^{\frac{d}{2}}+v+u^{\frac{d}{2}}v+(1+u-v)v^{\frac{d}{2}}\bigg)\notag\\
    &=\frac{1}{2}\bigg(\mathcal{H}_s^{(O_{d-1}O_{d-1}|d-2|O_{d-1}O_{d-1})}+\mathcal{H}_t^{(O_{d-1}O_{d-1}|d-2|O_{d-1}O_{d-1})}+\mathcal{H}_u^{(O_{d-1}O_{d-1}|d-2|O_{d-1}O_{d-1})}\bigg),
\end{align}
where we have set $a=\frac{1}{2}$ in \eqref{Hsddim}.

For $\Delta<d-1$ we obtain the unique $\tau=d-2$ s channel twist conformal block,
\begin{align}
    \mathcal{H}_s^{(O_{d-2}O_{d-2}|d-2|O_{d-2}O_{d-2})}=\frac{a}{x_{12}^{2\Delta}x_{34}^{2\Delta}}\bigg[u^{\frac{d-2}{2}}+\bigg(\frac{u}{v}\bigg)^{\frac{d-2}{2}}\bigg], a\in\mathbb{R},
\end{align}
which coincides with the results obtained earlier \cite{Alday_2017,Alday1_2017,Sleight_2018}. 
\subsection*{Five point functions}
We can also obtain TCBs for scalar five point functions in $d$ dimensions generalizing the result in \eqref{5ptTCB}. For five identical scalars with scaling dimension $\Delta$ in $d$ dimensions with twist $\tau=d-2$ exchange, we obtained the following s channel answer that solves Alday's equation\eqref{schannelAldayequation}.
\begin{align}
    &\mathcal{H}_s^{(O_{d-1}O_{d-1}|d-2|O_{d-1}O_{d-1}O_{d-1})}=\frac{a_1}{(x_{12}x_{13}x_{14}x_{15}x_{23}x_{24}x_{25}x_{34}x_{35}x_{45})^{\frac{d-2}{2}}(u_1 u_2 v_1 v_2 w)^{\frac{(d-2)}{4}}}\bigg((1+u_2^{\frac{d-2}{2}})u_2 w^2(v_1 v_2)^{\frac{d}{2}}\notag\\&+(u_2^{\frac{d}{2}}v_1 v_2+u_2 v_1^{\frac{d}{2}}v_2+(1+v_1^{\frac{d-2}{2}})v_1 u_2 v_2^{\frac{d}{2}})w^{d}\bigg).
\end{align}
Obtaining the other channel answers using \eqref{utvchannels} and setting $a_1=1$ we see that the free bosonic correlator is reproduced,
\small
\begin{align}
    &\langle OOOOO\rangle_{FB}=\frac{1}{2}\bigg(\mathcal{H}_s^{(O_{d-1}O_{d-1}|d-2|O_{d-1}O_{d-1}O_{d-1})}+\mathcal{H}_u^{(O_{d-1}O_{d-1}|d-2|O_{d-1}O_{d-1}O_{d-1})}+\mathcal{H}_t^{(O_{d-1}O_{d-1}|d-2|O_{d-1}O_{d-1}O_{d-1})}+\mathcal{H}_v^{(O_{d-1}O_{d-1}|d-2|O_{d-1}O_{d-1}O_{d-1})}\bigg)\notag\\
    &=\frac{2}{(x_{12}x_{13}x_{14}x_{15}x_{23}x_{24}x_{25}x_{34}x_{35}x_{45})^{\frac{d-2}{2}}(u_1 u_2 v_1 v_2 w^2)^{\frac{(d-2)}{4}}}\Bigg((1+u_2^{\frac{d-2}{2}})u_1 u_2 w^2(v_1 v_2)^{\frac{d}{2}}+u_1\bigg(u_2^{\frac{d}{2}}v_1 v_2+u_2 v_1^{\frac{d}{2}}v_2\notag\\&+(1+v_1^{\frac{d-2}{2}})v_1 u_2 v_2^{\frac{d}{2}}\bigg)w^{d}+u_1^{\frac{d}{2}}\bigg( (u_2 (v_1 v_2)^{\frac{d}{2}}+u_2^{\frac{d}{2}}((1+v_1^{\frac{d-2}{2}})v_1v_2+v_1 v_2^{\frac{d}{2}}))w^2+(1+u_2^{\frac{d-2}{2}})u_2 v_1 v_2 w^{d}\bigg)\Bigg).
\end{align}
\section{Details of solutions for spinning twist conformal blocks and another parity odd example}\label{sec:detailsofsolutionstoTCB}
\subsection*{Details of solution to $\mathcal{H}_s^{(JJ|1|O_1O_1)}$}
In this appendix we provide a step by step guide on how to obtain spinning twist conformal blocks by solving Alday's equation \eqref{3dAldayEquationSpinning} by presenting the explicit computational details for $\mathcal{H}^{\langle JJO_1O_1\rangle_{\tau=1}}$. We work in the $5$ dimensional embedding space throughout.
The most general ansatz for this TCB is,
\begin{align}\label{JJOOansatz}
    \frac{(X_3\cdot X_4)^2}{(X_1\cdot X_3)^3 (X_2\cdot X_4)^3}\bigg(W_1 W_2 f_1(u,v)+\Bar{W}_1\Bar{W}_2 f_2(u,v)+(W_1\Bar{W}_2-\Bar{W}_1 W_2)f_3(u,v)+H_{12}f_4(u,v)\bigg),
\end{align}
where the $(1\leftrightarrow 2)$ and the $(3\leftrightarrow 4)$ symmetry as appropriate for this correlator forces the following relations,
\begin{align}\label{acrossing}
    &f_i(\frac{u}{v},\frac{1}{v})=v^3 f_i(u,v),i\in\{1,2,4\}\notag\\
    &f_3(\frac{u}{v},\frac{1}{v})=-v^3 f_3(u,v).
\end{align}
We now take half integer polynomial ansätze for these functions,
\begin{align}
    f_i(u,v)=\sum_{\alpha,\beta\in\mathbb{Z}/2}c_{i,\{\alpha,\beta\}} u^\alpha v^\beta, c_{i,\{\alpha,\beta\}}\in\mathbb{R}.
\end{align}
The $c_{i,\{\alpha,\beta\}}$ by virtue of \ref{acrossing} are forced to obey,
\begin{align}\label{acrossing1}
    &c_{i,\{\alpha,\beta\}}=c_{i,\{\alpha,-(3+\alpha+\beta)\}},i\in\{1,2,4\}\notag\\
    &c_{4,\{\alpha,\beta\}}=-c_{4,\{\alpha,-(3+\alpha+\beta)\}}.
\end{align}
We further restrict the highest powers of $u$ and $v$ that can appear in the $f_i(u,v)$ by demanding consistency with the operator product expansion. The leading order term in the s,t and u channel OPEs is due to a $O_1$ scalar exchange, \footnote{ We ignore the contribution of the identity operator as it contributes to the disconnected part of correlators.}
\begin{align}
    &\textbf{s channel OPE:}\lim_{u\to 0,v\to 1}\mathcal{H}^{(JJ|1|O_1O_1)} \sim\frac{(X_3\cdot X_4)^2}{u^{5/2}(X_1\cdot X_3)^{5/2}(X_2\cdot X_4)^{7/2}} (Z_1\cdot X_2)(Z_2\cdot X_1)+\cdots\notag\\
    &\textbf{t channel OPE:}\lim_{u\to 1,v\to 0}\mathcal{H}^{(JJ|1|O_1O_1)}\sim\frac{(X_3\cdot X_4)^2}{v^{3/2}(X_1\cdot X_3)^{5/2}(X_2\cdot X_4)^{7/2}} (u-1)(Z_1\cdot X_4)(Z_2\cdot X_3)+\cdots\notag\\
     &\textbf{u channel OPE:}\lim_{u\to 1\infty,v\to 0\infty}\mathcal{H}^{(JJ|1|O_1O_1)}\sim\frac{v(X_3\cdot X_4)^2}{u^2(X_1\cdot X_3)^{5/2}(X_2\cdot X_4)^{7/2}}(Z_1\cdot X_3)(Z_2\cdot X_4)+\cdots.
\end{align}
This then restricts our ansatz even further,
\begin{align}\label{fiuv}
    f_i(u,v)=\sum_{\alpha=-\frac{5}{2}}^{-2}\sum_{\beta=-\frac{3}{2}}^{1}c_{i,\{\alpha,\beta\}}u^\alpha v^\beta,
\end{align}
with the $c_{i,\{\alpha,\beta\}}$ obeying \eqref{acrossing1}.We then substitute \eqref{fiuv} into \eqref{JJOOansatz}. The next step is to plug the updated ansatz into Alday's s channel twist conformal block eigen value equation \eqref{3dAldayEquation} to identify the s channel TCB. This results in the a complicated set of fourth order coupled partial differential equations for the $f_i(u,v),i=1,2,3,4$. After substituting \eqref{fiuv}, these PDEs become algebraic equations which we can solve order by order in $u$ and $v$. Once we do this we are left with the following expression that solves Alday's s channel equation \eqref{3dAldayEquation}:
\begin{align}\label{JJOOansatzupdated}
    \frac{(X_3\cdot X_4)^2}{(X_1\cdot X_3)^3 (X_2\cdot X_4)^3}\bigg(W_1 W_2 F_1(u,v)+\Bar{W}_1\Bar{W}_2 F_2(u,v)+(W_1\Bar{W}_2-\Bar{W}_1 W_2)F_3(u,v)+H_{12}F_4(u,v)\bigg),
\end{align}
with,
\begin{align}
    &F_1(u,v)=\frac{a_1v(1+v{1/2})+b_1(1+v^{5/2})}{u^{5/2}v^{3/2}}, F_2(u,v)=\frac{a_2(1+v^{3/2})+b_2(1+v^{5/2}}{u^{5/2}v^{3/2}}\notag\\
    &F_3(u,v)=\frac{b_1-b_2+(a_1-a_2)(v^{1/2}-1)v+(b_2-b_1)v^{5/2}}{2u^{5/2}v^{3/2}},F_4(u,v)=\frac{-2(b_1+b_2)+3c_1 v+3 c_1v^{3/2}-2(b_1+b_2)v^{5/2}}{3 u^{5/2}v^{3/2}}.
\end{align}
However, for this expression to represent a valid correlator we need to impose the conservation of the spin one currents at non-coincident operator insertions which in embedding space is given by the following equation:
\begin{align}
    \frac{\partial}{\partial X_{1}^A}D^{A}_{Z_{1}}\langle J(X_1,Z_1) J(X_2,Z_2)O_1(X_3)O_1(X_4)\rangle=0,
\end{align}
where $D_{Z_{1A}}$ is the Thomas Todorov derivative,
\begin{align}
    D^{A}_{Z}=\frac{1}{2}\frac{\partial}{\partial Z_A}+Z_{B}\frac{\partial^2}{\partial Z_B Z_A}-\frac{1}{2}Z^{A}\frac{\partial^2}{\partial Z_B\partial Z^B}.
\end{align}
This then yields the following constraints,
\begin{align}
    a_1=a_2=0,b_2=-b_1,c_1=4b_1.
\end{align}
which when substituted into \eqref{JJOOansatzupdated} and the expressions converted back to the three dimensional real space, yields the unique s channel $\tau=1$ TCB,
\begin{align}
     &\mathcal{H}^{(JJ|1|O_1O_1)}_s\propto \frac{x_{34}^4}{x_{13}^6 x_{24}^6 u^{5/2}v^{3/2}}\bigg((W_1 W_2-\Bar{W}_1\Bar{W}_2)(1+v^{5/2})+(W_1 \Bar{W}_2-\Bar{W}_1 W_2)(1-v^{5/2})+4 H_{12}v(1+\sqrt{v})\bigg),
\end{align}
which the is result that we presented in the main text \eqref{JJO1O1sTCB}. A similar analysis can be carried out in the t and u channels yielding \eqref{JJO1O1tTCB} and its $(3\leftrightarrow 4)$ exchanged expression respectively.
\subsection*{Another parity odd TCB: $\mathbf{\mathcal{H}^{\langle JJO_1O_1\rangle_{\tau=1,\text{odd}}}}$}
The most general ansatz for this TCB consistent with crossing symmetry is,
\begin{align}
    &\mathcal{H}^{\langle JJO_1O_1\rangle_{\tau=1,\text{odd}}}=\frac{1}{x_{12}^6 x_{34}^2}\Bigg[S_{1,234}\bigg(V_{2,13}h_1(u,v)-V_{2,14}h_1(\frac{u}{v},\frac{1}{v})\bigg)\notag\\&+S_{2,134}\bigg(V_{1,23}h_1(\frac{u}{v},\frac{1}{v})-V_{1,24}h_1(u,v)\bigg)+Y_{12,3}h_2(u,v)+Y_{1,24}h_2(u,v)\bigg],
\end{align}
where $h_2(u,v)$ satisfies,
\begin{align}
    h_2(\frac{u}{v},\frac{1}{v})=h_2(u,v).
\end{align}
Taking OPE consistency into account and then solving Alday's equation in the s channel \eqref{3dAldayEquationSpinning} we obtained the following s channel TCB,
\begin{align}\label{JJO1O1oddsTCB}
    &\mathcal{H}_{s,\text{odd}}^{(JJ|1|O_1O_1)}=\frac{1}{x_{12}^6 x_{34}^2}\Bigg[S_{1,234}\bigg(V_{2,13}f_1(u,v)-V_{2,14}f_1(\frac{u}{v},\frac{1}{v})\bigg)\notag\\&+S_{2,134}\bigg(V_{1,23}f_1(\frac{u}{v},\frac{1}{v})-V_{1,24}f_1(u,v)\bigg)+Y_{12,3}f_2(u,v)+Y_{1,24}f_2(u,v)\bigg],
\end{align}
where,
\begin{align}
    &f_1(u,v)=a_1\frac{u\bigg(540(u-1)+936(u-1)\sqrt{v}+1125v+458 v^{3/2}+(397+36 u)v^3-252 v^{7/2}-252 v^4\bigg)}{850 v^3}\notag\\&-a_2\frac{u(9-10\sqrt{v}+v^2)}{24 v^2}+a_3\frac{(87 u^2+u(169+30\sqrt{v}-1939 v)+250(\sqrt{v}-1)^2(1+18\sqrt{v}+17 v)}{250 u^{5/2}}\notag\\
    &-a_4\frac{5u(-11-3u+30\sqrt{v}+41v)}{250 u^{5/2}}\notag\\
    &f_2(u,v)=a_1\frac{\sqrt{u}}{425 v^{5/2}}\bigg(36-277v-72 u^2(1+v^{5/2})+v^{3/2}(425+425 v^{3/2}-277 v^2+36 v^3)\notag\\&+2u(18+5v+v^{5/2}(5+18 v))\bigg)+a_2\frac{\sqrt{u}}{12 v^{3/2}}\bigg(1+2u-5\sqrt{v}+12 v+2(6+u)v^{3/2}-5 v^2+v^{5/2}\bigg)\notag\\
    &+a_3\frac{1}{75 u^3}\bigg(75(\sqrt{v}-1)^4(\sqrt{v}+1)^2-4u^2(5+7\sqrt{v}+5v)+60 u(\sqrt{v}-2v+v^{3/2})\bigg)\notag\\&+a_4\frac{1}{15 u^2}\bigg(u(-5+2\sqrt{v}-5v)+15(v-1)^2\bigg).
\end{align}
It is interesting to note that we have obtained a four parameter family of solutions\footnote{ Just as was the case for $ \mathcal{H}^{\langle JJO_1O_1\rangle_{\tau=1,\text{even}}}$ we can also construct a t channel TCB and add to it a u channel TCB to form a valid TCB. A more careful analysis of parity odd TCBs taking into account the various degeneracies that accompany parity odd expressions could further constrain these four parameters that appear in \eqref{JJO1O1oddsTCB}. However, we defer such a systematic analyis of parity odd TCBs to a future work and present this example just as a testement of our method's applicability to the parity odd case.}. The CS matter theory contains a  $\langle JJO_1O_1\rangle_{\text{odd}}$ correlator and hence it would be interesting to see if our TCB is realized there in some capacity.
\section{Higher Spin Equations}\label{sec:HSE}
In this appendix, we ask the following question. "Given the higher spin symmetry of the free bosonic theory and the fact that the scalar four point correlator is given by a sum of $\tau=1$ TCBs, does it follow that the spinning correlators are also given by a sum of $\tau=1$ TCBS?\\

The higher spin equation that connects $\langle OOOO\rangle$ to $\langle TOOO\rangle$ in the free bosonic theory reads \cite{Jain:2020rmw},
\small
\begin{align}\label{HSE1}
&\bigg(\partial_{1-}^3+\partial_{2-}^3+\partial_{3-}^3+\partial_{4-}^3\bigg)\langle OOOO\rangle\notag\\&+\frac{9}{5}\bigg(\partial_{1-}\langle T_{--}OOO\rangle+\partial_{2-}\langle O T_{--}OO\rangle+\partial_{3-}\langle OOT_{--}O\rangle+\partial_{4-}\langle OOOT_{--}\rangle\bigg)=0,
\end{align}
where we work with the metric $ds^2=dx^{+}dx^{-}+dx^2$ and we have chosen all the free indices in the above equation to be $-$. We know that $\langle OOOO\rangle$ can be can be written as a sum of s,t and u channel $\tau=1$ TCBs \eqref{O1O1O1O1FBasTCB}.  Similarly, let us write $\langle TOOO\rangle$ as a sum of s,t and u channel contributions which at the moment we do not know if they are TCBs or not. \eqref{HSE1} then splits into a sum of s,t and u channel equations. For instance, the s channel equation reads,
\small
\begin{align}\label{HSE1s}
&\bigg(\partial_{1-}^3+\partial_{2-}^3+\partial_{3-}^3+\partial_{4-}^3\bigg)\bigg[\frac{1}{x_{12}^2 x_{34}^2}\bigg(\sqrt{u}+\sqrt{\frac{u}{v}}\bigg)\bigg]\notag\\&+\frac{9}{5}\bigg(\partial_{1-}\langle T_{--}OOO\rangle_s+\partial_{2-}\langle O T_{--}OO\rangle_s+\partial_{3-}\langle OOT_{--}O\rangle_s+\partial_{4-}\langle OOOT_{--}\rangle_s\bigg)=0.
\end{align}
If this equation is satisfied then then the fact that $\langle OOOO\rangle$ is a $\tau=1$ TCB will indeed imply that so is $\langle TOOO\rangle$. Before trying to solve this equation, we first substitute the known answer for $\langle TOOO\rangle_s$ that we computed earlier \eqref{TO1O1O1sTCB} into \eqref{HSE1s}. We find that the s channel HSE \eqref{HSE1s} is not satisfied independently and thus see that the HSE does not obviously imply that $\langle TOOO\rangle$ is a $\tau=1$ TCB.
\section{Degeneracy identities}\label{sec:misc}
While solving for the spinning twist conformal blocks we found the following degeneracies useful:
\begin{align}
    V_{1,23}-\frac{1}{v}V_{1,24}+\frac{u}{v}V_{1,34}=0.
\end{align}
\begin{align}
    \epsilon^{X_2 X_3 X_4 Z_1 Z_2}=-\frac{1}{X_1\cdot X_2}\bigg((Z_1\cdot X_2)\epsilon^{X_1 X_2 X_3 X_4 Z_2}-(X_2\cdot X_4)\epsilon^{X_1 X_2 X_3 Z_1 Z_2}+(X_2\cdot X_3)\epsilon^{X_1 X_2 X_4 Z_1 Z_2}\bigg).
\end{align}
\begin{align}\label{HSEschannel}
     \epsilon^{X_1 X_3 X_4 Z_1 Z_2}=-\frac{1}{X_1\cdot X_2}\bigg((Z_2\cdot X_1)\epsilon^{X_1 X_2 X_3 X_4 Z_1}+(X_1\cdot X_4)\epsilon^{X_1 X_2 X_3 Z_1 Z_2}-(X_1\cdot X_3)\epsilon^{X_1 X_2 X_4 Z_1 Z_2}\bigg).
\end{align}

\section{AdS contact diagrams}\label{sec:AdSContactdiagrams}
If we compute the $d+1$ dimensional AdS contact diagram for $O_{\Delta}$ scalars arising from a bulk $\phi^4$ interaction we get the following result \cite{Heemskerk_2009}:
\begin{align}
    \mathcal{A}_{\text{contact}}\propto \Bar{D}_{\Delta\Delta\Delta\Delta}(z,\Bar{z}),
\end{align}
where,
\small
\begin{align}
    &u=z\Bar{z},v=(1-z)(1-\Bar{z})\notag\\
    &\Bar{D}_{\Delta_1 \Delta_2 \Delta_3 \Delta_4}=\frac{\Gamma(\Delta_1)\Gamma(\Delta_2)\Gamma(\Delta_3)\Gamma(\Delta_4)}{\pi^h}\frac{x_{13}^{2(h-\Delta_4)}x_{24}^{2\Delta_2}}{x_{14}^{2(h-\Delta_1-\Delta_4)}x_{34}^{2(h-\Delta_3-\Delta_4)}}\int \frac{d^d x}{(x-x_1)^{2\Delta_1}(x-x_2)^{2\Delta_2}(x-x_3)^{2\Delta_3}(x-x_4)^{2\Delta_4}},
\end{align}
with $h=\frac{d}{2}$.
\normalsize
For instance for $d=3$ and $\Delta=1$ we get,
\begin{align}
    \Bar{D}_{1111}=\frac{1}{z-\Bar{z}}\bigg(\log(z \Bar{z})\log\bigg(\frac{1-z}{1-\Bar{z}}\bigg)+2 Li(z)-2 Li(\Bar{z})\bigg).
\end{align}
Contact diagrams for spinning correlators only get more complicated and in general have singularities of the form $\frac{1}{z-\Bar{z}}$, see section 3.3 in \cite{Silva_2021} for instance. For instance one of the bulk contact diagrams for $\langle TO_2O_2O_2\rangle$ is roughly given by,
\begin{align}
    \mathcal{A}_{\text{contact}}\propto D_{12}D_{13}(\Bar{D}_{3332}),
\end{align}
where the $D_{ij}$ are given in appendix \ref{sec:weightshiftingspinraisingoperators}. As we can see, they do not admit a simple closed form solution and even if they do, will be extremely complicated functions.
\section{OPE analysis}\label{sec:OPE}
The leading term due to the exchange of an operator $O_{\mu_1\cdots \mu_J}^{(\Delta,J)}$ with scaling dimension $\Delta$ and spin $J$ in the limit as $x_{12}\to 0$ in the OPE of two scalars $O_{\Delta_1}(x_1)$ and $O_{\Delta_2}(x_2)$ is given by,
\begin{align}\label{OOope}
    O_{\Delta_1}(x_1)O_{\Delta_2}(x_2)\sim(x_{12}^2)^{\frac{\sigma-\sigma_1-\sigma_2}{2}}\frac{x_{12}^{\mu_1}\cdots x_{12}^{\mu_J}}{(x_{12}^2)^{J}}O_{\mu_1\cdots \mu_J}^{(\Delta,J)}(x_2)+\cdots,
\end{align}
where $\sigma=\Delta+s$ as was introduced in subsection \ref{sec:spinningTCB}.  For example, if we have identical scalars with scaling dimension $\Delta_O$ which exchange conserved currents ($\sigma=2J+1$) we find,
\begin{align}
    O_{\Delta_O}(x_1)O_{\Delta_O}(x_2)\propto (x_{12}^2)^\frac{1-2\Delta_O}{2}.
\end{align}

If instead, we have a spinning operator and a scalar then we have,
\begin{align}\label{JsOope}
    J_{\mu_1\cdots \mu_{s_1}}(x_1)O_{\Delta_O}(x_2)\sim (x_{12}^2)^{\frac{\sigma-\sigma_1-\sigma_2}{2}}\frac{x_{12\mu_1}\cdots x_{12\mu_{s_1}}x_{12}^{\nu_1}\cdots x_{12}^{\nu_J}}{(x_{12}^2)^{J}}O_{\nu_1\cdots \nu_J}(x_2).
\end{align}
For instance consider $s=2$ and $\Delta_s=3$ (the stress tensor) with the exchanged operators as conserved currents. We have,
\begin{align}
    T_{\mu_1 \mu_2}(x_1)O_{\Delta_O}(x_2)\propto (x_{12}^2)^{\frac{-4-\Delta_O}{2}}.
\end{align}
Moving on to the OPE of two spinning operators we have the following behaviour as $x_{12}\to 0$,
\begin{align}\label{JsJsope}
    J_{\mu_1\cdots \mu_{s_1}}(x_1)J_{\nu_1\cdots \nu_{s_2}}(x_2)\sim (x_{12}^2)^{\frac{\sigma-\sigma_1-\sigma_2}{2}}\frac{x_{12\mu_1}x_{12\nu_1}\cdots x_{12\mu_{s_1}}x_{12\nu_{s_2}}x_{12\rho_1}\cdots x_{12\rho_J}}{(x_{12}^2)^{J}}O^{\rho_1\cdots \rho_J}(x_2).
\end{align}
For example consider the two external operators to both be conserved spin one currents and the exchanges to also be conserved currents. In that case we obtain,
\begin{align}
    J_{\mu_1}(x_1)J_{\nu_1}(x_2)\propto (x_{12}^2)^{\frac{-5}{2}}.
\end{align}
\section{The Double Discontinuity}\label{sec:dDisc}
Consider an arbitrary correlator (spinning or scalar), $\langle J_{s_1}J_{s_2}J_{s_3}J_{s_4}\rangle$. It's double discontinuity is simply the following double commutator,
\begin{align}\label{dDiscDoubleCommutator}
    &\mathbf{dDisc_s}(\langle J_{s_1}J_{s_2}J_{s_3}J_{s_4}\rangle)=-\frac{1}{2}\langle [J_{s_1},J_{s_2}][J_{s_3},J_{s_4}]\rangle\notag\\&~~~~~~~~~~~~~~~~~~~~~~~~~~~~~~~=-\frac{1}{2}\bigg(\langle J_{s_1}J_{s_2}J_{s_3}J_{s_4}\rangle-\langle J_{s_1}J_{s_2}J_{s_4}J_{s_3}\rangle-\langle J_{s_2}J_{s_1}J_{s_3}J_{s_4}\rangle+\langle J_{s_2}J_{s_1}J_{s_4}J_{s_3}\rangle\bigg).
\end{align}
where we had to Wick rotate from the Euclidean to Lorentzian signature to obtain the commutators.
The different orderings are characterized by the phase picked up by time-like separated distances \footnote{Please refer to say, section 3 in David Duffins' lecture notes on Dispersive Functionals available on his \href{http://theory.caltech.edu/~dsd/.}{Caltech home page} for more details.}and are obtained from the Euclidean correlator as follows (The subscript $E$ is for Euclidean),
\begin{align}
    &\langle J_{s_1}J_{s_2}J_{s_3}J_{s_4}\rangle=\langle  J_{s_1}J_{s_2}J_{s_3}J_{s_4}\rangle_{E}\bigg(x_{12}\to e^{\frac{i\pi}{2}}(-x_{12}^2)^{\frac{1}{2}},x_{34}\to e^{\frac{i\pi}{2}}(-x_{34}^2)^{\frac{1}{2}}\bigg),\notag\\
     &\langle J_{s_1}J_{s_2}J_{s_4}J_{s_3}\rangle=\langle  J_{s_1}J_{s_2}J_{s_3}J_{s_4}\rangle_{E}\bigg(x_{12}\to e^{\frac{i\pi}{2}}(-x_{12}^2)^{\frac{1}{2}},x_{34}\to e^{\frac{-i\pi}{2}}(-x_{34}^2)^{\frac{1}{2}}\bigg),\notag\\
      &\langle J_{s_2}J_{s_1}J_{s_3}J_{s_4}\rangle=\langle  J_{s_1}J_{s_2}J_{s_3}J_{s_4}\rangle_{E}\bigg(x_{12}\to e^{\frac{-i\pi}{2}}(-x_{12}^2)^{\frac{1}{2}},x_{34}\to e^{\frac{i\pi}{2}}(-x_{34}^2)^{\frac{1}{2}}\bigg),\notag\\
       &\langle J_{s_2}J_{s_1}J_{s_4}J_{s_3}\rangle=\langle  J_{s_1}J_{s_2}J_{s_3}J_{s_4}\rangle_{E}\bigg(x_{12}\to e^{-\frac{i\pi}{2}}(-x_{12}^2)^{\frac{1}{2}},x_{34}\to e^{\frac{-i\pi}{2}}(-x_{34}^2)^{\frac{1}{2}}\bigg).
\end{align}
One can also obtain analogous formulae in the t and u channels by $(2\leftrightarrow 4)$ and $(2\leftrightarrow 3)$ exchanges.
We will now compute the contribution to the double discontinuity of spinning correlators from generic operator exchanges. Let us first review the identical scalar case for pedagogical reasons.
\subsection*{External identical scalar operators}
Consider the correlator $\langle O_{\Delta_O}(x_1)O_{\Delta_O}(x_2)O_{\Delta_O}(x_3)O_{\Delta_O}(x_4)\rangle $. The s channel contribution to the correlator due to the exchange of an operator $O^{(\Delta)}_{\mu_1\cdots \mu_J}$ with spin $J$ and scaling dimension $\Delta$ is given by,
\begin{align}\label{idscalarsJdeltaex}
    (x_{12}^2)^{\frac{\Delta-J-2\Delta_O}{2}}(x_{34}^2)^{\frac{\Delta-J-2\Delta_O}{2}}x_{12}^{\mu_1}\cdots x_{12}^{\mu_J}x_{34}^{\nu_1}\cdots x_{34}^{\nu_J}\langle O^{(\Delta)}_{\mu_1\cdots \mu_J}(x_2)O^{(\Delta)}_{\nu_1\cdots \nu_J}(x_4)\rangle,
\end{align}
where we performed the OPE between $O_{\Delta_O}(x_1)$ and $O_{\Delta_O}(x_2)$ and also $O_{\Delta_O}(x_3)$ and $O_{\Delta_O}(x_4)$.
\begin{align}
    &O_{\Delta_O}(x_1)O_{\Delta_O}(x_2)\sim (x_{12}^2)^{\frac{\Delta-J-2\Delta_O}{2}}x_{12}^{\mu_1}\cdots x_{12}^{\mu_J}O^{(\Delta)}_{\mu_1\cdots \mu_J}(x_2)+\cdots\notag\\
     &O_{\Delta_O}(x_3)O_{\Delta_O}(x_4)\sim (x_{34}^2)^{\frac{\Delta-J-2\Delta_O}{2}}x_{34}^{\nu_1}\cdots x_{34}^{\nu_J}O^{(\Delta)}_{\nu_1\cdots \nu_J}(x_4)+\cdots.
\end{align}
Let us now compute the s channel double discontinuity due to \eqref{idscalarsJdeltaex}. Using \eqref{dDiscDoubleCommutator}, it is clear that it suffices to look only at the pre-factor in \eqref{idscalarsJdeltaex} as the rest of the terms do not contribute to the s channel double discontinuity. We see that,
\begin{align}\label{dDiscsOOOO}
    \mathbf{dDisc_S}\bigg( (x_{12}^2)^{\frac{\Delta-J-2\Delta_O}{2}}(x_{34}^2)^{\frac{\Delta-J-2\Delta_O}{2}}\bigg)=2\sin^2\bigg(\frac{\pi}{2}(\Delta-J-2\Delta_O)\bigg)(-x_{12}^2)^{\frac{\Delta-J-2\Delta_O}{2}}(-x_{34}^2)^{\frac{\Delta-J-2\Delta_O}{2}}.
\end{align}
For single trace contributions, say from conserved currents, we have $\Delta=J+1$ and \eqref{dDiscsOOOO} becomes,
\begin{align}\label{scalartau1dDisc1}
    2\cos^2(\pi\Delta_O)(-x_{12}^2)^{\frac{1-2\Delta_O}{2}}(-x_{34}^2)^{\frac{1-2\Delta_O}{2}},
\end{align}
which is non-zero for integer $\Delta_O$.\\
For double trace contributions of the form $O_{\Delta_O}\Box^n\partial_{\mu_1}\cdots\partial_{\mu_s}O_{\Delta_O}$ we have $\Delta=2\Delta_O+2n+J$ and hence their contribution to the double discontinuity is proportional to,
\begin{align}\label{scalartaudtdDisc1}
    2\sin^2(n\pi)=0,
\end{align}
since $n$ is an integer. Similarly, one can also check that the exchange of an arbitrary double trace operator schematically of the form $J_{s_1}\Box^n \partial\cdots\partial J_{s_2}$ where $J_{s_1}$ and $J_{s_2}$ are conserved currents does not contribute to the double discontinuity.
\subsection*{External non-identical scalar operators}
Consider the correlator $\langle O_{\Delta_1}(x_1)O_{\Delta_2}(x_2)O_{\Delta_3}(x_3)O_{\Delta_4}(x_4)\rangle $. The s channel contribution to the correlator due to the exchange of an operator $O^{(\Delta)}_{\mu_1\cdots \mu_J}$ with spin $J$ and scaling dimension $\Delta$ is given by,
\begin{align}\label{nidscalarsJdeltaex}
    (x_{12}^2)^{\frac{\Delta-J-\Delta_1-\Delta_2}{2}}(x_{34}^2)^{\frac{\Delta-J-\Delta_3-\Delta_4}{2}}x_{12}^{\mu_1}\cdots x_{12}^{\mu_J}x_{34}^{\nu_1}\cdots x_{34}^{\nu_J}\langle O^{(\Delta)}_{\mu_1\cdots \mu_J}(x_2)O^{(\Delta)}_{\nu_1\cdots \nu_J}(x_4)\rangle,
\end{align}
where we performed the OPE between $O_{\Delta_O}(x_1)$ and $O_{\Delta_O}(x_2)$ and also $O_{\Delta_O}(x_3)$ and $O_{\Delta_O}(x_4)$.
\begin{align}
    &O_{\Delta_1}(x_1)O_{\Delta_2}(x_2)\sim (x_{12}^2)^{\frac{\Delta-J-\Delta_1-\Delta_2}{2}}x_{12}^{\mu_1}\cdots x_{12}^{\mu_J}O^{(\Delta)}_{\mu_1\cdots \mu_J}(x_2)+\cdots\notag\\
     &O_{\Delta_3}(x_3)O_{\Delta_4}(x_4)\sim (x_{34}^2)^{\frac{\Delta-J-\Delta_3-\Delta_4}{2}}x_{34}^{\nu_1}\cdots x_{34}^{\nu_J}O^{(\Delta)}_{\nu_1\cdots \nu_J}(x_4)+\cdots.
\end{align}
Let us now compute the s channel double discontinuity due to \eqref{nidscalarsJdeltaex}. Using \eqref{dDiscDoubleCommutator}, it is clear that it suffices to look only at the pre-factor in \eqref{nidscalarsJdeltaex} as the rest of the terms do not contribute to the s channel double discontinuity. We see that,
\begin{align}\label{dDiscsOOOOn}
    \mathbf{dDisc_s}\bigg( (x_{12}^2)^{\frac{\Delta-J-\Delta_1-\Delta_2}{2}}(x_{34}^2)^{\frac{\Delta-J-\Delta_3-\Delta_4}{2}}\bigg)\propto 2\sin\bigg(\frac{\pi}{2}(\Delta-J-\Delta_1-\Delta_2)\bigg)\sin\bigg(\frac{\pi}{2}(\Delta-J-\Delta_3-\Delta_4)\bigg).
\end{align}
For single trace contributions, say from conserved currents, we have $\Delta=J+1$ and \eqref{dDiscsOOOOn} becomes proportional to,
\begin{align}
   2\cos\bigg(\frac{\pi}{2}(\Delta_1+\Delta_2)\bigg)\cos\bigg(\frac{\pi}{2}(\Delta_3+\Delta_4)\bigg),
\end{align}
which is non-zero when $\Delta_1+\Delta_2$ and $\Delta_3+\Delta_4$ are even. However when they are odd, this contribution vanishes. Thus in contrast to the identical scalar case, single trace contributions can have zero double discontinuity for correlators of non identical scalars, which is something we shall also see for the spinning case which we now turn to.
\subsection*{$\mathbf{\langle TO_{\Delta_O}O_{\Delta_O}O_{\Delta_O}\rangle}$}
The s channel contribution to the correlator due to the exchange of an operator $O^{(\Delta)}_{\mu_1\cdots \mu_J}$ with spin $J$ and scaling dimension $\Delta$ is given by,
\begin{align}\label{TOOOcontributiondDisc}
    (x_{12}^2)^{\frac{\Delta-J-\Delta_O-5}{2}}(x_{34}^2)^{\frac{\Delta-J-2\Delta_O}{2}}x_{12}^{\mu}x_{12}^{\nu}x_{12}^{\mu_1}\cdots x_{12}^{\mu_J}x_{34}^{\nu_1}\cdots x_{34}^{\nu_J}\langle O^{(\Delta)}_{\mu_1\cdots \mu_J}(x_2)O^{(\Delta)}_{\nu_1\cdots \nu_J}(x_4)\rangle.
\end{align}
To compute the contribution of this exchange to the double discontinuity, one just needs to focus on the pre-factor just as in the scalar case. Using \eqref{dDiscDoubleCommutator}, we find that the contribution of \eqref{TOOOcontributiondDisc} to the double discontinuity is equivalent to,
\begin{align}
        &\mathbf{dDisc_s}\bigg( (x_{12}^2)^{\frac{\Delta-J-\Delta_O-5}{2}}(x_{34}^2)^{\frac{\Delta-J-2\Delta_O}{2}}\bigg)\notag\\&=\Bigg(\sin(\frac{\pi\Delta_O}{2})+\sin((J-\Delta+\frac{3\Delta_O}{2})\pi)\Bigg)(-x_{12}^2)^{\frac{\Delta-J-\Delta_O-5}{2}}(-x_{34}^2)^{\frac{\Delta-J-2\Delta_O}{2}}.
\end{align}
Let us now consider the exchange of single trace operators (such as the conserved currents) with $\Delta=J+1$. The double discontinuity of such contributions equals,
\begin{align}\label{TOOOrtau1dDisc1}
    \bigg(\sin(\frac{\pi\Delta_O}{2})-\sin(\frac{3\pi\Delta_O}{2})\bigg)(-x_{12}^2)^{\frac{-\Delta_O-4}{2}}(-x_{34}^2)^{\frac{1-2\Delta_O}{2}}.
\end{align}
For $\Delta_O=1$ this quantity equals $\frac{2}{(-x_{12}^2)^{\frac{5}{2}}(-x_{34}^2)^{\frac{1}{2}}}$. However for $\Delta_O=2$, we find that it is identically $0$.

Double trace operators of the form $O_{\Delta_O}\Box^n\partial_{\mu_1}\cdots\partial_{\mu_J}O_{\Delta_O}$ which have spin $J$ and $\Delta=2\Delta_O+2n+J$ contribute to the double discontinuity as follows,
\begin{align}\label{OOdoubletraceTOOO}
\bigg(\sin(\frac{\pi\Delta_O}{2})-\sin(\frac{\pi(4n+\Delta_O)}{2})\bigg)(-x_{12}^2)^{\frac{2n+\Delta_O-5}{2}}(-x_{34}^2)^{2n},
\end{align}
which for $\Delta_O=1$ and $\Delta_O=2$ clearly vanishes. If we consider other double trace operators such as $T^{\mu_1 \mu_2}\Box^n \partial_{\mu_3}\cdots\partial_{\mu_{J-2}}O_{\Delta_O}$ or $T^{\mu_1\mu_2}\Box^n \partial_{\mu_3}\cdots \partial_{\mu_{J-2}}T_{\mu_{J-1}\mu_{J}}$ and compute their contribution to the double discontinuity we find that they too vanish.\\

Thus to summarize: For $\Delta_O=1$, the single trace contributions have non-zero double discontinuity whereas the double trace contributions have zero double discontinuity. However for $\Delta_O=2$, both the single and double trace contributions have zero double discontinuity.\\

Therefore, if we consider $\langle TO_1O_1O_1\rangle$ in the free bosonic theory, then in a given channel, say the s channel, it can be split into a sum of single trace and double trace contributions characterized by the fact that the double discontinuity of the full correlator entirely comes from the former. However, if we consider $\langle TO_2O_2O_2\rangle$ in the critical bosonic theory, then such a split is not possible as both single trace and double trace contributions have zero double discontinuity. A similar analysis can also be carried out for higher spin correlators such as $\langle J_4 O_{\Delta}O_{\Delta}O_{\Delta}\rangle$. We find exactly the same conclusion as we did for $\langle TO_{\Delta}O_{\Delta}O_{\Delta}\rangle$, that is that for $\Delta_O=1$, only single trace operators contribute to non zero double discontinuity but for $\Delta_O=2$, both single trace and double trace do not contribute to the double discontinuity at all. 
\section{Correlators in the critical bosonic theory}\label{sec: CBcorrelatorsappendix}
\subsection*{$\mathbf{\langle J_{s_1}O_2O_2O_2\rangle}$}
Let us now try to abstractly analyze the results of the $s_1=2,4$ cases (please see \eqref{TO2O2O2CBcorrelator} and \eqref{J4OOOCB} respectively) and try to generalize it to the case where the spin of the first operator is arbitrary.

We have (up to contact terms)
\begin{align}\label{Js1O2O2O2CB}
    \langle J_{s_1}(x_1)O_2(x_2)O_2(x_3)O_2(x_4)\rangle_{CB}=\int\frac{d^3 x_5 d^3 x_6 d^3 x_7}{x_{25}^4 x_{36}^4 x_{47}^4}\langle J_{s_1}(x_1)O_1(x_5)O_1(x_6)O_1(x_7)\rangle_{FB}
\end{align}
Decomposing both sides of this equation into s,t and u channel expressions and applying Alday's operator on say, the s channel \footnote{We use the fact that applying $\mathcal{A}_{12}$ and $\mathcal{A}_{34}$ on s channel TCBs should yield identical results.} yields,
\begin{align}
      &\mathcal{A}_{34}\langle J_{s_1}(x_1)O_2(x_2)O_2(x_3)O_2(x_4)\rangle_{CB,s}\notag\\&=\int\frac{d^3 x_5 d^3 x_6 d^3 x_7}{x_{25}^4 }\bigg(\frac{-12 x_{37}^2 x_{46}^2}{x_{36}^3 x_{47}^3}+\frac{6}{x_{36}^2 x_{47}^2}-\frac{24 x_{34}^2 x_{67}^2}{x_{36}^3 x_{47}^3}\bigg)\langle J_{s_1}(x_1)O_1(x_5)O_1(x_6)O_1(x_7)\rangle_{FB,s}
\end{align}
which is a non zero quantity. To be absolutely sure whether this is zero or not, one must explicitly carry out the integration which in general is a very difficult task. However, based on our experiences with the cases $s=2$ and $s=4$, it is highly likely that for general $s$ that this correlator is not a $\tau=1$ twist conformal block.
\subsection*{$\mathbf{\langle JJO_2O_2\rangle}$}
We have\footnote{For general spins $s_1$ and $s_2$ we have, 
\begin{align}
    &\langle J_{s_1}(x_1)J_{s_2}(x_2)O_2(x_3)O_2(x_4)\rangle_{CB}=\int \frac{d^3 x_5 d^3 x_6}{x_{25}^4 x_{36}^4}\langle J_{s_1}(x_1)J_{s_2}(x_2)O_1(x_3)O_1(x_4)\rangle_{FB}\notag\\
    &-\frac{1}{N}\bigg[\int \frac{d^3 x_5 d^3 x_6}{x_{56}^2}\langle J_{s_1}(x_1)O_2(x_4)O_2(x_5)\rangle_{CB}\langle J_{s_2}(x_2)O_2(x_3)O_2(x_6)\rangle_{CB}\bigg]
\end{align}
For $s_1=s_2=1$, $\langle JO_2O_2\rangle=0$ so the second line, which is the $O_2$ conformal block contribution, vanishes. However, for even spins such as $s_1=s_2=2$, we do have such a contribution. Since it comes from the $O_2$ exchange it is a $\tau=2$ TCB as can also easily be checked by substituting the explicit forms of the three point functions.},
\begin{align}\label{Js1Js2O2O2CB}
    &\langle J(x_1)J(x_2)O_2(x_3)O_2(x_4)\rangle_{CB}=\int \frac{d^3 x_5 d^3 x_6}{x_{35}^4 x_{46}^4}\langle J(x_1)J(x_2)O_1(x_3)O_1(x_4)\rangle_{FB}
\end{align}
We have already seen that the free bosonic $\langle JJO_1O_1\rangle_{FB}$ can be written as a sum of t and u channel $\tau=1$ TCBs \eqref{JJO1O1asTCB}. Let us see if such a decomposition exists for the CB correlator. Using \eqref{JJO1O1asTCB} we re-write \eqref{Js1Js2O2O2CB} as follows:
\small
\begin{align}\label{JJOOCBastu}
    &\langle J(x_1)J(x_2)O_2(x_3)O_2(x_4)\rangle_{CB}\notag\\&=\frac{1}{2}\int \frac{d^3 x_5 d^3 x_6}{x_{35}^4 x_{46}^4}\bigg(\mathcal{H}_t^{(JO_1|1|JO_1)}(x_1,z_1,x_2,z_2,x_5,x_6)+\mathcal{H}_u^{(JO_1|1|JO_1)}(x_1,z_1,x_2,z_2,x_5,x_6)\bigg)\notag\\&=\frac{1}{2}\bigg[\langle J(x_1)J(x_2)O_2(x_3)O_2(x_4)\rangle_{CB,t}+\langle J(x_1)J(x_2)O_2(x_3)O_2(x_4)\rangle_{CB,u}\bigg]
\end{align}
where we have identified the t channel part of the expression,
\begin{align}\label{JJOOCBtchannel}
   &\langle J(x_1)J(x_2)O_2(x_3)O_2(x_4)\rangle_{CB,t}= \int \frac{d^3 x_5 d^3 x_6}{x_{35}^4 x_{46}^4}\mathcal{H}_t^{(JO_1|1|JO_1)}(x_1,z_1,x_2,z_2,x_5,x_6)
\end{align}
and $\langle J(x_1)J(x_2)O_2(x_3)O_2(x_4)\rangle_{CB,u}=\langle J(x_1)J(x_2)O_2(x_3)O_2(x_4)\rangle_{CB,t}(3\leftrightarrow 4)$. We then apply Alday's t channel operator $\mathcal{A}_{14}$ on \eqref{JJOOCBtchannel}. By using the explicit form of the free bosonic t channel TCB $\mathcal{H}_t^{(JO_1|1|JO_1)}$ \eqref{JJO1O1tTCB}, we see that the result is not a $\tau=1$ TCB at least at the integrand level. Similarly, we find that the u channel critical bosonic expression is also not a $\tau=1$ TCB.
Thus we see that $\langle JJO_2O_2\rangle_{CB}$ is not entirely given by a sum of $\tau=1$ TCBs. This result can also be generalized to $\langle J_{s_1}J_{s_2}O_2O_2\rangle$.
\section{Twist Conformal Blocks vs Bulk Amplitudes and Euclidean bulk locality}\label{TCBvsAmplitudes}
In this appendix, we briefly discuss the differences between boundary twist conformal blocks and bulk amplitudes. More details can be found in \cite{Hijano_2016}.\\
There are two types of AdS isometry respecting diagrams that we can draw. These are the traditional Witten diagrams and the geodesic Witten diagrams.  The difference between them is that for the former, the
vertices are integrated over all of AdS whereas for the latter, the integration is only over the geodesics connecting the two pairs of boundary points. For the higher spin theories that are of interest to us we have the following equivalences. First, boundary CFT single trace twist conformal blocks of a fixed twist $\tau$, $\mathcal{H}^{\tau}$ are equal to a sum over all geodesic Witten diagrams with twist $\tau$ exchanges $\mathcal{W}^{\tau}$. For instance in the s channel we have,
\begin{align}
    \mathcal{H}^{\tau}_s=\mathcal{W}^{\tau}_s
\end{align}
Witten diagrams (denoted by $\mathcal{A}$) on the other hand are given by,
\begin{align}
    &\mathcal{A}_s\sim\mathcal{H}_s^{\tau}+\text{s channel double trace contributions}\notag\\
    &\mathcal{A}_t,\mathcal{A}_u,\mathcal{A}_{\text{contact}}\sim\text{s channel double trace contributions}
\end{align}
We can write down similar equations with the t or u channel decompositions on the CFT side, for instance,
\begin{align}
    &\mathcal{A}_t\sim\mathcal{H}_t^{\tau}+\text{t channel double trace contributions}\notag\\
    &\mathcal{A}_s,\mathcal{A}_u,\mathcal{A}_{\text{contact}}\sim\text{t channel double trace contributions}
\end{align}
Using these results, let us now briefly discuss the issue of locality of the higher spin theories following \cite{Neiman:2023orj}. It was shown that for Euclidean AdS that just identifying $\mathcal{A}_i$ as $\mathcal{H}_{i}$ with some double trace dressing, $i\in \{s,t,u\}$ and then concluding that the bulk amplitudes are non local is not the way to proceed. By an explicit bulk analysis, it was shown that the bulk s channel exchange diagram's singularity is absent in the t,u exchange diagrams and contact diagrams thus showing that the Euclidean AdS higher spin theory is local at distances greater than $R_{AdS}$. This analysis was for the identical scalar case. We expect a similar story to hold for the spinning correlators that we computed when taken as Euclidean AdS correlators but we defer such an analysis to a future work.

\bibliographystyle{JHEP}
\bibliography{biblio}
\end{document}